**Title**

AN OUTER PLANET BEYOND PLUTO AND ORIGIN OF THE TRANS-NEPTUNIAN BELT ARCHITECTURE

**Proposed Running Head**

AN OUTER PLANET BEYOND PLUTO

**Authors**

Patryk S. Lykawka[*] and Tadashi Mukai

Kobe University, Graduate School of Science - Earth and Planetary Sciences, 1-1 rokkodai-cho, nada-ku, Kobe, 657-8501, Japan

[*]Corresponding Author e-mail address: patryk@dragon.kobe-u.ac.jp

Pages total: 80
Figures: 24
Tables: 7

**Postal Address**

Patryk Sofia Lykawka
Kobe University
Graduate School of Science
Department of Earth and Planetary Sciences
1-1 rokkodai-cho, nada-ku, Kobe
657-8501
Japan
Phone number: +81 (0)78 803-6446
FAX: +81 (0)78 803-6446
E-mail: patryk@dragon.kobe-u.ac.jp

**Abstract**

Trans-Neptunian objects (TNOs) are remnants of a collisionally and dynamically evolved planetesimal disk in the outer solar system. This complex structure, known as the trans-Neptunian belt (or Edgeworth-Kuiper belt), can reveal important clues about disk properties, planet formation, and other evolutionary processes. In contrast to the predictions of accretion theory, TNOs exhibit surprisingly large eccentricities, $e$, and inclinations, $i$, which can be grouped into distinct dynamical classes. Several models have addressed the origin and orbital evolution of TNOs, but none have reproduced detailed observations, e.g., all dynamical classes and peculiar objects, or provided insightful predictions. Based on extensive simulations of planetesimal disks with the presence of the four giant planets and massive planetesimals, we propose that the orbital history of an outer planet with tenths of Earth's mass can explain the trans-Neptunian belt orbital structure. This massive body was likely scattered by one of the giant planets, which then stirred the primordial planetesimal disk to the levels observed at 40–50 AU and truncated it at about 48 AU before planet migration. The outer planet later acquired an inclined stable orbit ($\geq$100 AU; 20–40°) because of a resonant interaction with Neptune (an $r$:1 or $r$:2 resonance possibly coupled with the Kozai mechanism), guaranteeing the stability of the trans-Neptunian belt. Our model consistently reproduces the main features of each dynamical class with unprecedented detail; it also satisfies other constraints such as the current small total mass of the trans-Neptunian belt and Neptune's current orbit at 30.1 AU. We also provide observationally testable predictions.

**Subject Headings:** Kuiper Belt; solar system: formation; minor planets, asteroids; methods: N-body simulations

# 1. INTRODUCTION

Based on anomalies in the motion of Uranus and Neptune (which features were spurious, however; Standish 1993), Percival Lowell predicted the existence of a massive planet beyond Neptune ("Planet X"), the search for which led to the accidental discovery of (134340) Pluto in 1930 orbiting at a semimajor axis, $a$, 39.4 AU. Nevertheless, it was soon realized that Pluto was several orders of magnitude less massive than Planet X (Duncombe et al. 1968; Delsanti & Jewitt 2006 and references therein). In addition, Pluto was also too small to account for the outer solar system mass distribution. That is, considered in terms of surface density, σ, the mass distribution decreases with heliocentric distance, $R$, in a way that it can be described by a power law with the exponent $-3/2 \pm 1/2$ (Hayashi et al. 1985; Morbidelli & Brown 2004; Jewitt 2006). The extrapolation of this distribution to trans-Neptunian distances allowed the prediction of several Earth masses ($M_\oplus$) beyond Neptune. This motivated researchers to postulate the existence of a massive trans-Neptunian belt composed of a "large" number of small icy trans-Neptunian objects (TNOs) (Leonard 1930; Edgeworth 1949; Kuiper 1951). Because of this pioneer work, the trans-Neptunian belt is also commonly referred as the "Kuiper belt" or "Edgeworth-Kuiper belt." We use the most general nomenclature in this paper.

Although its discovery showed that the region beyond Neptune was not empty, Pluto was recognized as a planet and not a member of the hypothesized trans-Neptunian belt. The discovery of (15760) 1992 QB$_1$ in 1992 (Jewitt & Luu 1993) and subsequent discoveries of TNOs (Williams 1997; Jewitt 1999) confirmed the existence of the trans-Neptunian belt and also put Pluto into context as just one of the largest members of the belt. Because TNOs have had insufficient time or surface density to form large planets, it is now widely accepted that these icy bodies constitute the remnants of the primordial planetesimal disk, which represents a fingerprint of the planet formation era. TNOs thus offer several clues about the dynamical, collisional, and thermal evolution of the solar system over billions of years (Luu & Jewitt 2002). Furthermore, TNOs are also linked to other solar system minor body populations, namely the short-period comets (SPCs) and Centaurs (Fernández 1980; Duncan et al. 1988; Holman & Wisdom 1993; Duncan & Levison 1997; Horner et al. 2003; Fernández et al. 2004; Emel'yanenko et al. 2005).

As of June 2007, more than 1100 TNOs with diameters ($D$) typically larger than 100 km have been observed.[1] This sample represents only 1–2% of the expected population within the same size range (Trujillo et al. 2001a; Sheppard 2006). More than 700 TNOs have been observed during two or more oppositions; thus in principle, these bodies possess more reliable orbital elements (e.g., with small uncertainties). However, observations suffer from several biases, and TNO orbital distributions should be accepted with caution. Trans-Neptunian belt members exhibit near-circular to large eccentricity ($e$) orbits within $a \approx 48$ AU, whereas TNOs in the scattered disk reservoir of the belt have typically large eccentricities (usually set at $a > 48$ AU) (Duncan & Levison 1997; Luu et al. 1997; Morbidelli et al. 2004). In general, TNOs also present a wide inclination distribution, reaching up to inclination, $i \approx 48°$ (Fig. 1). Further details about the characterization of TNOs in element space is given in Morbidelli & Brown (2004), Gladman et al. (2007) and Lykawka & Mukai (2007b) (hereafter, LM07b).

---

[1] Orbital elements of TNOs are available at public domain databases:
Lowell Observatory, *ftp://ftp.lowell.edu/pub/elgb/astorb.html*
Minor Planet Center, *http://cfa-www.harvard.edu/iau/TheIndex.html*

The complex orbital structures of TNOs have revealed distinct dynamical classes: the classical, resonant, scattered, and detached TNOs (e.g., LM07b and references therein). Classical TNOs are nonresonant objects that orbit around 37 AU < $a$ < 48 AU (the classical region). Curiously, these bodies represent the superposition of two different subclasses, the cold and hot populations, which are defined by classical bodies with $i$ < 5° and $i$ > 5°, respectively (Morbidelli & Brown 2004 and references therein; LM07b; Chiang et al. 2007). This division is supported by studies of the distributions of their colors, sizes, inclinations, and dynamical origin/evolution (Brown 2001; Levison & Stern 2001; Doressoundiram et al. 2002; Hainaut & Delsanti 2002; Trujillo & Brown 2002; Gomes 2003b; Bernstein et al. 2004; Morbidelli & Brown 2004; Peixinho et al. 2004). Resonant TNOs are currently locked in resonances with Neptune, which we take to refer to any external mean motion resonance with Neptune described by $r$:$s$, where $r$ and $s$ are integers. Resonant populations have been observed at distances from 30 AU (Neptune Trojans) to almost 108 AU (27:4 resonance). Some resonances are notably populated, particularly the 3:2 ($a$ = 39.4 AU), 5:3 ($a$ = 42.3 AU), 7:4 ($a$ = 43.7 AU), 2:1 ($a$ = 47.8 AU), and 5:2 ($a$ = 55.4 AU) resonances (Fig. 1). Pluto is one of the 3:2 resonance members. Scattered TNOs are objects associated with gravitational scattering by Neptune (Duncan & Levison 1997). These bodies have no particular boundaries in the semimajor axis, but usually possess perihelion distances close to the giant planet, $q$ < 37–40 AU. A notable member of the scattered population is (136199) Eris, the largest and most massive TNO known thus far (Bertoldi et al. 2006; Brown et al. 2006a; Brown & Schaller 2007). Finally, detached TNOs are nonclassical objects that do not encounter Neptune; thus, they appear to be "detached" from the solar system. In fact, orbital integrations have shown that these bodies evolve in quasi-static orbits over 4–5 Gyr (Gladman et al. 2002; LM07b). Here, we consider TNOs with $a$ > 48 AU (i.e., beyond the 2:1 resonance) and $q$ > 40 AU as part of the detached population (Fig. 2). See LM07b for a detailed discussion of the identification of detached TNOs.

Classical and scattered TNOs should be roughly equally populated, whereas resonant TNOs should represent about 10% of the former (Trujillo et al. 2001a). However, the total mass of scattered bodies, in particular, is quite uncertain (Gomes et al. 2007). Moreover, the intrinsic fraction of resonant TNOs is likely to be higher because only the 3:2- and 2:1-resonant populations were taken into account by Trujillo and colleagues. The detached population was also unknown at that time. This means that the contribution of other resonant bodies and detached TNOs might reduce the intrinsic fraction of classical TNOs. Lastly, because of strong observational biases, there must be a substantial population of detached TNOs that could even surpass that of scattered bodies (Gladman et al. 2002; Allen et al. 2006).

According to accretion models, TNOs should have formed only if the primordial planetesimal disk was under very cold orbital conditions (i.e., orbits with $e \approx 0$ and $i \approx 0°$) (Stern & Colwell 1997; Kenyon & Luu 1998; Kenyon & Luu 1999; Kenyon & Bromley 2004a). In contrast, the trans-Neptunian belt is currently in an excited orbital state (i.e., large $e$ and/or $i$). Indeed, it is currently in an erosive regime; that is, the encounter velocities of TNOs are large enough to favor disruptive collisions in the region. How the trans-Neptunian belt acquired its intriguingly complex orbital structure is therefore a subject of great importance (Morbidelli & Brown 2004; Jewitt 2006).

What mechanisms could excite the trans-Neptunian belt? An obvious mechanism is gravitational perturbation by the planets. However, although the excited orbits of scattered TNOs can be well explained by continuous scattering by Neptune, the same is not true for other classes of TNOs (Morbidelli & Brown 2004). Resonance dynamics represents another important mechanism for TNO excitation. In particular, sweeping resonances played an essential role in capturing TNOs from the planetesimal disk during planet migration, an event that is well supported by several lines of evidence (Fernández & Ip 1984; Malhotra 1995; Liou & Malhotra 1997; Hahn & Malhotra 1999; Ida et al. 2000b; Levison & Stewart 2001; Gomes 2003b; Levison et al. 2007). According to adiabatic theory, captured bodies are transported outwards by resonances (i.e., *a* increases) with an increase in eccentricity (Peale 1976; Murray & Dermott 1999; Chiang et al. 2007). Depending on the resonance, a moderate excitation in inclination is also a common outcome. Resonance dynamics could also explain the origin of detached TNOs (Fig. 2). However, although a small fraction of scattered objects can be temporarily detached from the gravitational domain of Neptune through resonant interactions with the giant planet, this mechanism cannot account for the intrinsic total population of detached bodies (Gomes et al. 2005a; Chiang et al. 2007; LM07b).

In conclusion, the existence of distinct classes of TNOs is the outcome of several evolutionary processes that can sculpt the solar system. Some of these ceased long ago (e.g., planetary migration), whereas others are still active such as the gravitational perturbation and resonant effects of the planets. However, the unexplained excited *e*-distribution of cold classical TNOs and the existence of a substantial population of detached TNOs require a sculpting mechanism other than the foregoing ones. Alternative excitation mechanisms include passing stars, large (massive) planetesimals that existed in the past, giant molecular clouds, an unseen planet, and a temporarily eccentric Neptune (Kobayashi & Ida 2001; Brunini & Melita 2002; Morbidelli & Levison 2004; Gladman & Chan 2006; Morbidelli et al. 2007).

We next discuss the main observational constraints and some scenarios that are invoked to explain this complex structure. We introduce our own scenario in Section 3, which is based on the existence of a massive trans-Plutonian planet. In Section 4, we explain the methods used in our model. Sections 5 and 6 are devoted to the preliminary and main simulations of our scenario. The main results of the paper are shown in Section 7. We discuss additional implications of our scenario and the orbital/physical properties of a hypothetical trans-Plutonian planet in sections 8 and 9. We summarize the main achievements and predictions of this study in Section 10. Finally, caveats and future work are given in the last section.

## 2. THE TRANS-NEPTUNIAN BELT ARCHITECTURE AND MAIN SCENARIOS

2.1 Constraints from the Trans-Neptunian Belt
- *Cold and hot classical TNOs with different physical properties*. In general, whereas cold classical TNOs are fainter (i.e., smaller) and exhibit mostly red colors, their counterparts in the hot population are intrinsically brighter (i.e., larger) with a wide variety of colors. Apparent correlations of colors and sizes with inclinations have been found (Tegler & Romanishin 2000; Levison & Stern 2001; Trujillo & Brown 2002; Doressoundiram 2003 and references therein; McBride et al. 2003). Although noteworthy, these properties are not caused by mechanisms related to impacts (Morbidelli & Brown 2004). Distinct color, size, and inclination distributions can be interpreted as evidence that cold and hot classical TNOs formed at distinct places in the primordial planetesimal disk (Gomes 2003b). That is, the cold classical population would have formed in situ, thus representing relics of the planetesimal disk between about 35 AU and the original edge of the disk. The hot population would consist of planetesimals that formed in the inner solar system (15–35 AU) and were later deposited in the classical region via resonant interactions (Gomes 2003b). Intriguingly, the superposition of cold and hot populations is apparent only among classical bodies;

- *Orbital excitation of cold (in eccentricities) and hot (both in eccentricities and inclinations) classical populations*. Classical TNOs present an unexpected excitation in their eccentricities and inclinations (Figs. 1 and 3) that cannot be attributed to gravitational interactions with Neptune (assuming the current solar system architecture) or to mutual planetesimal gravitational stirring (Morbidelli 2005). The excitation of eccentricities also reveals an apparent lack of low-$e$ TNOs beyond about 45 AU (Fig. 3). This remarkable feature is probably not due to observational biases (Morbidelli & Brown 2004; Morbidelli 2005). Moreover, studies of the long-term evolution of classical TNOs have confirmed that the outer skirts of the trans-Neptunian belt are stable at low eccentricities, so the absence of low-$e$ TNOs beyond about 45 AU is unexpected (Holman & Wisdom 1993; Duncan et al. 1995; Kuchner et al. 2002; Lykawka & Mukai 2005c);

- *Resonant TNOs*. Prominent resonant populations that are stable over the age of the solar system are found throughout the entire trans-Neptunian region (Figs. 1–3; Lykawka & Mukai 2007a; LM07b). To account for the population of 5:2-resonant TNOs, Chiang et al. (2003) showed that a disk of planetesimals with initially excited orbits is necessary prior to planet migration. This result was recently confirmed (Hahn & Malhotra 2005). In addition, the origin of long-term resonant bodies in the scattered disk strongly suggests that the ancient trans-Neptunian belt was already excited in both its eccentricities and inclinations when the Neptunian resonances passed through the belt (Fig. 3; Lykawka & Mukai 2007a). A similar condition would also explain the formation of all 3:2-resonant TNOs (Wiegert et al. 2003);

- *Scattered TNOs*. There is a substantial population of TNOs evolving on orbits that suffer scattering by Neptune (Fig. 2). Scattered TNOs are thought to be primordial, and not sustained by TNOs coming from unstable regions of the trans-Neptunian belt (Morbidelli 2005);

- *Detached TNOs* (Fig. 3). This population is intriguing because the sole presence of the giant planets cannot explain their origin (Gomes et al. 2007). This is particularly evident in the case of extreme objects such as (148209) 2000 CR$_{105}$ ($a$ = 224.6 AU; $q$ = 44.1 AU) and (90377) Sedna ($a$ = 525.6 AU; $q$ = 76.2 AU), and low-$i$ detached TNOs. Indeed, the great majority of detached bodies produced by resonances have $i > 30°$ (Lykawka & Mukai 2004; Gomes et al. 2005a; Gallardo 2006a; Lykawka & Mukai 2006). Of all currently identified detached TNOs (nine objects) in LM07b, only one member is in resonance (8:3 resonance). In addition, all members but one have $i < 30°$;

- *Very high-$i$ TNOs* (objects with $i > 40°$; LM07b). There are only three members known thus far, i.e., 2004 DG$_{77}$ ($i$ = 47.6°), Eris ($i$ = 44.0°), and 2004 XR$_{190}$ ($i$ = 46.7°), because severe observational biases are involved. That is, most surveys focus near the ecliptic. Because very high-$i$ TNOs spend an extremely small time of their orbits near the ecliptic, their discovery is discriminated against. For a survey probing a sky region near the ecliptic, the probability of discovering a TNO with $i$

= 40° is approximately 20–50 times smaller than finding one with $i \approx 0$ (Trujillo et al. 2001; see also Jones et al. 2005). The apparent fraction of very high-$i$ TNOs is about 0.5% (3 out of 600–650 TNOs with longer-arc orbits), so we estimate the intrinsic fraction to be roughly ~10–25%. Because the results of simulations predict that 1% of objects would possess $i > 40°$ (LM07b, and references therein), resonances or the sole perturbation of Neptune cannot produce this subclass;

- *The outer edge of the trans-Neptunian belt at 48 AU.* The outer edge is characterized by the absence of near-circular TNOs beyond about 48 AU and the abrupt decrease in the number of TNOs with $R$, revealing a dearth of these objects at ~47–50 AU (Jewitt et al. 1998). Despite the observational capability to detect sufficiently large TNOs beyond 48 AU, several observations did not find them, thus supporting the existence of the edge (Gladman et al. 1998; Jewitt et al. 1998; Allen et al. 2001; Gladman et al. 2001; Trujillo et al. 2001a; Bernstein et al. 2004; Morbidelli & Brown 2004; Larsen et al. 2007). Simulations of the orbital evolution of objects in near-circular orbits around 45–55 AU have shown that this region is stable over the age of the solar system (e.g., Brunini 2002). The 2:1 resonance is also unable to produce the edge (Duncan et al. 1995; Lykawka & Mukai 2005c). Furthermore, a cold thin disk composed of bodies larger than 185 km beyond 48 AU and with a similar size distribution as in the classical region can be ruled out with 95% confidence for any disk with inclination $\leq 1°$ to the invariable plane (Allen et al. 2002). Alternative explanations for the edge such as extreme size/albedo effects (e.g., maximum size/albedo decreases with $R$), drop in $e$-distribution with $R$, and steeper ecliptic-plane surface density variation with $R$ have been ruled out (Trujillo & Brown 2001; Trujillo et al. 2001a). Possible explanations for the edge include external perturbations such as massive planetesimals, passing stars, or UV photoevaporation (Ida et al. 2000a; Brunini & Melita 2002; Adams et al. 2004); inward radial migration of bodies during the assembly of < 1 km-sized planetesimals (Weidenschilling 2003); or the outward transportation of bodies by the 2:1 resonance (Levison & Morbidelli 2003);

- *Current low total mass of the trans-Neptunian belt.* Estimates of the total mass are on the order of 0.1 $M_\oplus$ (Jewitt et al. 1998; Gladman et al. 2001; Bernstein et al. 2004; Chiang et al. 2007). Stern (1995) and Stern & Colwell (1997) demonstrated that 100 km-sized TNOs are unable to form via accretion in the current outer solar system in timescales comparable to the formation of Neptune. Indeed, the formation of Neptune is likely to preclude the growth of $\geq$ 50 km-sized TNOs because the planet induces eccentricities larger than 0.01 (i.e., higher collisional velocities), which favor nonaccreting collisions. This problem can be solved assuming that the trans-Neptunian belt carried much more mass in the past. Accretion/collisional models indicate that this would require at least ~10–30 $M_\oplus$ in an annulus from 35 AU to 50 AU (about 100 times more massive than now) to account for the formation of TNOs with sizes of 100–1000 km (Davis & Farinella 1997; Stern & Colwell 1997; Kenyon & Luu 1998; Kenyon & Luu 1999; Kenyon & Bromley 2004a). The formation of large rubble piles and satellites around planetary-sized TNOs requires a much more active collisional environment than now, and hence a more massive primordial belt (Jewitt & Sheppard 2002; Sheppard & Jewitt 2002; Brown et al. 2006b). If the ancient trans-Neptunian belt was more massive in the past, we have a new problem: because collisional grinding cannot account for all the mass loss, how was 99% of the mass lost?

In summary, a successful model for the trans-Neptunian region must explain (1) the excitation of the primordial planetesimal disk; (2) distinct classes of TNOs and their orbital structure, physical properties, and intrinsic ratios; (3) the creation of an edge at about 48 AU; and (4) the missing mass of the trans-Neptunian belt.

2.2 Scenarios for the Origin and Evolution of the Trans-Neptunian Belt

Several models have addressed the origin and orbital evolution of TNOs, but none have reproduced detailed observations of all dynamical classes or provided insightful predictions. However, we note that the intent of several models below was not to explain the entire trans-Neptunian belt structure. We included them because a more comprehensive model probably requires two or more mechanisms.

2.2.1 The Scattered Disk Model

Duncan & Levison (1997) proposed that scattering by Neptune could produce a population of scattered bodies. Morbidelli et al. (2004) revisited this scenario and showed it to be successful in explaining the origin and orbital structure of scattered TNOs. The inclination distribution of scattered TNOs also supports their origin via scattering by Neptune (Brown 2001). According to the scattered disk model, scattered TNOs would represent approximately 1% of the original population that was in Neptune-encountering orbits 4.5 Gyr ago. Recently, Morbidelli (2005) has convincingly confirmed the foregoing scenario, ruling out other possibilities for the origin of the scattered population such as the 3:2 resonance (via chaotic diffusion).

2.2.2 The Passing Star Model

Ida et al. (2000a) and Kobayashi & Ida (2001) proposed that a passing star could explain the orbital excitation in the trans-Neptunian belt. Assuming that the Sun formed in a dense cluster, a neighboring star might have encountered this system within a distance of about 80–200 AU during very early times, thus perturbing the outer regions of the cold primordial planetesimal disk. Kobayashi et al. (2005) revisited this scenario, focusing mainly on reproducing the dual character of the classical region.

The passing star model can excite the trans-Neptunian belt, deplete its total mass, and effectively truncate the disk (creating an outer edge). However, serious issues exist, including star passage timing problems (timescales too short, thus precluding the growth of TNOs), the preservation of the Oort cloud, and an orbital structure of TNOs incompatible with observations. The model also fails to preserve cold classical TNOs (Levison et al. 2004; Chiang et al. 2007). Finally, a close encounter at 80–200 AU is a low-probability event. To match the classical TNOs inclination distribution and satisfy the accretion timescale constraint (tens of Myr), the star must have passed near 100 AU and after 100 Myr, which has a much lower probability. At any rate, unless planet migration was considerably delayed, this particular "late" stellar encounter can be ruled out by the stability of 3:2- and other stable-resonant TNOs, which would otherwise have been destroyed (Gladman et al. 2001; Morbidelli & Brown 2004).

2.2.3 The Large Planetesimal Model

There are two variations of this scenario. The first case considers the temporary existence of massive planetesimals in the trans-Neptunian belt (rogue planets) (Morbidelli & Valsecchi 1997; Petit et al. 1999; Gladman & Chan 2006). The second suggests the existence of an as yet undiscovered planet orbiting at trans-Plutonian distances. We call this the "unseen planet scenario" (Matese & Whitmire 1986; Harrington 1988; Maran et al. 1997; Brunini & Melita 2002; Melita & Williams 2003; Melita et al. 2004).

Morbidelli & Valsecchi (1997) and Petit et al. (1999) showed that Neptune-scattered massive planetesimals could have sculpted the ancient trans-Neptunian belt and produced orbital excitation and dynamical depletion. It was recently shown that rogue planets can also create a population of detached bodies (Gladman & Chan 2006). However, in these scenarios, the massive planetesimals would break the stability of any primordial resonant TNOs, including the prominent 3:2-resonant population.

The distant resident planets postulated by Matese & Whitmire (1986), Harrington (1988), Maran et al. (1997), and other early studies (see references in Hogg et al. 1991) are very unlikely based on constraints from the motion of planetary orbits/spacecrafts and observations (Hogg et al. 1991; Morbidelli et al. 2002; Gaudi & Bloom 2005). Therefore, we did not include these models in our discussions below. Brunini & Melita (2002) proposed the existence of a Mars-like planet at about 60 AU to explain the excitation in the classical region and the trans-Neptunian outer edge. The unseen planet scenario was further explored by Melita & Williams (2003) and Melita et al. (2004), focusing on the detached population and the edge of the trans-Neptunian belt, respectively.

The advantages of the large planetesimal model include the excitation of the classical region, dynamical depletion, the creation of an outer edge, and an effective production of detached objects. However, the published orbital structure hardly resembles that of the classical TNOs. The hot classical population cannot be reproduced without simultaneously destroying the cold component. The resonant populations are also not obtained. Another problem is that in unseen planet scenarios, the planet would be at 50–60 AU on a near-circular and low-$i$ orbit. Under these circumstances, it should have already been discovered by any of the wide sky area surveys (Morbidelli et al. 2002). Finally, Melita et al. (2004) concluded that a resident planet as proposed in their model is unable to simultaneously satisfy the creation of the outer edge, excitation of classical TNOs, and the existence of 3:2-resonant TNOs.

2.2.4 The Resonance Sweeping Model

Resonance sweeping over a cold disk of planetesimals was first introduced to explain Pluto's orbit (Malhotra 1995). During planet migration, Jupiter, Saturn, Uranus, and Neptune suffered radial displacements on the order of –0.2 to –0.3, +0.7 to +0.9, +3 to +5, and +8.5 to +15 AU, respectively (Malhotra 1995; Fernández & Ip 1996; Chiang & Jordan 2002; Chiang et al. 2003; Gomes 2003b; Gomes et al. 2004; Hahn & Malhotra 2005). At the same time, all of Neptune's resonances swept the planetesimal disk, eventually capturing many bodies. This process can be understood by monitoring the location of an $r$:$s$ resonance, given by $a_{res} = a_N (r/s)^{\frac{2}{3}}$, where $a_N$ is the semimajor axis of Neptune. In addition, according to model parameters, the intrinsic ratio of 3:2- to 2:1-resonant TNOs and the spatial distribution of the latter resonants would provide important constraints on the migration history of Neptune (Ida et al. 2000b; Melita & Brunini 2000; Chiang & Jordan 2002; Murray-Clay & Chiang 2005).

Although resonant TNOs are mostly reproduced within 48 AU in this model, the stable resonant populations in the scattered disk are not. In addition, the model does not produce enough excitation in the classical region and it fails to produce detached objects. A variant of the model was proposed to explain high-$i$ 3:2-resonant and hot classical TNOs (Gomes 2003a, 2003b, 2006). Nevertheless, the production efficiency of hot classicals is too low and possibly in conflict with observations (Chiang et al. 2007). Moreover, there is a clear deficiency of detached objects in Gomes' model (see Section 2.1).

Hahn & Malhotra (2005) performed simulations of planet migration over both cold and hot (excited) planetesimal disks. A merit of using excited disks is the ability to produce resonant populations in the scattered disk (e.g., Lykawka & Mukai 2007a). Nevertheless, the assumption of an initially stirred planetesimal disk is unjustified.

In general, the resonance sweeping models have problems explaining the outer edge and low total mass of the trans-Neptunian belt. Moreover, independent of model parameters, the obtained detached population outnumbers the intrinsic observed one by at least several times.

2.2.5 The Solar Companion Model

This scenario would necessitate a massive body (a few to hundreds of $M_\oplus$) orbiting the Sun at some $10^3$–$10^4$ AU, which has unconstrained eccentricity and inclination. The models proposed in Murray (1999) and Collander-Brown et al. (2000) are inconclusive concerning any effects on TNOs. More recently, Matese et al. (2005) and Gomes et al. (2006) presented an improved model in which the main achievement is the production of extreme detached TNOs, especially Sedna-like objects.

2.2.6 The Packed-Ice Giants Model

Recently, Chiang and colleagues have argued that the solar system may have begun as a compact and crowded system consisting of Jupiter, Saturn, and as many as five ice giant planets of Neptune mass at $R < 25$ AU (Chiang et al. 2007; Ford & Chiang 2007). Because such systems are initially unstable, these ice giants can stir each other and suffer strong perturbations from the larger gas giant planets. In general, their models indicate that the surviving ice giants could have strongly perturbed the primordial planetesimal disk, and the latter could have circularized the orbits of the analogs of Uranus and Neptune. However, see Levison & Morbidelli (2007) for a critical analysis of this scenario.

Although their model can markedly excite classical and scattered objects, the lack of high-$i$ bodies in the resulting systems is an important limitation. Furthermore, because little emphasis was given to the formation of the trans-Neptunian belt, it is unclear whether this scenario can reproduce the fine structure of the classical region, form the scattered, resonant, and detached populations, and satisfy the other constraints discussed above.

2.2.7 The Nice Model

This model is currently the most comprehensive scenario for the outer solar system. In the Nice model, the planetesimal disk is assumed to be truncated at 30–35 AU. During the early stages, the mutual 1:2 resonance lock of Jupiter with Saturn would have created a large instability, and Uranus and Neptune were scattered outwards, severely perturbing the disk. The orbits of Uranus and Neptune would have been circularized by dynamical friction, and slowly migrated outwards. Focusing on the trans-Neptunian region, the merits of the model include plausible explanations for the outer edge at $a \approx$ 48 AU (based on the assumption that it coincides with the current 2:1 resonance location), the missing mass problem, and Neptune "naturally" migrating until 30 AU, among others (Gomes et al. 2005b; Morbidelli et al. 2005; Tsiganis et al. 2005). This model has been recently reviewed by Levison et al. (2007) and Morbidelli et al. (2007).

An uncomfortable situation with the Nice model is that an extensive comparison of detailed results with the populations of TNOs and their general properties has not yet been published. For this reason, it is not clear if the model can consistently explain the main constraints mentioned above (Section 2.1). In any case, the resonant structure in the scattered disk (>50 AU), cold classical TNOs, the detached population, TNOs with $i > 40°$, and peculiar TNOs represent critical tests for the model.

## 3. THE UNSEEN PLANET SCENARIO REVISITED

### 3.1 Planet Formation and the Existence of a Large Population of Massive Planetesimals

The most comprehensive planet formation models are based on the accretion of planetesimals and can be divided into three main stages: planetesimal formation, runaway growth, and oligarchic growth (Kenyon 2002; Kokubo & Ida 2002; Rafikov 2003; Goldreich et al. 2004b). First, there is the build up of kilometer-sized bodies. After this stage, the largest bodies in the swarm of planetesimals grow much faster than their smaller neighbors. Finally, several large planetesimals (or planetary embryos) are obtained, a few of which can eventually collide to form larger planets or the cores of giant planets (Pollack et al. 1996; Kenyon & Luu 1999; Goldreich et al. 2004a; Kenyon & Bromley 2004a; Rafikov 2004).

Accretion models agree that planet formation is an inefficient process, despite the different parameters and assumptions used in these models (Gladman 2005). In the late stages of planet formation, giant planets clean their neighborhoods by scattering disk planetesimals that move through the system, many of which are massive. A huge number of planetesimals carrying a substantial mass have been scattered by the planets (Fernández & Ip 1984, 1996; Pollack et al. 1996, Jewitt 1999; Petit et al. 1999). In addition, a natural consequence of these scattering events is planet migration (Fernández & Ip 1984; Hahn & Malhotra 1999; Gomes et al. 2004). In the end, the fate of scattered planetesimals is one of the following: (a) ejection from the solar system; (b) collision with a planet or the Sun; (c) placement in the scattered disk; or (d) placement at very large distances (formation of the Oort cloud) (Oort 1950; Brasser et al. 2006).

Several lines of evidence in the present solar system support the existence of a substantial population of massive planetesimals in the past: Pluto and other multiple systems, the high tilts of Uranus and Neptune, the retrograde orbit of Triton, and the discovery of very large TNOs (e.g., Eris). The origin of the Pluto system is better understood as a giant impact of embryos during the early solar system (Stern 1992, 1998; Canup 2005; Stern et al. 2006; Weaver et al. 2006). The formation of Eris and (136108) 2003 $EL_{61}$ satellite systems and the tilts of Uranus and Neptune are also explained by giant collisions during the late stages of planet formation (Stern 1991 and references therein; Brunini &

Melita 2002; Brunini et al. 2002 and references therein; Barkume et al. 2006; Brown et al. 2006b; Lee et al. 2007). Further evidence includes the Neptunian satellites Triton and Nereid. The former was possibly captured from the trans-Neptunian region because of its retrograde orbit (Luu & Jewitt 2002; Agnor & Hamilton 2006). Nereid may also be a captured TNO (Brown 2000; Schaefer & Schaefer 2000). In summary, the above examples are best explained by giant impact or close encounter events of massive planetesimals, which must have existed in large numbers to permit such events (Stern 1991; Brown 2002). Indeed, Stern (1991) predicted hundreds of planetary bodies with ~0.1–1.0 $M_\oplus$ and a few very large ones with ~1–5 $M_\oplus$ during the early history of the solar system.

Finally, we note that current observations are still incomplete. For instance, the existence of still other Pluto-sized objects in the trans-Neptunian region is predicted (Trujillo et al. 2001a). This expectation has been fulfilled recently, in particular with the discovery of planetary-sized TNOs such as Eris, 2003 $EL_{61}$, and (136472) 2005 $FY_9$ (Brown et al. 2004, 2005a; Brown et al. 2006b). Thus, it is perfectly possible that planetary bodies larger than Pluto or Eris exist in more distant resonances or far away in the scattered disk (Jewitt 2003; Gladman 2005).

### 3.2 Primordial Planetesimal Disk Orbital Conditions

The results of Lykawka & Mukai (2007a) suggest that the trans-Neptunian belt had a substantial population of planetesimals in excited orbits before planet migration (Fig. 3) and the radius of the planetesimal disk was at least 45–50 AU. However, planetesimals in very excited orbits (above the gray region in Fig. 3) or in near-circular and low-$i$ orbits may have existed. Nevertheless, nonresonant objects with $q < 37$ AU are likely to be removed from the solar system through scattering by Neptune, implying that even if very excited planetesimals existed, they are long gone. Therefore, we think that the pre-migration ancient trans-Neptunian belt extended to 50 AU or more and was composed of objects with eccentricities and inclinations ranging from near zero (up to ~45–50 AU) to the excited values illustrated by the gray region in Fig. 3. This hypothesis is supported by the current distribution of TNOs in which we observe classical TNOs in near-circular orbits up to 45 AU and an excited portion (45–50 AU) overlapping with the gray region. This would imply that the original cold planetesimal disk suffered an excitation from outside prior to planet migration, leading to the observed perturbed orbital distribution (Fig. 3). We propose that a massive planetesimal could be the main agent that excited the disk.

### 3.3 Building a Coherent Hybrid Scenario

Here, we argue that the orbital history of a massive planetesimal with tenths of $M_\oplus$ can explain the orbital structure in the trans-Neptunian belt. This massive body was likely scattered by one of the giant planets and managed to survive in the scattered disk. We refer to this body as the "outer planet," "trans-Plutonian planet," or "planetoid," denoted by a $_P$ subscript. We built the model intending to self-consistently satisfy the main constraints discussed in Section 2.1.

For a given orbital configuration, the planetoid should be massive enough to efficiently perturb the outer regions of the trans-Neptunian belt and truncate it at 48 AU. However, this excitation should not last over excessive timescales, otherwise the inner trans-Neptunian belt would be excessively depleted. Likewise, for a certain fixed timescale, the outer planet should be located in a convenient orbital configuration (neither too close nor too far away) for its perturbation to affect classical TNOs.

More importantly, because trans-Plutonian planets in near-circular and low-inclination orbits have been ruled out (Morbidelli et al. 2002), our hypothetical planet must be inclined (>10°) and relatively distant (e.g., >60–70 AU). A Neptune-scattering mechanism probably provided such orbital conditions for this planet. However, because gravitational scattering occurs at approximately fixed positions in space (Gladman et al. 2002), the perihelion of the outer planet could remain near the orbit of Neptune over prohibitively long periods of time, thus destroying the entire trans-Neptunian belt. Alternatively, dynamical friction could raise the planetoid's perihelion before the disruption of the belt (Del Popolo et al. 1999). However, this process could lead to a near-circular and very low-$i$ orbit, which would be in conflict with the observational constraint above.

Investigations of resonances in the scattered disk offer an appealing possibility. First, the strongest and dominant resonances beyond 50 AU are distributed mainly at $a < 250$ AU and have the lowest argument $s$, in particular those of type $r$:1 or $r$:2 (i.e., 3:1, 4:1, ..., 24:1; 5:2, 7:2, ..., 35:2) (Gallardo 2006a, 2006b; Lykawka & Mukai 2006, 2007c). Furthermore, the increase in perihelion during resonance captures is particularly important for the resonances cited above (Fig. 4). The capture probability in $r$:1 or $r$:2 resonances is also higher than in other resonances in the scattered disk (Lykawka & Mukai 2007c). In conclusion, we postulate that after being scattered outwards by a giant planet, the outer planet interacted with a Neptunian $r$:1 or $r$:2 resonance. Therefore, capture in a strong scattered disk resonance represents an excellent mechanism to modify the outer planet's orbit, helping to protect the classical region from disruption.

## 4. METHODS

We performed extensive simulations placing many thousands of disk planetesimals under the gravitational influence of the Sun, the four giant planets, and the outer planet, most over the age of the solar system. We used the hybrid symplectic second-order integrators EVORB and MERCURY (Chambers 1999; Brunini & Melita 2002; Fernández et al. 2002). The mass of the terrestrial planets was added to the Sun. The giant planets were fully considered as massive perturbers in a self-consistent way with initial orbital elements obtained from Jet Propulsion Laboratory ephemeris files (DE405), and minor bodies suffered only perturbation from massive bodies. Disk planetesimals were treated as massive and massless bodies in the MERCURY and EVORB integrations, respectively. The integrations completely accounted for collisions and close encounters with the planets and the Sun; particles that collided with a massive body were removed from the integration. The time step used was 6 months in self-consistent planet migration simulations, and typically 12 months in the 4-Gyr runs. This time step is sufficiently small to reliably integrate the orbits of trans-Neptunian bodies (Wisdom & Holman 1991; Duncan et al. 1998; Chambers 1999; Horner et al. 2004).

In most investigations, the ancient trans-Neptunian belt was represented by a cold planetesimal disk ($e \sim 0.001$ and $i \sim 0.2°$), usually extended from 17 to 50–100 AU. The giant planets were initially considered in both static (at their current orbits) and pre-migration compact orbital configurations (within ~17–20 AU). The planetoids were in typical Neptune-scattered and distant orbits (~40–160 AU), and with $i_P = 10$–40°. In most cases the masses of the outer planets ranged from 0.1 to 1.0 $M_⊕$. This mass range is well within the upper limits provided by several theoretical and observational constraints, namely 1–3 $M_⊕$ inside 60–70 AU (Hogg et al. 1991; Melita et al. 2004 and references therein; see also Fig. 6 in Gaudi & Bloom 2005). The initial argument of perihelion $ω_P$, longitude of ascending node $Ω_P$, and mean anomaly $Λ_P$ (angular orbital elements) of the outer planet were chosen randomly in the simulations.

Initially, we performed preliminary simulations to test the feasibility of our scenario. Later, after obtaining important constraints from the results of these simulations, we conducted major simulations to investigate the outcomes of the hybrid trans-Plutonian planet model (Section 6). Details of the simulations, as well several complementary runs, are given in the relevant sections below.

## 5. PRELIMINARY SIMULATIONS AND RESULTS

We performed more than 700 runs to investigate the excitation of an outer planet on the trans-Neptunian belt. These systems were evolved over varied timescales (from tens of Myr to 4 Gyr), depending on the purpose of the investigation.

### 5.1 Static Outer Planets

We first explored the influence of outer planets in fixed orbits, varying the initial orbital and mass parameters of the planet in systems with the giant planets at their current orbits. The main goal was to see if a massive planet with an orbit near a resonance location in the scattered disk could reproduce the excitation in the classical region and form the edge at 48 AU. The time span of the simulations was 4 Gyr. A summary of the runs is given in Table 1.

We found that an outer planet, provided that it is massive enough and possesses a convenient perihelion distance, is able to disrupt the local cold disk at 48 AU, in particular the removal of objects with $e < 0.1$. This was possible for planetoids with 50 AU $< q_P <$ 56 AU and having $i_P <$ 20–25° and a few tenths of $M_⊕$ when located within about 90 AU. At larger distances, more massive outer planets yielded better results (0.5–1.0 $M_⊕$) (Fig. 5). Planetoids with smaller perihelia yielded too much excitation in the trans-Neptunian belt, whereas those with larger perihelia and/or higher inclinations did not produce an edge at all. The disruption of the cold disk beyond 48 AU occurred on timescales of 0.5–1 Gyr for planetoids at 63–88 AU (3:1, 4:1, and 5:1 resonances), although better removal of objects with low $e$ requires more than 2–3 Gyr. In the 10:1 resonance, these timescales were approximately two times longer. Another outcome of these runs was an excitation of inclinations for objects orbiting the outer planet crossing region.

We also tested the presence of a less massive planetary body in the 2:1 resonance using $M_P =$ 0.01–0.05 $M_⊕$ over 1.5 Gyr (SIM 1–3, Table 1). The planetoids excited local planetesimals to large eccentricities and $i$ ~10–20° in only a few hundred Myr, creating narrow and sharp peaks in $a$–$e$ and $a$–$i$ element space distributions. Nevertheless, 2:1-resonant planetoids either led to too little perturbation in the classical region and beyond or excessively depleted the inner classical region. Thus, a Moon-like or more massive body in the 2:1 resonance cannot satisfy the trans-Neptunian belt observational constraints discussed previously (Section 2.1).

A drawback of static planets is that the observed orbital distribution in the classical region is not reproduced (e.g., Fig. 5). Indeed, in general, particles survived billion of years in near-circular orbits within 50 AU. When the classical region is excited by the planet (usually when $q_P <$ 50 AU), the final distribution is incompatible with observations. In conclusion, static planets cannot form the edge of the trans-Neptunian belt and at the same time reproduce the orbits of TNOs in the classical region. A similar conclusion was found by Melita et al. (2004). However, one could argue that the fine-tuning of simulations or a much larger number of runs with higher resolution could yield results that are in agreement with observations. However, this is not the case, as we will show in Section 5.3.

5.2 Migrating Outer Planets

Considering the difficulties of static outer planets in consistently explaining the excitation of the classical region and the trans-Neptunian belt edge, we decided to investigate systems with planet migration. The main goal was to verify whether a planetoid initially crossing the classical region could provide the appropriate excitation to the latter and remove bodies beyond 48 AU while migrating outwards. When not mentioned, the simulation time span was 100–200 Myr. A summary of the runs is provided in Table 2.

The giant planets are assumed to have formed in a more compact configuration than at present, in line with migration models (Morbidelli et al. 2007 and references therein). We assumed the planetoid had initial perihelion $q_{P0} < (a_{N0} + 10$ AU$)$, where $a_{N0}$ is the initial semimajor axis of Neptune before migration. Thus, the planetoid represented a "recent" Neptune-scattered body. We set $a_{P0}$ as a free parameter using arbitrary values beyond 40 AU (sometimes close to the initial position of a resonance of interest) and $10° \leq i_{P\,0} \leq 30°$ and tested different masses for the outer planet. We performed simulations in which, along with the giant planets, the planetoid was forced to migrate outwards obeying a predefined radial displacement (in some cases according to the position of a distant and strong $r$:1 resonance with Neptune). We imposed an artificial force to induce planet migration as described in Malhotra (1995) and Hahn & Malhotra (2005). For simplicity, we neglected disk interactions and other stochastic phenomena during migration, so most outer planets ended the simulations with perihelia around 50–80 AU. We stress that the intent of this simple migration model was to test the viability of the scenario in which a planetoid initially perturbs the classical region and is later transported outwards, ending the migration phase in a stable distant orbit near/in a resonance in the scattered disk. Thus, the final outcomes acquired by the outer planet here resemble the initial conditions conjectured in Section 5.1.

The giant planets and the outer planet migrated according to
$$a_n(t) = a_n - \Delta a_n \exp(-t/\tau) \quad (1),$$
where $\Delta a_n$ (in AU) is the total radial displacement of the planet (indicated by the subscript $n$), $a_n$ is the current planet's semimajor axis, and $\tau$ is the migration timescale. We prepared three sets where $\Delta a_n = \{-0.2, 0.8, 3.0, 7.0\}$, $\{-0.2, 0.9, 4.0-4.3, 9.9-10.2\}$, and $\{-0.2, 0.9, 4.5-5.0, 12.0-13.0\}$ for Jupiter ($J$), Saturn ($S$), Uranus ($U$), and Neptune ($N$), respectively (i.e., Neptune started at 23.1, ~20, and 17.1–18.1 AU). We tested various migration timescales, but preferentially used $\tau = 5–15$ Myr. These values are constrained by several past studies (Hahn & Malhotra 1999; Gomes 2000; Ida et al. 2000b; Melita & Brunini 2000; Friedland 2001; Chiang et al. 2007).

In general, the inclusion of a migrating planetoid resulted in the simultaneous formation of resonant populations (<50 AU), excitation of the classical region (mostly in eccentricities), and the disruption of the disk beyond about 48 AU well before the long-term stage (Section 5.1). Moreover, as the outer planet moved to large distances, the stability of the classical region was also guaranteed. A didactic example is shown in Fig. 6, and some of the best runs are illustrated in Fig. 7. Fast migrations (i.e., using smaller $\tau$) yielded too little excitation in the disk, whereas incursions over several Myr at <50 AU caused too much perturbation in the region. Finally, even for final orbits as close as $a_P = 60$ AU, low-mass outer planets ($\leq 0.2$ $M_\oplus$) had difficulties in correctly reproducing the classical region excitation and the creation of the trans-Neptunian belt edge.

Although a migrating outer planet could explain some of the main constraints posed by observations, the formation of hot classical TNOs is problematic. We performed a few simulations to investigate the formation of the hot classical component following the approach of Gomes (2003b). We set two cold disks, i.e., 18.6–30.1 AU and 27.6–30.1 AU, with 9900 particles each, and evolved these systems over 200–250 Myr using the same model parameters of some of the best runs (Table 2). Hot classical particles were obtained in most cases. However, the efficiency is quite low at less than 0.1–0.2%. Also, the smallest eccentricities of hot objects appear to be somewhat higher than those of hot classical TNOs. We also noticed that the efficiency depends on the migration timescale. In test simulations, we found that the production of hot classical objects decreases with increasing migration timescales. Thus, because most of the best runs have $\tau \geq 5$ Myr, the low efficiency problem cannot be solved using models with fast migration in our scenario.

Another serious issue is the formation of long-term resonant populations beyond 50 AU. In fact, none of the runs produced any obvious resonant populations in the 9:4 ($a = 51.7$ AU), 5:2 ($a = 55.4$ AU), and 8:3 ($a = 57.9$ AU) resonances. The reason for this negative result is easy to understand. The formation of TNOs with 9:4, 5:2, and 8:3 resonances requires that a population of planetesimals had $e >$ 0.1–0.3 before the resonances swept the region (Fig. 3; Lykawka & Mukai 2007a). In Fig. 6, when the resonances are passing through the classical region at 5 Myr (panel b), the eccentricities are less than 0.05. Later on, although the disk acquires $e \sim 0.1$–0.12 near the resonances at 20 Myr (panel c), it is still quite below the minimum necessary to account for observed resonant populations (Fig. 3). In other runs we noted very similar outcomes, some of which had even smaller stirred eccentricities. In summary, there are two major issues. First, the excitation of the trans-Neptunian belt occurs too late when distant resonances have already swept the disk. Second, the resulting excitation is not high enough inside 50 AU, even in the case of $M_P > 0.5$ $M_\oplus$ or longer timescales for passages through the disk by the planetoid.

If the planetesimal disk were initially excited at an appropriate level, would the planetoids preclude resonant capture in resonances beyond 50 AU? We re-ran a few runs of migrating outer planets using excited disks with broad initial eccentricities beyond 45 AU. The outcomes after 100 Myr indicate that 5:2 and other resonant bodies beyond 50 AU are obtained.

5.3 Survival of 3:2, 2:1, 5:2, and Other Resonant TNOs

In most of the successful runs discussed above, the perihelia of the outer planet were within 50–60 AU. In such circumstances, the planetoid could threaten the stability of 3:2, 5:3, 7:4, 2:1, and other observed resonant populations over 4 Gyr. That is, eccentric resonant bodies near aphelion could encounter the trans-Plutonian planet during its perihelion approach. These perturbations are likely to dislodge bodies from the resonances in question. Because these bodies will no longer be protected from close encounters with Neptune by the libration mechanism (e.g., Peale 1976; Malhotra 1998; Murray & Dermott 1999), they will be ultimately scattered by the giant planet for large eccentricities ($q < 37$ AU). For lower eccentricities, former members of resonances will become fossilized in orbits near their parent location, $a_{res}$. Because the majority of TNOs in the 3:2, 2:1, 5:2, and other resonances are known to be stable over the age of the solar system (i.e., with orbital motion deeply inside the resonances), could they survive under the perturbing influence of a massive outer planet as considered above?

To determine the survival of the resonant populations of interest, we initially evolved 1000 particles inside the 3:2, 2:1, and 5:2 resonances over 4 Gyr with initial $i < 30°$, and $0.05 < e < 0.36$, $0.1 < e < 0.42$, and $0.35 < e < 0.45$, respectively. These ranges cover the observed orbital distributions. Angular orbital elements were chosen at random within 0–360°. Only the four giant planets were considered in this particular investigation. We then obtained three populations stable over the age of the system for the 3:2 (166 particles), 2:1 (147 particles), and 5:2 (260 particles) resonances within about the same eccentricity and inclination ranges. Finally, we checked the evolution of the stable resonant populations with the inclusion of a massive outer planet (see also Table 1).

First, we report on the survival of 3:2- and 2:1-resonant populations for outer planets with $q_P = 50$–$60$ AU. As expected, resonant bodies were strongly perturbed by the planet (Fig. 8). The 3:2 resonance was depleted at ~30–90% levels (Table 3). In particular, the destabilization of very large $e$-3:2 bodies ($>0.25$–$0.3$) was evident, often leading to their total disruption. In the case of the 2:1 resonance, the situation was much more severe (Figs. 8 and 9). In fact, the 2:1-resonant population was completely devastated in almost all runs and yielded resonant survivors with $e < 0.26$ (Table 3). Therefore, the general tendency was fast destruction of the 2:1-resonant population at high eccentricities (500 Myr to 1 Gyr timescales) and significant depletion of the 3:2, 5:3, and 7:4 resonances over 4 Gyr. These results remain valid even if the outer planet's inclination is as high as 40° because higher inclinations just increased the timescales for the dynamical depletion in these resonances (but still <<4 Gyr). Thus, the existence of stable resonant populations provides an invaluable constraint on the perihelion of the trans-Plutonian planet. That is, in our simulations, a fraction of resonant objects can survive 4 Gyr if their aphelion, $Q$, is about 2–3 AU smaller than the outer planet's perihelion for $M_P = 0.1$–$0.5$ $M_\oplus$. This condition translates into $q_P > Q + 2$ or $+ 3$ AU.

We conclude that the observed 3:2- and 2:1-resonant populations can survive over the age of the solar system only if $q_P > 53$ AU and $q_P > 67$ AU, respectively (Figs. 8 and 9 and Table 3). More importantly, a planetoid with perihelion around 50–60 AU cannot create the outer edge of the trans-Neptunian belt and excite the classical region without destroying most (if not the whole) of the 2:1-resonant population. Moreover, even if a few percent of 2:1-resonant TNOs could survive 4 Gyr, their maximum eccentricities would be in clear conflict with observations (Fig. 9 and Table 3). It turns out that resident outer planets as shown in panels a–c of Fig. 7 cannot hold for 4 Gyr because their final perihelia are too small. On the contrary, outer planets that acquired $q_P > 70$–$80$ AU after migration could allow the survival of resonant populations in the trans-Neptunian belt (<50 AU; see Fig. 7, panel d). These results also rule out resident outer planets as proposed by Matese & Whitmire (1986), Harrington (1988), Brunini & Melita (2002), Melita & Williams (2003), and Melita et al. (2004) because they generally have 50 AU $< q_P <$ 60 AU.

The survival of 5:2-resonant TNOs offers another important constraint. We tested the long-term survival of 2:1- and 5:2-resonant populations with planetoids analogous to those in Fig. 7 (see also Table 1). In general, we found that 5:2-resonant bodies can survive 4 Gyr if $q_P$ is about the same or slightly greater than their aphelia ($i_P$ = 10–40° and 0.3–1.0 $M_\oplus$). This result implies that any resident trans-Plutonian planet must have $q_P$ > 80 AU, otherwise we would not observe stable 5:2 resonants with their present orbital distributions (Fig. 9 and 10). This result is also independent of the semimajor axes (60–140 AU) used for the outer planet. Consequently, the unseen planet at ~70 AU proposed by Brown et al. (2004) to explain Sedna's orbital characteristics is also ruled out.

## 6. MAIN SIMULATIONS: HYBRID MIGRATING OUTER PLANET MODEL

In an aim to provide an early excitation to the planetesimal disk to account for the formation of 9:4-, 5:2-, and 8:3-resonant TNOs (Fig. 3), we improved the model discussed in Section 5.2 by allowing the outer planet to perturb the cold planetesimal disk before planet migration. In other words, we proposed that after the outer planet was scattered by Neptune; it managed to temporarily excite the disk to appropriate ranges of $e$ and $i$ and the trans-Plutonian planet was transported outwards by resonance sweeping with Neptune and/or scattering by the giant planet. In this manner, the migrating outer planet model with delayed migration corresponds to our hybrid scenario. This is consistent with a late start of planet migration (Gomes et al. 2005b).

The hybrid outer planet model is divided into three phases. Our time $t$ = 0 is set at the late stages of giant planet formation, tens of Myr after the birth of the solar system (e.g., see Montmerle et al. 2006).

*I. Pre-migration excitation of the planetesimal disk* ($t$ = 0 until $t \approx$ 30–100 Myr). In pre-migration systems, the four giant planets, located within ~17–20 AU, and the scattered outer planet perturb the primordial cold planetesimal disk. Inspired by the best results of previous sections, we preferentially used planetoids with $a_{P0}$ = 50–80 AU, $M_P$ = 0.3–1.0 $M_\oplus$, and $q_{P0}$ a few AU distant from Neptune. The planetesimal disks were represented by several hundred to a few thousand small-mass particles (seeds) in uniform distributions. No migration was imposed on any of the planets.

*II. Planet migration* ($t \approx$ 30–100 Myr until $t \approx$ 150–200 Myr). We took the excited disks at the end of pre-migration runs (1000–2000 particles up to 60–65 AU) as initial conditions for the planet migration simulations. Next, we assumed that the outer planet was quickly captured in an $r$:1 (or $r$:2) resonance during or at the end of planet migration. To satisfy the required $q_P$ > 80 AU (Section 5.3), two potential mechanisms can pump up the perihelion: dynamical friction with the planetesimals disk and the Kozai resonance (KR), which is defined as the libration of the argument of perihelion rather than circulation (Kozai 1962; Wan & Huang 2007). Because the KR is quite common inside these resonances (Gomes et al. 2005a; Gallardo 2006a, 2006b; Lykawka & Mukai 2007c), the outer planet likely exhibited decreased $e_P$ (increased $q_P$) at the expense of increased $i_P$. This effect was modeled in a way that approximately conserved vertical angular momentum, $\sqrt{1-e_P^2}\cos(i_P)$. Alternatively, the outer planet could have acquired large $q_P$ and moderate $i_P$ even if the KR was not active (e.g., moderate dynamical friction and $r$:1-resonant effects). Ultimately, the planetoid acquired a detached and highly inclined orbit, satisfying several constraints (Hogg et al. 1991; Morbidelli et al. 2002; Melita et al. 2004 and references therein).

In the simulations, the outer planet acquired about $q_P = 80$–$85$ AU and $i_P = 30$–$45°$ in timescales comparable to planet migration itself, ~100–200 Myr. Similar KR timescales have been reported in the literature (Gomes et al. 2005a; Lykawka & Mukai 2006). It is worth noting that for diverse plausible outer planet's initial eccentricities and inclinations, further evolution associated with KR dynamics limits the final $i_P$ to within 30–50° and $e_P$ to values near the minimum to maintain resonance lock in an $r$:1 resonance. In non-KR cases, smaller inclinations would be expected (~10–30°).

*III. Long-term sculpting by the planets* ($t$ ~150–200 Myr until present). The long-term gravitational perturbation by the giant planets and the outer planet would have sculpted the solar system for more than 4 Gyr. To explore this phase, we extended some of the planet migration runs to 4 Gyr. In the end, the trans-Plutonian planet should currently be located near an $r$:1 (or $r$:2) resonance.

In summary, we introduce a new hybrid model that combines the main mechanisms invoked in the scattered disk, large planetesimal, and resonance sweeping models. This combination means that some basic and well-supported aspects of the solar system history are taken into account: scattering of large planetary bodies by the newly formed giant planets, planet migration, and resonance capture.

6.1 Pre-migration Perturbation

We executed 300 runs of pre-migration systems with scattered planetoids over tens of Myr to provide a convenient $e$–$i$ excitation in the outer part of the planetesimal disk (Sections 5.2 and 6). The wandering of the outer planets in the semimajor axis was never greater than 5 AU over the same timescales as a result of their perihelia ~5–10 AU away from the initial position of Neptune.

A scattered outer planet represents an excellent stirring mechanism. In addition to satisfying the required excitation for the formation of distant resonant TNOs on reasonable timescales (typically tens of Myr), an interesting and unexpected feature is that the perturbed disks presented orbital distributions very similar to observations (nonresonant populations), especially those observed in the 40–50-AU region (Fig. 11).

For reference, we show some outcomes of stirred disks in pre-migration systems for different initial masses and orbital parameters of the outer planet in Fig. 12. Although the heliocentric location of perturbations in the disk depends essentially on the initial position of the outer planet, the shape and degree of excitement of the disk depends on the outer planet's mass and time span.

6.2 Some Detailed Investigations

In certain large-scale simulations conducted in the migration phase, the planetesimal disk was modeled with a few thousand particles following an $R^{-1}$ distribution. In these simulations, outer planets of 0.4 and 0.5 $M_\oplus$ were transported outwards following the location of the 6:1 and 9:1 resonances, respectively. The giant planets and the outer planet were forced to migrate using $\tau = 10$ Myr (Table 4). We also confirmed that the results are independent of the timing of the onset of KR for the outer planet (SIM1,2 in Table 4). We integrated all migrating systems for 100 Myr. A few outcomes after planet migration are illustrated in Fig. 13. In the long-term stage, we created clones of each object up to 54 AU in high-resolution runs, obtaining an excited planetesimal disk composed of several thousand particles (up to 17000). We evolved the system of four giant planets plus the outer planet plus the disk over 4 Gyr.

**7. MAIN RESULTS**

Except in Section 7.9, the results presented below are based mainly on a series of large-scale simulations extended to 4 Gyr. However, we note that the model used in these simulations has limitations (e.g. smooth planet migration and disks of "massless" planetesimals).

7.1 Depletion of the Inner Trans-Neptunian Belt

Unsurprisingly, the migration of Neptune was responsible for the dynamical depletion of the inner region of the disk until 39 AU, confirming the results of previous studies (Morbidelli & Brown 2004). The removal of objects in this region occurred through the triple action of Neptune's gravitational scattering (quickly clearing a region up to 35 AU), overlapping of resonances giving rise to unstable chaotic behaviors, and the capture of local bodies by strong first-order sweeping resonances (e.g., 5:4, 4:3, 3:2). Therefore, the lack of TNOs in this particular region is evidence of the outward migration of Neptune, supporting our current understanding of the theory of planet migration.

7.2 Classical Region Structure

7.2.1 Cold and Hot Classical Populations

To compare our results with current observations, we considered cold ($i \leq 5°$) and hot ($i > 5°$) classical populations as nonresonant particles with $a < 50$ AU and $q > 37$ AU at the end of simulations. The results confirm the standard picture for the origin of cold and hot classical TNOs (Section 2.1), although it differs in some points, as discussed below.

The cold population was probably formed in situ beyond $a \sim 37$ AU and was perturbed early by the planetoid (Sections 6 and 7.2.2). The final eccentricities of classical bodies were remarkably similar to observed values (cf. Figs. 14 and 3). Our obtained median $e$ and $i$ of cold populations were 0.056–0.077 and 2.4–3.2°, respectively, comparable to the observed values of 0.056 and 2.4°. Furthermore, we found that some local (initially cold) classical bodies acquired $i \sim 5$–15°. In addition, recalling that resonance excitation is active in the classical region (Nesvorný & Roig 2001; Lykawka & Mukai 2005a; LM07b), possibly a significant part of the observed cold population was promoted to the hot one, especially in the 5–15° range of inclinations.

The hot population should have originated from two distinct sources, i.e., the local planetesimal disk and the inner solar system, between ~15–35 AU. Interestingly, hot classical TNOs formed by the excitation of local planetesimals (mechanism above) should be moderately inclined (5–15°), whereas bodies that entered the classical region from inner regions would constitute hot classical TNOs with moderately high inclinations (10–35°) (Fig. 15).

Consequently, distinct physical properties (i.e., color and size distributions) would become more evident among classical TNOs with low inclinations and those with $i > 10$–15°. Indeed, this finding is supported by the distribution of inclinations with spectral slopes of classical TNOs (see Fig. 4 of Chiang et al. 2007) in which the differences in these distributions become clearer for classical TNOs with higher inclinations.

Finally, the above picture also offers an explanation for the intrinsic fraction of cold and hot populations, estimated at unity (Morbidelli & Brown 2004). This fraction is in serious conflict with Gomes' scenario (Gomes 2006 and references therein) because of the too-low efficiency to produce hot classicals. According to our results, an important fraction of cold classical objects was promoted to the hot population, helping to balance the estimated ratio of the populations. The ratios of cold to hot classical bodies from the large-scale simulations can vary as much as 0.7–5.0.

### 7.2.2 An Excited Classical Region

In our scenario, despite the shortage of high-$i$ classical bodies (>15–20°), the planetoid can explain the excitation of eccentricities and inclinations in the classical region quite well, including the lack of low-$e$ objects beyond 45 AU (Figs. 13 and 16). This excited orbital distribution is a consequence of the perturbation from the outer planet during the pre-migration phase (see Figs. 11–13; see also Section 6.1). The time span necessary to reproduce the observed excitation was about 50–100 Myr for $M_P = 0.3$–$0.5\ M_\oplus$, although it can be as small as 20–30 Myr in the case of more massive planetoids ($M_P = 0.7$–$1.0\ M_\oplus$). Finally, the perturbation of the planetoid on the region during the migration phase was not significant.

In summary, our results suggest that the global excitation of the classical region was created during the first tens of Myr after planet formation (4.5 Gyr ago) and ended when planet migration ceased. Consequently, the excitation of the classical region would provide an important constraint on the orbital evolution and elementary properties of the planetoid.

### 7.3 Resonant Structure in the Trans-Neptunian Region

We identified objects locked in various resonances in the trans-Neptunian region (Figs. 14, 16, and 17). The great majority of the resonant population was originally distributed between 30 and 50 AU during resonance capture. After 4 Gyr, the main occupied resonances, sorted by distance from the Sun, are: 5:4, 4:3, 7:5, 3:2, 8:5, 5:3, 7:4, 9:5, 11:6, 2:1, 13:6, 11:5, 9:4, 7:3, 5:2, 8:3, 11:4, and 3:1. We also found 4-Gyr members in all of these resonances. More importantly, the stable resonant populations in the scattered disk agree well with the existence of long-term 9:4-, 5:2-, and 8:3-resonant TNOs (see also Section 5.2). Symmetric and asymmetric 2:1-resonant TNOs were also reproduced (see Murray-Clay & Chiang 2005 for details about the 2:1 resonance).

The model is able to reproduce resonant TNOs quite well in eccentricities and inclinations (Fig. 16 and Table 5). Moreover, the eccentricities of scattered disk resonant TNOs were better matched for disks without bodies beyond ~51–54 AU in low- to intermediate-eccentricity orbits. Nevertheless, we stress that the scattered disk resonant structure is still not well known (i.e., low number statistics), so only future observations will better constrain model results. Concerning inclination distributions, our resonant objects had typically $i < 20$–25°, whereas the observed high-$i$ component of 3:2-resonant TNOs ($i > 25°$) was obtained only in specific runs via the mechanism proposed in Gomes (2003b).

Concerning resonance properties, the distribution of the amplitudes of resonant angles (libration amplitudes), which are a measure of Neptune to TNOs' relative distance during resonant motion (e.g., Murray & Dermott 1999), are in good agreement with those derived from observations (Fig. 18). Libration amplitudes ranged from a few to 120–160° for most occupied resonances (Table 5). In addition, stable populations in the 9:4, 5:2, and 8:3 resonances yielded a wide range of libration amplitudes, which is compatible with values determined from TNOs. Lastly, particles experiencing KR were found among resonant populations in the 5:4, 4:3, 7:5, 3:2, 8:5, 5:3, 7:4, 2:1, 7:3, 5:2, 8:3, and 3:1 resonances. The relative fraction of Kozai librators inside these resonances is reasonably compatible with observations, except for the underabundance in the 5:3 and 2:1 resonances (compare Table 5 with Table 6 of LM07b).

### 7.3.1 Stability of 5:2, 3:1, and Other Distant Resonant Populations

The maximum eccentricities of 5:2-, 8:3-, 11:4-, and 3:1-resonant TNOs can give clues to the perihelion approaches of the outer planet (Fig. 14 and Table 5) because primordial populations of high-$e$ bodies in these resonances were removed from resonance beyond critical eccentricities (Section 5.3). For the same reason, long-term resonant populations were almost completely absent beyond the 3:1 resonance.

In conclusion, the maximum eccentricity of distant resonant bodies obeyed the relation $Q \leq q_P$, implying that the future identification of long-term populations in these resonances will provide an important constraint on the existence of the proposed planet.

### 7.3.2 Neptune Trojans

We found a stable Neptune Trojan body with $e \approx 0.06$ and $i \approx 11°$ at the end of 4 Gyr in one of the large-scale simulations (Fig. 16). Although this object was initially scattered by Neptune before migration, it was quickly captured in the sweeping 1:1 resonance during the first Myr of planet migration when Neptune was at 24.5 AU. In addition, the outcome of extra simulations of planet migration yielded the formation of Neptune Trojans with wide distributions of eccentricity ($e < 0.14$) and inclination ($i < 20°$). These results suggest that the observed Neptune Trojan population could arise within the framework of the resonance-sweeping model used in our scenario. We will address the formation of Neptune Trojans in more detail in the near future.

### 7.4 Scattered Population

There is a clear population of scattered objects at the end of the simulations (Fig. 17), found in typical eccentric and low-to-high-inclination orbits (<50°), as a result of gravitational encounters with Neptune (mainly) and the planetoid. This is in agreement with observations. Although their identification is more difficult, scattered bodies are also found in the classical region (Fig. 14). Even with the inclusion of an outer planet, the formation of scattered objects over billions of years proceeds in a very similar fashion to that proposed by Duncan & Levison (1997). For instance, the fraction of scattered bodies is on the order of 1–2% of the original population on Neptune-encountering orbits 4 Gyr ago.

Interestingly, the early excitation of the trans-Neptunian belt allowed the region beyond 40 AU to contribute significantly as a source of scattered bodies. Excluding objects leaving the main resonances (<10%), about half of the scattered TNOs would have originated somewhere around 40–50 AU, <5% in the path transversed by the migrating Neptune (~20–30 AU), and the remaining (a few tens of percent) from the region sculpted by Neptune over Gyr, namely 30–40 AU. Table 6 shows the statistics of scattered populations during their evolution as compiled from several independent runs.

Finally, scattered bodies in orbits close to the outer planet suffered significant excitation in inclinations, leading to $i = 40–50°$ (Figs. 17, 19, and 20). The model also produced analogs of Eris (Section 8.3).

7.5 Detached Population

A large population of detached objects was obtained in several simulations (Figs. 17, 19, and 20). Detached bodies originated mostly from Neptune-scattered orbits and later acquired larger perihelia and/or inclinations caused by perturbation from the outer planet over typically several hundred Myr. At the end of 4 Gyr, the bulk of the detached population resulted in $q = 40–60$ AU and $i < 60°$ (the two were weakly correlated), a result compatible with observational constraints (Morbidelli & Levison 2004). We observed this tendency in all runs with $M_P \leq 0.5\ M_\oplus$. In runs in which larger values of $M_P$ were used, the distribution of perihelia tended to reach higher values, revealing a more irregular pattern (Fig. 20). More massive planetoids were also more efficient in creating detached objects with larger semimajor axes and perihelia. Alternatively, detached bodies could also have originated from the effect of the planetoid's perturbation on originally cold objects of the disk, which would have acquired higher eccentricities but still kept $q > 40$ AU.

Furthermore, the detached populations are comparable to or a few times larger than the final population of scattered objects, which is in agreement with unbiased observational estimates (Gladman et al. 2002; Allen et al. 2006; Chiang et al. 2007). It is worth noting that simulations using only the four giant planets (static and migrating) result in ratios of scattered to detached populations higher than 4.0 (see Table 6). Finally, similar to the formation sites of scattered objects, there was no obvious preference for the source region of detached bodies.

The outer planet's influence could account for the existence of detached TNOs with low $i$ (Figs. 19 and 20), a feature not achieved in scenarios that rely just on resonant interactions (Gomes et al. 2005a; Lykawka & Mukai 2006; LM07b). We also report the production of some extreme bodies with $a = 500–800$ AU and usually $40\ \text{AU} < q < 50$ AU. The model also produced analogs of typical detached TNOs such as 2004 XR$_{190}$, 2000 CR$_{105}$, and Sedna, although the Sedna population was small (see Section 8.3).

Finally, the long-term residence of the planetoid can leave observable orbital signatures in the scattered disk. For instance, scattered and detached objects with orbits near that of the outer planet acquired the largest inclinations and perihelia (Figs. 19 and 20). In conclusion, a better characterized orbital distribution of TNOs in the scattered disk may reveal the perturbing signatures of the outer planet.

7.6 Very High-$i$ Population

Unlike standard models (e.g., Duncan & Levison 1997; LM07b), the trans-Plutonian planet can produce very high-$i$ objects, predicting a non-negligible fraction of TNOs with $i > 40°$ beyond 50 AU. In particular, for $M_P \leq 0.5\ M_\oplus$, scattered and detached bodies acquired inclinations up to 50° and 60°, respectively (Sections 7.4 and 7.5). More massive outer planets can excite the inclinations of some objects to values as high as 90°. The $i$-excitation has a broad distribution in the semimajor axis, with a peak near the location of the outer planet (Figs. 19 and 20). Another mechanism that can promote TNOs to the very high-*i* subclass is the KR (Section 7.3). Therefore, it is possible that the outer planet perturbation and the KR mechanism may together explain this highly inclined subpopulation.

We discuss the production of analogs of very high-*i* TNOs in Section 8.3.

7.7 The Trans-Neptunian Belt Outer Edge

During simulations, the outer planet was very effective in truncating the trans-Neptunian belt at approximately 48 AU, thus reproducing not only the absence of low-*e* bodies, but also the abrupt decrease of the number density of TNOs with *R* beyond 45 AU (e.g., Figs. 13, 16, and 21). This finding is connected to planetoid-induced orbital excitation at 40–50 AU (Figs. 11 and 12), a feature that arose before planet migration, when the planetoid was still on a typical Neptune-scattered orbit. Because these features exhibited shifting in heliocentric distance according to the planetoid's initial location, planetoids at 60–80 AU, with $q_P$ = 20–30 AU and $i_P$ < 15°, yielded the best outcomes before migration. This takes into account uncertainties in the initial position of Neptune at 15–20 AU.

The location of the outer edge depends on the mass, initial orbital conditions, and timing of the perturbation of the planet (Fig. 12). The subsequent migration of the planetoid plays a minor role. In summary, the outer edge of the trans-Neptunian belt was possibly created during the first tens of Myr of solar system existence, before planet migration. Thus, similar to the conclusions of Section 7.2.2, the outer edge would tell us about the origin and orbital history of the trans-Plutonian planet.

7.8 Losing 99% of the Ancient Trans-Neptunian Belt Total Mass

This scenario could solve the problem of the missing mass in the trans-Neptunian belt and offers two particular advantages. First, the planetesimal disk was dynamically depleted during pre-migration stirring (scattering of objects by Neptune) and remained active over the age of the solar system. For instance, in general, 15%–40% of the planetesimals remained in the system after 4 Gyr for disks of ~51–54 AU radius. Second, with the outer planet's perturbation, a large fraction of planetesimals initially on very cold orbits were excited to levels at which random velocities render fragmentation rather than accretion (Stern & Colwell 1997; Kenyon & Bromley 2004a). Random velocity is given by

$$v_{rnd} = \sqrt{v_{disp}^2 + v_{esc}^2} \quad (2),$$

where $v_{disp}$ and $v_{esc}$ are the dispersion and the planetesimal's escape velocities, respectively. The former is given by

$$v_{disp} = v_K \sqrt{e^2 + i^2} \quad (3),$$

where $v_K$ = 29.8 $a^{-\frac{1}{2}}$ (km s$^{-1}$) is the Keplerian velocity, and the orbital elements refer to the planetesimal.

By using the eccentricities and inclinations of excited planetesimals found in our pre-migration systems (e.g., Figs. 11 and 12), equation (2) yields catastrophic collisional velocities. Consequently, the trans-Neptunian belt experienced quite intense collisional grinding during this early period, and possibly after migration as well (Fig. 13). In other words, the outer planet was the smoking gun enhancing total mass loss in the disk.

By combining dynamical depletion (60–85%) (this work) and collisional grinding (92–97%) (Kenyon & Bromley 2004a), we obtain (0.15–0.4) (0.03–0.08) ≈ 0.5–3% as the remainder of the original belt mass. This is certainly an upper limit, though. An overabundance of stable resonant populations results in the underestimation of dynamical depletion fractions. That is, stochastic migration (not modeled here) suggests a smaller probability of resonance capture (e.g., Zhou et al. 2002). Moreover, collisional grinding estimates are based on planetesimal disks under much less excited conditions because no external perturbers such as the outer planet proposed here were considered in these studies.

7.9 Nature of Neptune's Migration

Neptune's migration is fed by planetesimals that can be scattered (transferring angular momentum), which follows the relation $\dot{a}_N \propto \eta(t)$, where $\eta(t)$ is a function describing the total mass in planet-encountering orbits. This behavior has been observed in simulations with massive disks (Levison et al. 2007). Past studies also claimed that Neptune should migrate beyond 30 AU for massive disks extending out to 50 AU or beyond (Gomes et al. 2004). To solve this problem, Gomes et al. (2004) proposed that Neptune would have stopped at 30 AU because it reached the original planetesimal disk outer edge at ~30–35 AU.

In our scenario, could Neptune stop at 30 AU in a 50–60 AU-sized disk with embedded planetoids? First, we considered only the four giant planets in compact orbital configurations. We prepared disks composed of 10000–20000 equal-mass bodies with an inner edge at ~10–20 AU and extended to 50–60 AU (27 runs). In 88 runs, we included a 0.5-$M_\oplus$ planetoid in diverse initial orbits (e.g., near Neptune or already eccentric with $i$ = 10° at 50–70 AU). Finally, in 16 other runs we included 5–10 planetoids embedded in each disk. In all of these runs, we used massive disks with objects in near-circular orbits and $i \approx 0$. All disks followed an $R^{-g}$ distribution with $g = 3/2 \pm 1/2$ and total mass corresponding to 1, 0.9, 0.8, 0.5, 0.3, 0.25, and 0.1 times the minimum mass solar nebula (MMSN). The surface densities used at 40 AU were σ = 0.14, 0.125, 0.11, 0.07, 0.04, 0.035, and 0.014 g cm$^{-2}$, respectively. We considered less massive disks (<1 MMSN) as an expected result of collisional grinding during the pre-migration epoch (see Kenyon et al. 2007 and references therein). The outcomes of these simulations resulted in Neptune migrating over 30 AU for most 1-MMSN disks. However, we found that in almost all 0.9–0.5-MMSN disks, Neptune reached ~25–30 AU after a few hundred million years (Fig. 22). Conversely, the giant planet ended at smaller orbital radii for less massive disks. These results are valid for disks with and without planetoids.

Therefore, provided that the disks evolved collisionally before Neptune's migration, acquiring total masses <1 MMSN, Neptune can stop at 30 AU without requiring the disk to be truncated at 30–35 AU. It is worth noting that this can be true even in 50–60 AU-sized 1-MMSN disks suffering run-out of feeding planetesimals (damped migration), which can be caused by asymmetries in the planetesimal disk. In addition, a close encounter or giant collision with a massive ancient planetesimal during migration could also change the migration behavior of the planet, leading to damped migration or even forcing the planet to stop migrating (Ida et al. 2000b; Gomes et al. 2004; Murray-Clay & Chiang 2006; Chiang et al. 2007; Levison et al. 2007). Such giant impacts involving large planetesimals and giant planets are the most accepted theory to explain the obliquities of Neptune and Uranus (Brunini et al.

2002; Lee et al. 2007). In support of these hypotheses, Neptune acquired 30 AU in one of our 1-MMSN disks due to the lack of planet-encountering planetesimals (Fig. 22). In addition, giant collisions of planetoids with Uranus and Neptune occurred in two other runs within the first 100 Myr. Due to the impact events, the planets were dislodged by ~0.5–1 AU, significantly affecting their migration behavior. Finally, mutual resonance crossings between giant planets also altered the migration evolution of Uranus and Neptune due to quick radial displacements of both planets.

## 8. DISCUSSION

The hybrid outer planet model can consistently explain the main features of all observed populations and satisfy other constraints (Section 2.1), indicating the robustness of our proposed unseen planet scenario. The results and the likelihood of this scenario (e.g., there were more less massive planetesimals in the past) tend to favor planetoids with 0.3–0.7 $M_\oplus$, but 1-$M_\oplus$ planetoids are also considered for completeness. Finally, the dependence of results on the initial planetoid's angular elements or the integrator used was negligible.

### 8.1 Resonances and Their Role in the Scattered Disk

Resonant configurations are a natural outcome after all dissipating forces are gone (e.g., gas drag, tidal effects, disk torques). Indeed, planets, satellites, and other minor bodies have a preference for commensurabilities in the solar system (Dermott 1973). Giant planets are in near mutual resonances, namely 2:5 (Saturn–Jupiter), 1:3 (Uranus–Saturn), and 1:2 (Neptune–Uranus) (Malhotra 1998). Prominent resonance structures do exist in the asteroid belt and the trans-Neptunian region (Nesvorný et al. 2002; LM07b). Pluto is in 3:2 resonance with Neptune. Resonant configurations have been found in extrasolar systems too (Udry et al. 2007). Therefore, as discussed in Section 6, the possibility of the outer planet occurring inside a distant resonance with Neptune is in line with the above reasoning.

The scenario presented here relies essentially on the capture of the outer planet in an $r$:1 (or $r$:2) resonance (Sections 3.3 and 6). In addition, according to the best results, the planetoid must have spent at least tens of Myr within about 60–80 AU before migration (see also Section 7.7). Which $r$:1 resonances were located within this distance range in the past? There is not yet consensus about Neptune's location at the beginning of planet migration. Thus, if the giant planet was initially located within 15–24 AU prior to migration, the best candidate $r$:1 resonances to trap the planetoid range from 6:1 to 12:1. Alternatively, those resonances or even farther $r$:1 resonances—in particular, similarly strong 13:1 and 14:1 resonances—could lock the planetoid after planet migration (Table 7. See also Section 9.2). What would be the likelihood of the capture of the planetoid into these resonances?

Recent studies using planet migration have shown that the capture probability of initially scattered bodies in the 9:4, 7:3, 5:2, 8:3, and 3:1 resonances are 0.5–1%, 0.5–1.2%, 2–3.5%, 0.5–1.2%, and 2–3%, respectively (Lykawka & Mukai 2007a). In supplementary simulations, we found capture probabilities of 1.5–2.6% and 1.3–2.2% in the 3:1 and 4:1 resonances, respectively (using $\tau$ = 1–10 Myr). Capture probabilities are linked to resonance strength and stickiness, both of which can be determined quantitatively (e.g., Gallardo 2006b; Lykawka & Mukai 2007c). Based on these quantities, we calculated capture probabilities in the 6:1, 7:1, ..., 14:1 resonances to be within a factor of ~0.25–1 times that found for the 3:1 and 4:1 resonances, finding roughly 0.5–3% for the $r$:1 resonances considered in our scenario (smaller for farther resonances) (Table 7). These capture probabilities are quite low. Nevertheless, there are at least two important points favoring captures in an $r$:1 resonance. First, during pre-migration stages, the giant planets were in a more compact orbital configuration. Consequently, all of Neptune's resonances were mutually closer. For instance, for initial locations of Neptune at 15–20 AU, the $r$:1 resonances of interest are spaced only 5–10 AU apart (or about half if we include $r$:2 resonances). Therefore, a scattered planetoid would have traversed a distance range (e.g., 60–80 AU) with a higher density of $r$:1 and $r$:2 resonances. Second, the probability of capture in these high-order resonances is maximized at high eccentricity, a region of space in which resonance width is wider (e.g., Murray & Dermott 1999). Thus, because the outer planet was initially scattered by Neptune, its eccentric orbit would favor resonance capture (see also Fig. 23).

In conclusion, the capture of the outer planet in an $r$:1 (or $r$:2) resonance during the early solar system appears likely. Moreover, recalling that large populations of massive planetesimals have appeared in the past, the likelihood of this scenario becomes plausible and consistent.

8.2 Primordial Planetesimal Disk Size and the Existence of TNOs Formed in situ Beyond 48 AU

On the basis of orbital distributions and the properties of objects at the end of our simulations (especially distant resonant populations and detached bodies around 50–60 AU), we found that the ancient trans-Neptunian belt appeared to have a radius of at least 50 AU. Moreover, disk planetesimals suffered eccentricity excitation with negligible radial changes. This disk size thus not only supports the formation of the observed distant resonant populations (Lykawka & Mukai 2007a), but also suggests that the planetesimal disk extended beyond the apparent outer edge at $a \approx 48$ AU. Further observational evidence is the existence of two detached TNOs, 2003 UY$_{291}$ ($a = 49.3$ AU; $q = 41.3$ AU) and (48639) 1995 TL$_8$ ($a = 52.6$ AU; $q = 40.0$ AU). Interestingly, their stable orbits and proximity to the outer trans-Neptunian belt suggest that these objects represent the continuation of the classical region beyond 48 AU. Therefore, 2003 UY$_{291}$ and 1995 TL$_8$ could be members of the primordial planetesimal disk prior to eccentricity pumping. If true, the minimum size of the ancient trans-Neptunian belt would be about 53 AU.

The future identification of long-term populations in distant resonances can also constrain the extension of the primordial planetesimal disk. According to our scenario, we found that the number of bodies, orbital elements, and other properties in the 5:2, 8:3, 11:4, and 3:1 resonances were different for disks extended to ~51–54 AU and 60 AU (see Table 5).

In more extended trans-Neptunian belts (>53 AU), local planetesimals perturbed by the outer planet acquired moderately eccentric orbits and remained after 4 Gyr, forming excited wings beyond 50 AU (Figs. 11, 13, and 14). In principle, these orbital structures are detectable by surveys. However, the timescale for accretion growth is proportional to $R^3$. Considering that the largest cold classical TNO at 45 AU has a diameter of about 600 km (assuming an albedo $p = 0.05$), its size at 55 and 65 AU would be a factor of 0.55 (330 km) and 0.33 (200 km) for the same growth time span, respectively. If planetesimals were perturbed by the planetoid during their formation era, the excited eccentricities and inclinations would lead to nonaccreting collisions (see eq. [2] and Section 7.8). Therefore, TNOs formed in situ beyond 48 AU may have suffered a growth cutoff. As a result, these distant local TNOs may be intrinsically smaller, perhaps with maximum diameters within 150–200 km, thus justifying their present nondetection by surveys (Trujillo et al. 2001a; Allen et al. 2002). Lastly, the characterization of an anomalous size distribution among some TNOs at $a > 45$–60 AU would also support the idea of interrupted growth. Indeed, the distribution of the largest (i.e., brightest) classical TNOs showing an apparent preference at $a < 45$ AU appear to support the above scenario (see also Lykawka & Mukai 2005b). The results of state-of-the-art collisional models and their further comparison with physical properties of TNOs provide additional evidence for external perturbers in stirring the primordial planetesimal disk (Kenyon et al. 2007).

### 8.3 TNOs of Interest

In this section, we briefly discuss the production of some TNOs of interest according to our scenario.

*(148209) 2000 CR$_{105}$ and (90377) Sedna.* These objects are currently the most distant detached TNOs, and their origin is of great importance for understanding the history of the solar system (Gladman et al. 2002; Brown et al. 2004). The outer planet model produced several analogs of 2000 CR$_{105}$ in almost all runs, even with different initial conditions for the planet (Figs. 19 and 20). The 2000 CR$_{105}$-like bodies were usually first scattered by Neptune and later detached from the giant planet's domain by external perturbation from the outer planet. Although the model had difficulties in creating Sedna-like bodies with less massive planetoids ($\leq 0.5\ M_\oplus$) at $a_P < 160$ AU (Fig. 19 and 20), more "Sednas" were commonly obtained in runs using planetoids with $M_P = 0.7\ M_\oplus$ and $1\ M_\oplus$, and/or located at $a_P = 200$–250 AU (the outermost boundaries for resonance interactions to work. See also Fig. 20). The analogs of Sedna originated in two possible ways: (a) near-circular objects located at 120–210 AU, which were later scattered by the outer planet (Sedna's accretion would be possible at these distances; Kenyon & Bromley 2004b; Stern 2005); and (b) objects initially scattered by Neptune that were detached by the influence of the planetoid. In both cases, long timescales (>1 Gyr) were a common requirement. Lastly, we stress that our model is able to produce more extreme detached TNOs with $q$ ~100–200 AU and various inclinations. Alternative scenarios to produce Sedna-like objects include a passing star (Kenyon & Bromley 2004b; Morbidelli & Levison 2004), formation of the solar system in a dense star cluster (Brasser et al. 2006), massive (1–2 $M_\oplus$) rogue planets that may have roamed the solar system at hundreds of AU in the past (Gladman & Chan 2006), and a solar companion model (Matese et al. 2005).

*2004 XR$_{190}$.* This is a very high-$i$ object in a detached orbit located close to the planetary system ($a = 57.5$ AU; $q = 51.2$ AU). The discovery of a TNO with such orbital properties was unexpected (Allen et al. 2006). In our model, analogs of 2004 XR$_{190}$ were obtained in several runs due to the planetoid's perturbation. A few are visible in Figs. 19 and 20. It is noteworthy that all 2004 XR$_{190}$-like objects originated from Neptune-scattered orbits, suggesting that this TNO was a scattered body in the past. A similar orbital history has been suggested previously (Allen et al. 2006; Gladman & Chan 2006). Alternatively, being close to the 8:3 resonance, 2004 XR$_{190}$'s small $e$ and very high $i$ could arise through the KR, which is expected to induce such $e$–$i$ anticorrelated configurations. However, the likelihood of being an 8:3-resonant TNO is only 4–5% (Allen et al. 2006; LM07b). On the other hand, the TNO could have left the 8:3 resonance in its very high-$i$ phase during planet migration (Gomes et al. 2007).

*(136199) Eris.* This planetary-sized very high-$i$ body is classified as a scattered TNO (LM07b). We obtained some analogs of Eris at $a < 100$ AU in our simulations ($i = 40$–$45°$). Although the outer planet excited inclinations, Eris-like objects managed to remain with $q < 40$ AU at the end of the runs (e.g., Fig. 19). Therefore, a probable scenario for the history of Eris would be formation near Neptune, further scattering by the giant planet, and extra $i$ excitation by the outer planet. It is worth noting that this agrees with the idea that a small fraction of Pluto-sized objects that were initially located near the giant planets were deposited in the trans-Neptunian belt (Stern 1991).

*(134340) Pluto*. The largest TNO known thus far in the 3:2 resonance was fully reproduced in our model. Pluto-like objects were obtained in substantial number in the 3:2 resonance, reproducing not only Pluto's orbital elements, but also its libration amplitude and the locking in KR over the age of the solar system (Fig. 16 and 18). This supports the idea that Pluto was initially located in a near-circular orbit closer to the Sun during the capture by the sweeping 3:2 resonance (Malhotra 1995; Hahn & Malhotra 2005).

*(136108) 2003 $EL_{61}$*. Apart from the Pluto quadruple system, 2003 $EL_{61}$ is currently the only other known multiple system in the trans-Neptunian belt, with two satellites (Brown et al. 2005b; Brown et al. 2006b). In addition, 2003 $EL_{61}$ is currently locked in the 12:7 resonance ($a = 43.1$ AU) (LM07b). According to our results, there were four 12:7-resonant objects, but only two of them survived over 4 Gyr. However, the orbits of these objects are not compatible with that of 2003 $EL_{61}$. Interestingly, 2003 $EL_{61}$ appears to be part of a collisional family (Brown et al. 2007), supporting the giant impact theory for the creation of its satellite system. Therefore, we think that 2003 $EL_{61}$ was probably captured in the 12:7 resonance after the collision event; otherwise, it would have been dislodged from the resonance with such a quite energetic impact.

*2003 $UY_{291}$* and *(48639) 1995 $TL_8$*. The outer planet model can reproduce the orbital elements of these detached TNOs quite easily (e.g., Fig. 16). Our results suggest that both of these objects are part of the excited outer region of a primordial planetesimal disk of at least 53 AU radius (Section 8.2).

8.4 Clues from the Distribution of Trans-Neptunian Binary/Multiple Systems

Several TNOs have been identified as binary or multiple systems (Noll et al. 2007). Interestingly, there is a statistically significant preference for binarity among classical TNOs with $i < 10°$ compared with the other TNO populations (approximately a 3–4 times greater fraction) (Stephens & Noll 2006). Because close gravitational encounters are likely to disrupt weakly bound binaries, this preference can be interpreted as evidence that low-$i$ classical TNOs suffered little perturbation from the giant planets, in contrast to the active dynamical evolution of high-$i$ classical, resonant, scattered, and possibly detached populations. Alternatively, early conditions in the planetesimal disk may have favored binary formation in specific regions of the disk. Thus, more binary systems may have formed at about 35–50 AU than in the inner regions of the disk, so that the binarity preference among low-$i$ classical TNOs would represent a primordial feature (Noll et al. 2006; Noll et al. 2007 and references therein). In any case, either possibility is consistent with the view that cold classical TNOs formed in situ (Section 2.1).

According to our findings (Sections 7.2 and 8.2), the bulk of the population of classical bodies with $i < 10$–$15°$ formed in situ and extended to at least $a = 50$ AU. These bodies were perturbed solely by the outer planet before migration (e.g., Fig. 11) and later survived for 4 Gyr essentially unaffected by any of the planets. Thus, the preference for binarity among low-$i$ classical objects could result from the lack of continuous perturbation from the giant planets and the planetoid over 4 Gyr and/or formation in situ in the classical region, where supposedly more binary systems were formed during the pre-migration epoch. The binary TNOs 1995 TL$_8$ ($a$ = 52.6 AU; $i$ = 0.2°) and 2002 GZ$_{31}$ ($a$ = 50.2 AU; $i$ = 1.1°) provide additional evidence for our scenario. That is, both TNOs have orbital properties compatible with those obtained for objects at the excited wings around 50–53 AU in our simulations (Figs. 12–14), suggesting that both TNOs are members of the primordial planetesimal disks that extended to 53 AU (see also Section 8.2).

8.5 Remarks on the Outer Planet Orbital Dynamical History

In accordance with the results of Melita et al. (2004), we found that the strength of the planetoid's perturbation depends essentially on $a_P$, its relative longitudes with disk planetesimals, and $M_P$ (e.g., Fig. 12). According to our self-consistent simulations (Section 7.9), the dynamical evolution of a planetoid in a massive planetesimal disk is complex and stochastic. The planetoid's initially scattered orbit and the properties of the disk such as its total mass and size can lead to diverse evolutionary paths. If the disk is relatively small, the effect of dynamical friction is reduced, implying that the outer planet could survive a much longer time in a typical scattered disk orbit. In extended disks (e.g. 100 AU), dynamical friction is stronger and the timescales for circularization and flattening of the orbit ($i$ near the plane of the disk) become smaller.

Figure 23 illustrates two representative examples of the evolution of scattered planetoids through massive disks. These bodies spent a few tens of Myr under the control of $r$:1 resonances during different stages of the pre-migration systems. Although not exhaustive, these examples show that the early capture of a large planetesimal in resonances would be possible.

## 9. PROSPECTS FOR THE EXISTENCE OF A TRANS-PLUTONIAN PLANET

Is a resident planetoid necessary to explain the observations? How is it possible to detect this massive outer planet? What are the orbital and physical properties of a hypothetical planetoid orbiting at a large distance?

9.1 A Resident Trans-Plutonian Planet: What do the Observations Tell us?

Below, we provide evidence supporting the current existence of a trans-Plutonian planet as described in our scenario.

- *A prominent population of detached TNOs*. To explain the detached population and in particular, its large total population number (e.g., its intrinsic ratio to the scattered population is ≥1.0), a quite broad inclination distribution (currently from a few degrees to almost 50°), and peculiar members (e.g., 2004 XR$_{190}$, 2000 CR$_{105}$, Sedna), the perturbation of the planetoid over 4 Gyr is probably the most plausible mechanism (see Sections 2.1, 2.2, and 7.5). Furthermore, other distant and detached TNOs were reproduced only when the gravitational influence of the outer planet lasted over billions of years. Note also that even very massive outer planets (1 $M_\oplus$) are apparently unable to explain all these features in timescales less than 1 Gyr (Table 6);

- *A population of TNOs with inclinations larger than 40°.* As demonstrated in Section 2.1, the existence of a significant fraction of very high-*i* TNOs cannot be explained according to previously published models. On the other hand, in our scenario, apart from gravitational scattering by Neptune, the planetoid provides an extra perturbation that, on average, increases the inclinations of objects in the whole trans-Neptunian region (see Fig. 19). Finally, to better match the estimated intrinsic fractions of the very high-*i* subclass, the outer planet's perturbation had to last longer than 3 Gyr (see also Sections 7.4–7.6);

- *Lack of long-term resonant populations beyond the 8:3 resonance.* Although 3:1-, 7:2-, and 4:1-resonant populations stable over 4 Gyr (located beyond the 8:3 resonance) could occur in the trans-Neptunian region according to recent models (Hahn & Malhotra 2005; Lykawka & Mukai 2007a), no evidence exists of such populations (LM07b). Despite the low numbers of TNOs in the scattered disk, this intriguing feature is unexplained by other scenarios. However, here the absence of long-term resonant TNOs at the 3:1 (at high eccentricities) and greater resonances can be well explained by the gravitational perturbation of an outer planet with $q_P$ = 80–90 AU in timescales comparable to the age of the solar system (e.g., see Sections 5.3 and 7.3.1).

Taking together all the aforementioned observational evidence, we conclude that a massive undiscovered trans-Plutonian planet is required to explain them consistently.

9.2 Orbital and Physical Properties

The possible capture in a distant resonance suggests $a_P \sim$ 100–175 AU near current locations of 6:1–14:1 resonances. It is noteworthy that large *e–i* variations and long-term resonant capture during the planetoid's orbital evolution would have been more likely to occur in those resonances (Lykawka & Mukai 2007c and references therein). However, the planetoid may have left its resonant configuration during the stochastic migration of Neptune, after dynamically diffusing at low eccentricity or by dynamical friction (Hahn & Malhotra 1999; Zhou et al. 2002; Gomes et al. 2007). Indeed, stochastic jumps in semimajor axis caused by continuous scattering may have allowed the planetoid to encounter resonances beyond 175 AU. Nevertheless, the presence of sufficiently strong resonances is limited at $a \leq 250$ AU (Gallardo 2006a, 2006b; Lykawka & Mukai 2007c). Finally, as discussed in Sections 5.3 and 6, the model suggests $q_P >$ 80 AU and $i_P =$ 10–50°, although 20–40° were apparently more commonly obtained in self-consistent simulations.

Concerning the physical properties of the outer planet, ideally we would like to constrain these from the distributions of planetary-sized TNOs. However, the sample of objects with determined physical properties is still very small (Luu & Jewitt 2002; Delsanti & Jewitt 2006; Delsanti et al. 2006; Cruikshank et al. 2007).

Simultaneous measurements of reflected and emitted light from the object of interest permit us to calculate albedos and diameters using well-established formulas (Russell 1916; Jewitt & Sheppard 2002; Jewitt 2006),

$$p\left(\frac{D}{2}\right)^2 \Phi(\alpha) = 2.25 \cdot 10^{16} R^2 \psi^2 10^{0.4(m_* - m)} \quad (4),$$

where *D* is in kilometers, $\Phi(\alpha)$ is the phase function, *R* is in AU, $\psi$ is the geocentric distance (AU), and *m* is the apparent magnitude of the object ($m_*$ represents that of the Sun). For the sake of simplicity, we assume that the phase function is negligible for our purposes.

A massive trans-Plutonian planet would be very probably a differentiated body with a rocky interior and layers composed of ices (e.g., De Sanctis et al. 2001; Merk & Prialnik 2006). In such a case, we can expect mean densities of ~2–3 g cm$^{-3}$. This density range is consistent with 2 g cm$^{-3}$, 2.3 g cm$^{-3}$, and 2.6 g cm$^{-3}$ densities determined for Pluto, Eris, and 2003 EL$_{61}$, respectively (Stern 1992; Rabinowitz et al. 2006; Brown & Schaller 2007; Lacerda & Jewitt 2007). We then estimated the planetoid's diameter assuming a spherical shape, $M_P = 0.3$–$0.7$ $M_\oplus$, and a mean density of 2–3 g cm$^{-3}$, finding a $D_P$ of ~10000–16000 km. Using the equation of density as a function of diameter given in Jewitt (2006), we obtain 2.1–2.4 g cm$^{-3}$ for a 10000–16000-km-sized body.

Considering the planetoid's orbital properties given above ($a_P \geq 100$ AU; $q_P > 80$ AU), we assumed an albedo $p_P = 0.1$–$0.3$ for the outer planet, which agrees with expected values for hypothetical distant icy planetoids (Stern 1991). For instance, a planet with $q_P \approx 100$ AU would be too far to hold an atmosphere, so surface darkening would be expected (~0.03–0.3). The lower albedos of these distant bodies may be a consequence of long exposure to space weathering because icy surfaces are expected to get darker from continuous UV irradiation and bombardment of solar and galactic ions (Thompson et al. 1987; Moroz et al. 1998; Cooper et al. 2003; Brunetto et al. 2006). Other potential reasons include very low temperatures that inhibit the sublimation of volatiles and negligible collisional resurfacing effects. Furthermore, our albedo assumption above also agrees with Sedna's albedo, which was constrained in the range 0.1–0.3 (Emery et al. 2007 and references therein). Sedna is the best-known representative of distant large icy bodies.

The planetoid possibly accreted near Uranus and Neptune during the pre-migration period (probably at the 10–20 AU region), so it would be composed mainly of $CH_4$, $CO$, or $N_2$, which are the most common outer solar system volatiles. Given its appreciable mass, the outer planet would probably have large reservoirs of icy species on the surface. For instance, recent spectral measurements support the existence of large amounts of $CH_4$, $H_2O$, and $N_2$ on the surface of planetary-sized TNOs (Brown et al. 2005a; Barkume et al. 2006; Licandro et al. 2006a; Licandro et al. 2006b; Rabinowitz et al. 2006; Dumas et al. 2007; Tegler et al. 2007; Trujillo et al. 2007). Besides, due to the orbital characteristics and large diameter of the planetoid, it should have retained its surface ice over the age of the solar system (Schaller & Brown 2007). Therefore, we should expect an inactive surface and the presence of heavily space-processed ice on the planetoid.

Finally, using equation (4) with the derived ranges of $D_P$ and $p_P$ above, we found $m_P$ ~15–17 mag for a planetoid with $q_P = 80$–$90$ AU during perihelion approach. The predicted apparent magnitude is uncertain because the planetoid's orbital properties and assumed albedo cannot be more strictly constrained.

9.3 Plausibility of Detection by Surveys

First, it is important to understand some physical quantities derived from surveys and the strategies used in the surveys. The cumulative luminosity function (CLF) provides the cumulative number of TNOs per sky surface area at a given limiting magnitude. The CLF is given by $\log[\Sigma(m)] = b(m - m_0)$, where $\Sigma(m)$ is the number of objects per square degree brighter than the apparent magnitude $m$, $b$ is a constant defining the slope of the curve, and $m_0$ is a constant for which $\Sigma = 1$ deg$^{-2}$. Recent studies suggest that $b = 0.6$–$0.8$, and $m_0 \approx 23$ mag (Trujillo et al. 2001b; Bernstein et al. 2004). The size distribution is usually derived from the CLF, described by a power law with exponent $u = 5b + 1$ (Bernstein et al. 2004 and references therein).

A straightforward interpretation of the CLF implies smaller sky densities for brighter bodies (bigger TNOs), and larger sky densities for fainter ones (smaller TNOs). Therefore, large TNOs have a greater probability of being discovered in volume-limited surveys (i.e., wide-area surveys) than in flux-limited surveys. Regarding the outer planet, we estimate that there would be one object every $\sim 10^4$–$10^5$ square degrees as bright as $m_P \sim 15$–$17$ mag (at perihelion); hence, it would be rare in the sky.

Surveys usually concentrate the search near opposition when a TNO has the highest relative sky motion due to parallax. Indeed, the probability of discovery of large TNOs is also affected by their apparent rate of movement across the sky, which can be determined using

$$\Theta \cong \frac{148}{R + \sqrt{R}} \text{ (arcsec h}^{-1}\text{)} \quad (5),$$

where $R$ is in AU. For typical trans-Neptunian belt orbits (30–60 AU), TNOs move at a rate of 2–5 arcsec h$^{-1}$. Because the majority of surveys has used the strategy of finding TNOs that move at such apparent sky motions, they are sensible at most to 1.5 arcsec h$^{-1}$ (Brown et al. 2004). For example, when Eris and Sedna were discovered (the most distant TNOs discovered thus far), they were moving at similar critical rates, i.e., 1.42 and 1.75 arcsec h$^{-1}$, respectively (Trujillo & Brown 2003; Brown et al. 2005a). In our scenario, for an outer planet with $q_P = 80$–$90$ AU, we obtain $\Theta_P = 1.5$–$1.7$ arcsec h$^{-1}$ during perihelion approach, near the lower limit discussed above. In addition, having a moderately large $e_P$ (Table 7), the outer planet would also spend more time near aphelion during its orbit, thus implying smaller apparent rates (i.e., more difficult to detect). Only a few recent wide-area surveys have probed apparent motion below these critical values (Larsen et al. 2007).

Past wide-area surveys have searched for very bright bodies in the outer solar system ($m < 17$ mag). Tombaugh (1961) concluded that there were no objects brighter than $m = 15.5$ mag in the entire sky north of a declination $-40°$, and no bodies with $m < 17.5$ mag for any object near the ecliptic (see also Trujillo et al. 2001b). Kowal (1989) surveyed the sky near the ecliptic for objects with $m < 20$ mag and found no TNOs. However, photographic plates have poor sensitivity for slowly moving objects and can present surface defects (Trujillo et al. 2001b). Except for Pluto, only recent planetary-sized TNOs have been discovered by dedicated surveys (Brown et al. 2004, 2005a). For reference, Pluto, Eris, 2003 EL$_{61}$, 2005 FY$_9$, and Sedna had apparent magnitudes of 14, 18.8, 17.5, 16.8, and 21 at discovery, respectively (Brown et al. 2006b). Other wide-area surveys have not found any very bright TNOs within about 10° (Sheppard et al. 2000; Trujillo & Brown 2003; Jones et al. 2006; Larsen et al. 2007) (see also Fig. 24).

In addition, recalling that the planetoid would have an inclined orbit (20–40°), we stress that (a) high-$i$ objects are approximately four times more likely to be discovered at ecliptic latitudes, $\beta \approx i$, than in the ecliptic ($\beta = 0°$); (b) $\Sigma(m)$ is smaller at higher $\beta$ and drops off quickly (for instance, the sky density at 20° is about one order of magnitude smaller than that near the ecliptic); and (c) among high-$i$ objects (20–40°), the fraction of an orbit spent near the ecliptic ($\beta = 0$–$10°$) is only $\sim 1.5$–$4\%$ (Trujillo et al. 2001a). In general, because wide-area surveys have searched near the ecliptic, the discovery of objects with higher inclinations has been less likely, even in surveys probing very slowly moving objects (Tombaugh 1961; Luu & Jewitt 1988; Kowal 1989; Jewitt & Luu 1995; Jewitt et al. 1998; Sheppard et al. 2000; Gladman et al. 2001; Trujillo et al. 2001a; Trujillo et al. 2001b; Trujillo & Brown 2003; Jones et al. 2006; Larsen et al. 2007).

Therefore, a massive trans-Plutonian planet has probably escaped detection because it is currently either far from the Sun (with sky motion below survey sensibility) or away from the nodal crossing point with the ecliptic. Apart from the planet's properties, the probability of nondetection should be ~10–15% because surveys avoid the region near the galactic plane (Chiang & Jordan 2002; Trujillo & Brown 2003). Other potential biases that may have prevented the detection of massive bodies in distant orbits are discussed in Bernstein & Khushalani (2000) and Horner & Evans (2002).

## 10. CONCLUSIONS AND SUMMARY

We performed extensive numerical simulations to study the origin and evolution of the entire trans-Neptunian region (>30 AU) with the presence of massive planetoids (0.01–1.0 $M_\oplus$) in trans-Plutonian orbits covering a wide range of orbital parameters ($a_P$, $q_P$, and $i_P$). The results allowed us to clarify the effects of a putative outer planet within the solar system.

First, we found that a static trans-Plutonian planet is unable to reproduce the fine architecture of the trans-Neptunian belt, in particular the excited orbital structure of classical TNOs and the outer edge of the belt. Note that the main orbital element connected to this conclusion is the planetoid's perihelion distance, implying that the semimajor axis and inclination play a negligible role. Moreover, a putative Moon-like or more massive body in the 2:1 resonance also cannot satisfy the constraints of Section 2.1.

The stability and orbital properties of currently known trans-Neptunian resonant populations provide important constraints on the presence of massive planetoids. Indeed, we concluded that the existence of 3:2-, 2:1-, and 5:2-resonant TNOs imply that any hypothetical resident distant planet must have an overall $q_P > 80$ AU. Therefore, we believe we can rule out *all* previous modern models based on the unseen planet scenario (Brunini & Melita 2002; Melita & Williams 2003; Brown et al. 2004; Melita et al. 2004) because they suggest resident planets with $q_P \leq 70$ AU. Earlier models (e.g., Matese & Whitmire 1986; Harrington 1988; Hogg et al. 1991; Maran et al. 1997 and references in those papers) are ruled out by the same reasoning and also on observational grounds (Hogg et al. 1991; Morbidelli et al. 2002; Gaudi & Bloom 2005).

In addition, in attempting to satisfy the constraints described above, we found that a migrating trans-Plutonian planet represents an excellent way to reconcile the excitation of the trans-Neptunian belt and the preservation of the latter, in particular its resonant populations. However, this modified model with a migrating planetoid does not form distant resonant populations (beyond 50 AU), in particular in the 9:4, 5:2, and 8:3 resonances (e.g., see Lykawka & Mukai 2007a). We solved this problem by allowing the planetoid to perturb the trans-Neptunian belt well before planet migration.

Therefore, we constructed a refined model divided into three main phases: pre-migration excitation of the belt (30–100 Myr), planet migration (100 Myr), and long-term sculpting (4 Gyr). We assumed the outer planet was originally a Uranus- or Neptune-scattered massive planetesimal, thus probably a member of a large population of bodies that formed in the vicinity of the giant planets at 10–20 AU during the late stages of planet formation (e.g., Stern 1991; Kenyon 2002; Goldreich et al. 2004a, 2004b). In fact, our results showed that planetesimal disks excited early by a scattered outer planet were remarkably similar to current observations in the 40–50 AU region. Moreover, the planetoid's perturbation also provided the necessary trans-Neptunian belt orbital conditions to satisfy the formation of distant resonant populations. In the migration phase, recalling the important role of strong $r$:1 and $r$:2 resonances in the scattered disk and the common outcome of the Kozai resonance inside those resonances, the outer planet was probably transported outwards by resonant interactions with and gravitational scattering by Neptune. Thus, these mechanisms emplaced the planetoid somewhere around $a_P$ = 100–175 AU (or less likely at 175–250 AU) and changed its orbital elements in a few hundred Myr (i.e., increase of perihelion and inclination), guaranteeing the stability of the trans-Neptunian belt over the age of the solar system.

After integrating tens of planetesimal disks and following a few of the obtained systems over 4 Gyr, we found that our model naturally explains the orbital characteristics of currently known TNOs and satisfies a large number of other constraints in unprecedented detail. Our scenario also suggests that the outer planet *did not* form in situ near its semimajor axis. That is, although theoretically possible (e.g., Kenyon & Bromley 2004b; Stern 2005), in situ formation would require the planet to undergo a very complex orbital history to satisfy all observational constraints discussed in this paper (e.g., inward migration to perturb the classical region followed by an outward transport to satisfy $q_P > 80$ AU).

Finally, we conclude that the orbital excitation at 40–50 AU and the truncation near 48 AU in the trans-Neptunian belt probably represent fossilized signatures of the outer planet, whereas the detached population and perhaps TNOs with $i > 40°$ resulted from the planetoid's perturbation over billions of years. Indeed, the existence of a substantial population of detached and very high-$i$ TNOs appears to require the presence of a resident distant planet within the solar system. In summary, our scenario reproduces all main aspects of trans-Neptunian belt architecture and offers insightful predictions (Section 10.3).

10.1 Tentative Chronological History of the Outer Solar System

- The planets, planetary embryos, and other TNOs formed in <100 Myr (Pollack et al. 1996; Kenyon 2002; Kenyon & Bromley 2004a). Uranus and Neptune formed within ~10–20 AU. The planetesimal disk was dynamically very cold (nearly circular and $i \approx 0$) and at least 53 AU in size;

- During the late stages of planet formation, a large planetesimal (a planetoid with 0.3–0.7 $M_\oplus$) was scattered by Neptune in a few Myr, experiencing an $a_P$ ~60–80 AU, $q_P$ near the location of Neptune, and probably a moderate $i_P$ (e.g., 10–15°);

- The outer planet excited the outer region of the planetesimal disk for several tens of Myr. This produced the first hot classical TNOs (moderate $i$). At the same time, the outer planet's perturbation also disrupted the disk at 48 AU;

- During planet migration, all resonant TNOs and the remaining hot classical TNOs were created;

- During or just after the end of planet migration, the outer planet was captured by a distant $r$:1 resonance (e.g., 6:1). The outer planet likely evolved into a KR orbit, which forced the planet to decrease its eccentricity and increase the inclination in 100–200 Myr ($q_P > 80$ AU, $i_P = 20$–$40°$). The trans-Plutonian planet was detached from the solar system, remaining in a stable orbit at ~100–175 AU (the exact location depends on the current specific $r$:1 resonance location);

- The giant planets and the planetoid gravitationally sculpted the trans-Neptunian belt over billions of years. The belt's outer edge appeared as unstable objects left the region around 45–53 AU. Scattered and detached TNO populations were produced mainly during the long-term phase. Cold classical TNOs, formed in situ, represent relics of the local disk planetesimals. The region beyond 30 AU lost 99% of its initial total mass by dynamical depletion and enhanced collisional grinding.

10.2 Main Achievements of the Hybrid Outer Planet Model

- Explains the lack of TNOs up to 39 AU (Figs. 13 and 14);
- Explains the dual nature of the classical region (cold and hot populations), including their orbital excitation and distinct physical properties. The final distributions in eccentricities are remarkably similar to observations (Fig. 16);
- Reproduces the resonant structure in the entire trans-Neptunian region, including eccentricities, inclinations, dynamical properties (e.g., libration amplitudes), and KR members (Figs. 14, 16, and 18 and Table 5);
- Reproduces the population of scattered TNOs with their distribution of orbital elements, including analogs of Eris (Figs. 16 and 17);
- Produces a substantial population of detached TNOs, including analogs of 2004 $XR_{190}$, 2000 $CR_{105}$, and Sedna. The obtained detached population is in agreement with intrinsic fraction estimates for this population in the scattered disk (Figs. 17, 19, and 20 and Table 6);
- Produces low-$i$ detached TNOs (Figs. 19 and 20);
- The very high-$i$ population can be produced by the simultaneous perturbation of the outer planet and the action of the KR mechanism inside resonances (Figs. 19 and 20);
- Produces the trans-Neptunian belt outer edge at 48 AU without threatening stable resonant populations, namely those in the 3:2, 2:1, and 5:2 resonances (Figs. 13, 14, and 16 and Table 3);
- Reproduces the abrupt decrease in number density of TNOs beyond 45 AU (Fig. 21);
- The problem of the missing mass of the trans-Neptunian belt can be solved by the simultaneous action of enhanced collisional grinding induced by the planetoid and dynamical depletion;
- Explains Neptune's current orbit at 30.1 AU (Fig. 22).

10.3 Predictions of the Hybrid Outer Planet Model

The following predictions are testable, so future observations will help to confirm and/or improve the model.

10.3.1 Outer Solar System

- Cold and hot classical bulk populations concentrated at $i < 5°$ and $i > 10$–$15°$, respectively, and mixed at intermediate inclinations. If cold and hot classical TNOs formed at different places in the planetesimal disk, their supposedly distinct physical properties will appear more evident within the respective inclination ranges above;

- Resonant TNOs stable over the age of the solar system in the following resonances: 1:1 ($a = 30.1$ AU), 5:4 ($a = 34.9$ AU), 4:3 ($a = 36.5$ AU), 7:5 ($a = 37.7$ AU), 10:7 ($a = 38.2$ AU), 3:2 ($a = 39.4$ AU), 8:5 ($a = 41.2$ AU), 5:3 ($a = 42.3$ AU), 12:7 ($a = 43.1$ AU), 7:4 ($a = 43.7$ AU), 9:5 ($a = 44.5$ AU), 11:6 ($a = 45.1$ AU), 13:7 ($a = 45.5$ AU), 2:1 ($a = 47.8$ AU), 13:6 ($a = 50.4$ AU), 11:5 ($a = 50.9$ AU), 9:4 ($a = 51.7$ AU), 16:7 ($a = 52.2$ AU), 7:3 ($a = 53.0$ AU), 12:5 ($a = 54.0$ AU), 17:7 ($a = 54.4$ AU), 5:2 ($a = 55.4$ AU), 8:3 ($a = 57.9$ AU), 11:4 ($a = 59.1$ AU), and 3:1 ($a = 62.6$ AU) (Fig. 14);
- Quite anemic resonant populations within 39 AU (Fig. 14);
- If there were favorable conditions for the formation of Gyr-resident TNOs in the 8:3 and other farther resonances, their orbits will be conditioned to $Q \leq q_P$ (Fig. 16);
- Kozai resonant TNOs in the following resonances: 5:4, 4:3, 7:5, 3:2, 8:5, 5:3, 7:4, 2:1, 7:3, 5:2, and 3:1 (Table 5);
- Detached TNOs beyond 48 AU with bulk $q = 40$–$60$ AU. A smaller fraction of detached TNOs with larger perihelia is also predicted. The detached population $i$-distribution would be 0–60° (up to 90° for more massive outer planets, $M_P = 1.0\ M_\oplus$), and weakly correlated with perihelion. In addition, objects with the highest inclinations will be concentrated near the planet's semimajor axis (Figs. 19 and 20);
- Very high-$i$ scattered and detached TNOs with $i = 40$–$50°$ and 40–60° (up to 90° for more massive outer planets, $M_P = 1.0\ M_\oplus$), respectively (Fig. 19 and 20);
- A relatively small population of Sedna-like bodies for $M_P \leq 0.5\ M_\oplus$;
- The existence of a massive resident outer planet in a distant orbit in the scattered disk.

10.3.2 Tentative Properties of the Resident Trans-Plutonian Planet

- Orbital elements: $a_P = 100$–$175$ AU (less probable, but also possible at 175–250 AU), $q_P > 80$ AU, $i_P = 20$–$40°$;
- Probably locked in or near a strong $r$:1 resonance with Neptune;
- Mass: 0.3–0.7 $M_\oplus$;
- Apparent sky motion: 0.35–1.7 arcsec h$^{-1}$ (eq. [5] and Fig. 24);
- Visual apparent magnitude: 14.8–17.3 at perihelion ($q_P = 80$–$90$ AU) (eq. [4] and Fig. 24).

**11. CAVEATS AND FUTURE WORK**

The unseen planet scenario presented here cannot explain all observational constraints; thus, unsolved issues remain. First, the mutual ratios of the populations of TNOs are not completely reproduced. Indeed, the model produced too many resonant objects. Because stochastic migration can yield a smaller total resonant population (Hahn & Malhotra 1999; Zhou et al. 2002), we intend to implement the effects of stochasticity in planet migration in the future. Another issue is the difficulty in obtaining hot classical objects with $i > 15$–$20°$, causing an overrepresentation of classicals at lower inclinations. Finally, if the real population of "Sednas" is large, the efficiency of our scenario might be too low to produce such extreme objects. We will further explore the production of Sedna-like bodies in upcoming refinements of our scenario.

To better constrain the proposed orbital elements of the outer planet, we plan to investigate the dynamics of comets, recalling that a group of comets could show peculiar features caused by an unseen distant planet (Horner & Evans 2002 and references therein; Kuz'michev & Tomanov 2006).

Regarding the dynamical evolution of the outer planet, we are aware that more extensive simulations with massive disks are necessary to better understand important physical phenomena not well explored here such as dynamical friction and the behavior of embedded planetoids (e.g., the probability of planetoids placement in distant and inclined orbits). We intend to address the nature of scattered planetoids by performing a much larger number of similar simulations to draw firmer conclusions and more detailed statistics. Furthermore, one could ask if a planetoid (or a system of planetoids) with lifetime less than the age of the solar system could satisfy the main observational constraints (Section 2.1). Although we showed several lines of evidence in favor of a resident planet in this paper, we cannot currently rule out such alternative variants of the scenario.

We stress the predictions of our model should not be regarded as "exact" because inevitable model uncertainties are involved, in particular the unexplored conditions in parameter space ($a_P$, $q_P$, $i_P$, $M_P$), lack of statistics about the behavior of large planetesimals in realistic self-consisting simulations (e.g., using much larger populations of planetesimals following mass distribution), and the absence of a fragmentation code to deal consistently with the collisional evolution of disk planetesimals.

Finally, an important unexplained event in our scenario is the Late Heavy Bombardment (LHB) (Chapman et al. 2007 and references therein). We intend to address the LHB in future investigations within the framework of our scenario.

**Acknowledgements**

We thank an anonymous referee for insightful and helpful comments. We also thank D. C. Jewitt for suggestions. This study was supported by The 21st Century COE Program of the Origin and Evolution of Planetary Systems of the Ministry of Education, Culture, Sports, Science and Technology, Japan (MEXT), and a MEXT scholarship. P. S. Lykawka is also grateful to a fellowship from the Japan Society for the Promotion of Science (JSPS).

**References**

Adams, F. C., Hollenbach, D., Laughlin, G., & Gorti, U. 2004, ApJ, 611, 360.

Agnor, C. B., & Hamilton, D. P. 2006, Nature, 441, 192.

Allen, R. L., Bernstein, G. M., & Malhotra, R. 2001, ApJ, 549, L241.

Allen, R. L., Bernstein, G. M., & Malhotra, R. 2002, AJ, 124, 2949.

Allen, R. L., Gladman, B., Kavelaars, J. J., Petit, J-M., Parker, J. Wm., & Nicholson, P. 2006, ApJ, 640, L83.

Barkume, K. M., Brown, M. E., & Schaller, E. L. 2006, ApJ, 640, L87.

Bernstein, G. M., & Khushalani, B. 2000, AJ, 120, 3323.

Bernstein, G. M., Trilling, D. E., Allen, R. L., Brown, M. E., Holman, M., & Malhotra, R. 2004, AJ, 128, 1364.

Bertoldi, F., Altenhoff, W., Weiss, A., Menten, K. M., & Thum, C. 2006, Nature, 439, 563.

Brasser, R., Duncan, M. J., & Levison, H. F. 2006, Icarus, 184, 59.

Brown, M. E. 2000, AJ, 119, 977.

Brown, M. E. 2001, AJ, 121, 2804.

Brown, M. E. 2002, Annu. Rev. Earth. Planet. Sci., 30, 307.

Brown, M. E., & Schaller, E. L. 2007, Science, 316, 1585.

Brown, M. E., Trujillo, C., & Rabinowitz, D. 2004, ApJ, 617, 645.

Brown, M. E., Trujillo, C. A., & Rabinowitz, D. L. 2005a, ApJ, 635, L97.

Brown, M. E., et al. 2005b, ApJ, 632, L45.

Brown, M. E., Schaller, E. L., Roe, H. G., Rabinowitz, D. L., & Trujillo, C. A. 2006a, ApJ, 643, L61.

Brown, M. E., et al. 2006b, ApJ, 639, L43.

Brown, M. E., Barkume, K. M., Ragozzine, D., & Schaller, E. L. 2007, Nature, 446, 294.

Brunetto, R., Barucci, M. A., Dotto, E., & Strazzulla, G. 2006, ApJ, 644, 646.

Brunini, A. 2002, A&A, 394, 1129.

Brunini, A., & Melita, M. D. 2002, Icarus, 160, 32.

Brunini, A., Parisi, M. G., & Tancredi, G. 2002, Icarus, 159, 166.

Canup, R. M. 2005, Science, 307, 546.

Chambers, J. E. 1999, MNRAS, 304, 793.

Chapman, C. R., Cohen, B. A., & Grinspoon, D. H. 2007, Icarus, 189, 233.

Chiang, E. I., Jordan, A. B. 2002, AJ, 124, 3430.

Chiang, E., Lithwick, Y., Murray-Clay, R., Buie, M., Grundy, W., & Holman, M. 2007, in Protostars and Planets V Compendium, ed. Reipurth, B., Jewitt, D., & Keil, K. (Tucson: Univ. Arizona Press), 895.

Chiang, E. I., et al. 2003, AJ, 126, 430.

Collander-Brown, S., Maran, M., & Williams, I. P. 2000, MNRAS, 318, 101.

Cooper, J. F., Christian, E. R., Richardson, J. D., & Wang, C. 2003, EM&P, 92, 261.

Cruikshank, D. P., Barucci, M. A., Emery, J. P., Fernandez, Y. R., Grundy, W. M., Noll, K. S., & Stansberry, J. A. 2007, in Protostars and Planets V Compendium, ed. Reipurth, B., Jewitt, D., & Keil, K. (Tucson: Univ. Arizona Press), 879.

Davis, D. R., & Farinella, P. 1997, Icarus, 125, 50.

De Sanctis, M. C., Capria, M. T., & Coradini, A. 2001, AJ, 121, 2792.

Del Popolo, A., Spedicato, E., & Gambera, M. 1999, A&A, 350, 685.

Delsanti, A., & Jewitt, D. 2006, in Solar System Update, ed. Blondel, Ph., Mason, J. (Berlin: Springer-Praxis), 267.

Delsanti, A., Peixinho, N., Boehnhardt, H., Barucci, A., Merlin, F., Doressoundiram, A., & Davies, J. K. 2006, AJ, 131, 1851.

Dermott, S. F. 1973, Nature, 244, 18.

Doressoundiram, A., Peixinho, N., de Bergh, C., Fornasier, S., Thébault, P., Barucci, M. A., & Veillet, C. 2002, AJ, 124, 2279.

Doressoundiram, A. 2003, EM&P, 92, 131.

Dumas, C., Merlin, F., Barucci, M. A., de Bergh, C., Hainault, O., Guilbert, A., Vernazza, P., & Doressoundiram, A. 2007, A&A, 471, 331.

Duncan, M. J., & Levison, H. F. 1997, Science, 276, 1670.

Duncan, M., Quinn, T., & Tremaine, S. 1988, ApJ, 328, L69.

Duncan, M. J., Levison, H. F., & Budd, S. M. 1995, AJ, 110, 3073; 3187.

Duncan, M. J., Levison, H. F., & Lee, M. H. 1998, AJ, 116, 2067.

Duncombe, R. L., Klepczynski, W. J., & Seidelmann, P. K. 1968, Science, 162, 800.

Edgeworth, K. E. 1949, MNRAS, 109, 600.

Emel'yanenko, V. V., Asher, D. J., & Bailey, M. E. 2005, MNRAS, 361, 1345.

Emery, J. P., Dalle Ore, C. M., Cruikshank, D. P., Fernández, Y. R., Trilling, D. E., & Stansberry, J. A. 2007, A&A, 466, 395.

Fernández, J. A. 1980, MNRAS, 192, 481.

Fernández, J. A., & Ip, W.-H. 1984, Icarus, 58, 109.

Fernández, J. A., & Ip, W.-H. 1996, Planet. Space Sci., 44, 431.

Fernández, J. A., Gallardo, T., & Brunini, A. 2002, Icarus, 159, 358.

Fernández, J. A., Gallardo, T., & Brunini, A. 2004, Icarus, 172, 372.

Ford, E. B., & Chiang, E. I. 2007, ApJ, 661, 602.

Friedland, L. 2001, ApJ, 547, L75.

Gallardo, T. 2006a, Icarus, 181, 205.

Gallardo, T. 2006b, Icarus, 184, 29.

Gaudi, B. S., & Bloom, J. S. 2005, ApJ, 635, 711.

Gladman, B. 2005, Science, 307, 71.

Gladman, B., & Chan, C. 2006, ApJ, 643, L135.

Gladman, B., Marsden, B. G., & VanLaerhoven, C. 2007, in The Kuiper Belt, ed. Barucci, M. A., Boehnhardt, H., Cruikshank, D., & Morbidelli, A. (Tucson: Univ. Arizona Press), in press.

Gladman, B., Kavelaars, J. J., Nicholson, P. D., Loredo, T. J., & Burns, J. A. 1998, AJ, 116, 2042.

Gladman, B., Kavelaars, J. J., Petit, J.-M., Morbidelli, A., Holman, M. J., & Loredo, T. 2001, AJ, 122, 1051.

Gladman, B., Holman, M., Grav, T., Kavelaars, J., Nicholson, P., Aksnes, K., & Petit, J.-M. 2002, Icarus, 157, 269.

Goldreich, P., Lithwick, Y., Sari, R. 2004a, ARA&A, 42, 549.

Goldreich, P., Lithwick, Y., & Sari, R. 2004b, ApJ, 614, 497.

Gomes, R. S. 2000, AJ, 120, 2695.

Gomes, R. 2003a, EM&P, 92, 29.

Gomes, R. S. 2003b, Icarus, 161, 404.

Gomes, R. 2006, in ACM Symp. 229, Asteroids, Comets, and Meteors, ed. Lazzaro, D., Ferraz-Mello, S., & Fernández, J. A (Cambridge, University Press), 191.

Gomes, R. S., Morbidelli, A., & Levison, H. F. 2004, Icarus, 170, 492.

Gomes, R., Matese, J. J., & Lissauer, J. J. 2006, Icarus, 184, 589.

Gomes, R. S., Gallardo, T., Fernandez, J. A., & Brunini, A. 2005a, Celest. Mech. Dynam. Astron., 91, 109.

Gomes, R., Levison, H. F., Tsiganis, K., & Morbidelli, A. 2005b, Nature, 435, 466.

Gomes, R. S., Fernandez, J. A., Gallardo, T., & Brunini, A. 2007, in The Kuiper Belt, ed. Barucci, M. A., Boehnhardt, H., Cruikshank, D., & Morbidelli, A. (Tucson: Univ. Arizona Press), in press.

Hahn, J. M., & Malhotra, R. 1999, AJ, 117, 3041.

Hahn, J. M., & Malhotra, R. 2005, AJ, 130, 2392.

Hainaut, O. R., & Delsanti, A. C. 2002, A&A, 389, 641.

Harrington, R. S. 1988, AJ, 96, 1476.

Hayashi, C., Nakazawa, K., & Nakagawa, Y. 1985, in Protostars and Planets II, ed. Black, D. C. & Matthews, M. S. (Tucson: Univ. Arizona Press), 1100.

Hogg, D. W., Quinlan, G. D., & Tremaine, S. 1991, AJ, 101, 2274.

Holman, M. J., & Wisdom, J. 1993, AJ, 105, 1987.

Horner, J., & Evans, N. W. 2002, MNRAS, 335, 641.

Horner, J., Evans, N. W., Bailey, M. E., & Asher, D. J. 2003, MNRAS, 343, 1057.

Horner, J. Evans, N. W., & Bailey, M. E. 2004, MNRAS, 354, 798.

Ida, S., Larwood, J., & Burkert, A. 2000a, ApJ, 528, 351.

Ida, S., Bryden, G., Lin, D. N. C., & Tanaka, H. 2000b, ApJ, 534, 428.

Jewitt, D. C. 1999, Annu. Rev. Earth. Planet. Sci., 27, 287.

Jewitt, D. C. 2003, EM&P, 92, 465.

Jewitt, D. 2006, in Saas Fee Lectures 2005, ed. Thomas, N., & Benz, W. (Murren: ?), in press.

Jewitt, D. C., & Luu, J. X. 1993, Nature, 362, 730.

Jewitt, D. C., & Luu, J. X. 1995, AJ, 109, 1867; 1935.

Jewitt, D. C., & Sheppard, S. S. 2002, AJ, 123, 2110.

Jewitt, D., Luu, J., & Trujillo, C. 1998, AJ, 115, 2125.

Jones, D. C., Williams, I. P., & Melita, M. D. 2005, EM&P, 97, 435.

Jones, R. L., et al. 2006, Icarus, 185, 508.

Kenyon, S. J. 2002, PASP, 114, 265.

Kenyon, S. J., & Bromley, B. C. 2004a, AJ, 127, 513.

Kenyon, S. J., & Bromley, B. C. 2004b, Nature, 432, 598.

Kenyon, S. J., & Luu, J. X. 1998, AJ, 115, 2136.

Kenyon, S. J., & Luu, J. X. 1999, AJ, 118, 1101.

Kenyon, S. J., Bromley, B. C., O'Brien, D. P., & Davis, D. R. 2007, in The Kuiper Belt, ed. Barucci, M. A., Boehnhardt, H., Cruikshank, D., & Morbidelli, A. (Tucson: Univ. Arizona Press), in press.

Kobayashi, H., & Ida, S. 2001, Icarus, 153, 416.

Kobayashi, H., Ida, S., & Tanaka, H. 2005, Icarus, 177, 246.

Kokubo, E., & Ida, S. 2002, ApJ, 581, 666.

Kowal, C. T. 1989, Icarus, 77, 118.

Kozai, Y. 1962, AJ, 67, 591.

Kuchner, M. J., Brown, M. E., & Holman, M. 2002, AJ, 124, 1221.

Kuiper, G. P. 1951, in Astrophysics, ed. J. A. Hynek (New York: McGraw-Hill), 357

Kuz'michev, V. V., & Tomanov, V. P. 2006, Astronomy Letters, 32, 353.

Lacerda, P., & Jewitt, D. C. 2007, AJ, 133, 1393.

Larsen, J. A., et al. 2007, AJ, 133, 1247.

Lee, M. H., Peale, S. J., Pfahl, E., & Ward, W. R. 2007, Icarus, 190, 103.

Leonard, F. C. 1930, Astronomical Society of the Pacific Leaflets, 1, 121.

Levison, H. F., & Stern, S. A. 2001, AJ, 121, 1730.

Levison, H. F., & Stewart, G. R. 2001, Icarus, 153, 224.

Levison, H. F., & Morbidelli, A. 2003, Nature, 426, 419.

Levison, H. F., & Morbidelli, A. 2007, Icarus, 189, 196.

Levison, H. F., Morbidelli, A., & Dones, L. 2004, AJ, 128, 2553.

Levison, H. F., Morbidelli, A., Gomes, R., & Backman, D., 2007, in Protostars and Planets V Compendium, ed. Reipurth, B., Jewitt, D., & Keil, K. (Tucson: Univ. Arizona Press), 669.

Licandro, J., Grundy, W. M., Pinilla-Alonso, N., & Leisy, P. 2006a, A&A, 458, L5.

Licandro, J., Pinilla-Alonso, N., Pedani, M., Oliva, E., Tozzi, G. P., & Grundy, W. M. 2006b, A&A, 445, L35.

Liou, J.-C., & Malhotra, R. 1997, Science, 275, 375.

Luu, J. X., & Jewitt, D. C. 1988, AJ, 95, 1256.

Luu, J. X., & Jewitt, D. C. 2002, ARA&A, 40, 63.

Luu, J., Marsden, B. G., Jewitt, D., Trujillo, C. A., Hergenrother, C. W., Chen, J., & Offutt, W. B. 1997, Nature, 387, 573.

Lykawka, P. S., & Mukai, T. 2004, BAAS, 36, 1102.

Lykawka, P. S., & Mukai, T. 2005a, Planet. Space Sci., 53, 1175.

Lykawka, P. S., & Mukai, T. 2005b, Planet. Space Sci., 53, 1319.

Lykawka, P. S., & Mukai, T. 2005c, EM&P, 97, 107.

Lykawka, P. S., & Mukai, T. 2006, Planet. Space Sci., 54, 87.

Lykawka, P. S., & Mukai, T. 2007a, Icarus, 186, 331.

Lykawka, P. S., & Mukai, T. 2007b, Icarus 189, 213. (LM07b)

Lykawka, P. S., & Mukai, T. 2007c, Icarus, in press.

Malhotra, R. 1995, AJ, 110, 420.

Malhotra, R. 1998, in ASP Conf. Ser. 149, Solar System Formation and Evolution, ed. Lazzaro, D., Vieira Martins, R., Ferraz-Mello, S., Fernández, J., & Beaugé, C. (?:ASP), 37.

Maran, M. D., Collander-Brown, S. J., & Williams, I. P. 1997, Planet. Space Sci., 45, 1037.

Matese, J. J., & Whitmire, D. P. 1986, Icarus 65, 37.

Matese, J. J., Whitmire, D. P., & Lissauer, J. J. 2005, EM&P, 97, 459.

McBride, N., Green, S. F., Davies, J. K., Tholen, D. J., Sheppard, S. S., Whiteley, R. J., & Hillier, J. K. 2003, Icarus, 161, 501.

Melita, M. D., & Brunini, A. 2000, Icarus, 147, 205.

Melita, M. D., & Williams, I. P. 2003, EM&P, 92, 447.

Melita, M. D., Williams, I. P., Brown-Collander, S. J., & Fitzsimmons, A. 2004, Icarus, 171, 516.

Merk, R., & Prialnik, D. 2006, Icarus, 183, 283.

Montmerle, T., Augereau, J.-C., Chaussidon, M., Gounelle, M., Marty, B., & Morbidelli, A. 2006, EM&P, 98, 39.

Morbidelli, A. 2005, in Lectures on Comet Dynamics and Outer Solar System Formation. Origin and Dynamical Evolution of Comets and their Reservoirs, in press.

Morbidelli, A., & Valsecchi, G. B. 1997, Icarus, 128, 464.

Morbidelli, A., & Brown, M. E. 2004, in Comets II, ed. Festou, M. C., Keller, H. U., & Weaver, H. A. (Tucson: Univ. Arizona Press), 175.

Morbidelli, A., & Levison, H. F. 2004, AJ, 128, 2564.

Morbidelli, A., Jacob, C., & Petit, J.-M. 2002, Icarus, 157, 241.

Morbidelli, A., Emel'yanenko, V. V., & Levison, H. F. 2004, MNRAS, 355, 935.

Morbidelli, A., Levison, H. F., & Gomes, R. 2007, in The Kuiper Belt, ed. Barucci, M. A., Boehnhardt, H., Cruikshank, D., & Morbidelli, A. (Tucson: Univ. Arizona Press), in press.

Morbidelli, A., Levison, H. F., Tsiganis, K., & Gomes, R. 2005, Nature, 435, 462.

Moroz, L. V., Arnold, G., Korochantsev, A. V., & Wasch, R. 1998, Icarus, 134, 253.

Murray, J. B. 1999, MNRAS, 309, 31.

Murray, C. D., & Dermott, S. F. 1999, Solar System Dynamics (Princeton: Princeton Univ. Press).

Murray-Clay, R. A., & Chiang, E. I. 2005, ApJ, 619, 623.

Murray-Clay, R. A., & Chiang, E. I. 2006, ApJ, in press.

Nesvorný, D., & Roig, F. 2001, Icarus, 150, 104.

Nesvorný, D., Ferraz-Mello, S., Holman, M., & Morbidelli, A., 2002, in Asteroids III, ed. Bottke Jr., W. F., Cellino, A., Paolicchi, P., & Binzel, R. P. (Tucson: Univ. Arizona Press), 379.

Noll, K. S., Levison, H. F., Grundy, W. M., & Stephens, D. C. 2006, Icarus, 184, 611.

Noll, K. S., Grundy, W. M., Chiang, E. I., Margot, J.-L., & Kern, S. D. 2007, in The Kuiper Belt, ed. Barucci, M. A., Boehnhardt, H., Cruikshank, D., & Morbidelli, A. (Tucson: Univ. Arizona Press), in press.

Oort, J. H. 1950, Bull. Astron. Inst. Netherlands, 11, 91.

Peale, S. J. 1976, ARA&A, 14, 215.

Peixinho, N., Boehnhardt, H., Belskaya, I., Doressoundiram, A., Barucci, M. A., & Delsanti, A. 2004, Icarus, 170, 153.

Petit, J.-M., Morbidelli, A., & Valsecchi, G. B. 1999, Icarus, 141, 367.

Pollack, J. B., Hubickyj, O., Bodenheimer, P., Lissauer, J. J., Podolak, M., & Greenzweig, Y. 1996, Icarus, 124, 62.

Rabinowitz, D. L., Barkume, K., Brown, M. E., Roe, H., Schwartz, M., Tourtellotte, S., &

Trujillo, C. 2006, ApJ, 639, 1238.

Rafikov, R. R. 2003, AJ, 125, 942.

Rafikov, R. R. 2004, AJ, 128, 1348.

Russell, H. N. 1916, ApJ, 13, 173.

Schaefer, B. E., & Schaefer, M. W. 2000, Icarus, 146, 541.

Schaller, E. L., & Brown, M. E. 2007, ApJ, 659, L61.

Sheppard, S. S. 2006, in ASP Conf. Ser. 352, New Horizons in Astronomy: Frank N. Bash Symposium, ed. Kannappan, S. J., Redfield, S., Kessler-Silacci, J. E., Landriau, M. & Drory, N. (Austin:ASP), 3.

Sheppard, S. S., & Jewitt, D. C. 2002, AJ, 124, 1757.

Sheppard, S. S., Jewitt, D. C., Trujillo, C. A., Brown, M. J. I., & Ashley, M. C. B. 2000, AJ, 120, 2687.

Standish, E. M. 1993, AJ, 105, 2000.

Stephens D. C., & Noll, K. S. 2006, AJ, 131, 1142.

Stern, S. A. 1991, Icarus, 90, 271.

Stern, S. A. 1992, ARA&A, 30, 185.

Stern, A. S. 1995, AJ, 110, 856.

Stern, A. S. 1998, in ASSL Ser. 227, Solar System Ices, ed. Schmitt, B., de Bergh, C., & Festou, M. (Netherlands: Kluwer), 685.

Stern, S. A. 2005, AJ, 129, 526.

Stern, S. A., & Colwell, J. E. 1997, AJ, 114, 841; 881.

Stern, S. A., et al. 2006, Nature, 439, 946.

Tegler, S. C., & Romanishin, W. 2000, Nature, 407, 979.

Tegler, S. C., Grundy, W., Romanishin, W., Consolmagno, G., Mogren, K., & Vilas, F. 2007, AJ, 133, 526.

Thompson, W. R., Murray, B. G. J. P. T., Khare, B. N., & Sagan, C. 1987, J. Geophys. Res., 92, 14933.

Tombaugh, C. 1961, in Planets and Satellites, ed. Kuiper, G. P., & Middlehurst, B. (Chicago: Univ. Chicago Press), 12.

Trujillo, C. A., & Brown, M. E. 2001, ApJ, 554, L95.

Trujillo, C. A., & Brown M. E. 2002, ApJ, 566, L125.

Trujillo, C. A., & Brown, M. E. 2003, EM&P, 92, 99.

Trujillo, C. A., Jewitt, D. C., & Luu, J. X. 2001a, AJ, 122, 457.

Trujillo, C. A., Luu, J. X., Bosh, A. S., & Elliot, J. L. 2001b, AJ, 122, 2740.

Trujillo, C. A., Brown, M. E., Barkume, M. E., Schaller, E. L., & Rabinowitz, D. L. 2007, ApJ, 655, 1172.

Tsiganis, K., Gomes, R., Morbidelli, A., & Levison, H. F. 2005, Nature, 435, 459.

Udry, S., Fischer, D., & Queloz, D. 2007, in Protostars and Planets V Compendium, ed. Reipurth, B., Jewitt, D., & Keil, K. (Tucson: Univ. Arizona Press), 685.

Wan, X.-S., & Huang, T.-Y. 2007, MNRAS, 377, 133.

Weaver, H. A., et al. 2006, Nature, 439, 943.

Weidenschilling, S. J. 2003, Lunar Planet. Sci. Conf., 34, abstract 1707.

Wiegert, P., Innanen, K., Huang, T.-Y., & Mikkola, S. 2003, AJ, 126, 1575.

Williams, I. P. 1997, Reports on Progress in Physics 60, 1.

Wisdom, J., & Holman, M. 1991, AJ, 1528.

Zhou, L.-Y., Sun, Y.-S., Zhou, J.-L., Zheng, J.-Q., & Valtonen, M. 2002, MNRAS, 336, 520.

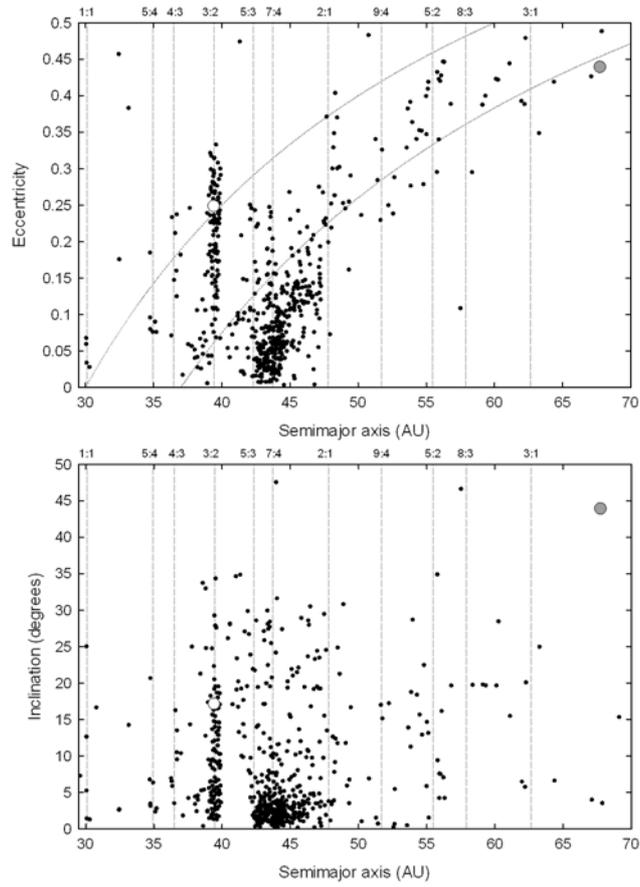

Figure 1. Orbital distribution of TNOs. Observational data were taken from the Lowell Observatory database on 12 September 2006. Only objects with long-arc orbits are plotted (≥2 oppositions). Vertical dashed lines indicate the positions of resonances with Neptune. Dotted curves represent the perihelia of 30 and 37 AU (upper panel). Pluto and Eris are shown as white and gray large circles, respectively.

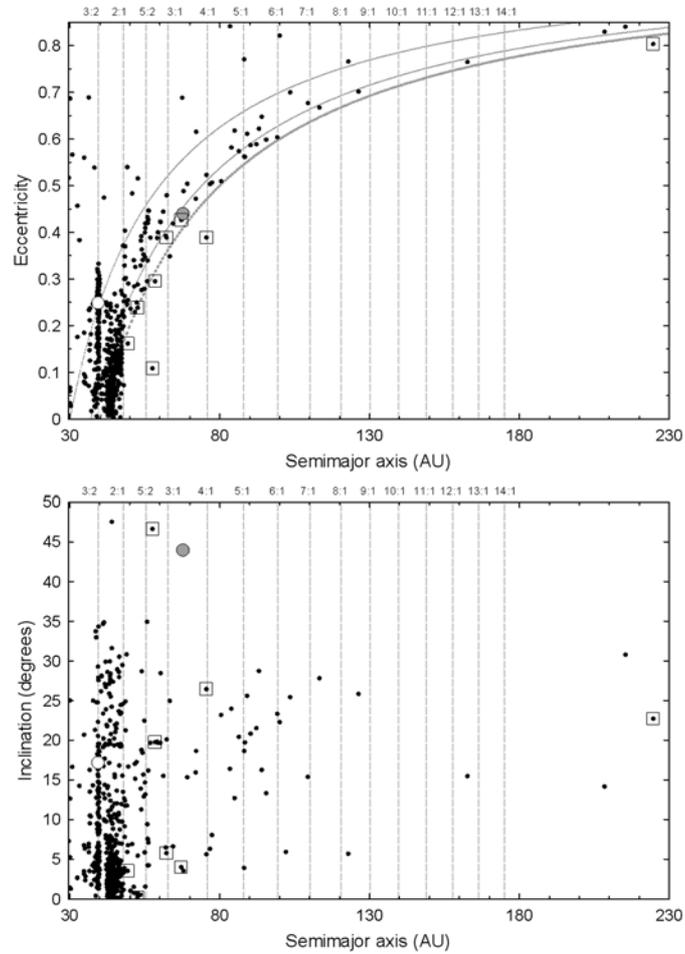

Figure 2. Orbital distribution of TNOs up to *a* = 230 AU. Observational data were taken from the Lowell Observatory database on 12 September 2006. Only objects with long-arc orbits are plotted (≥2 oppositions). Vertical dashed lines indicate the positions of resonances with Neptune. Beyond 50 AU, the strongest resonances are of the *r*:1 and *r*:2 type (some of interest are shown). Dotted curves represent the perihelia of 30, 37, and 40 AU (from the top in the upper panel). Pluto and Eris are shown as white and gray large circles, respectively. Detached TNOs are enclosed with squares, after their identification in LM07b. Another detached body, Sedna, is out of the range of this figure (*a* = 525.6 AU; *q* = 76.2 AU; *i* = 11.9°).

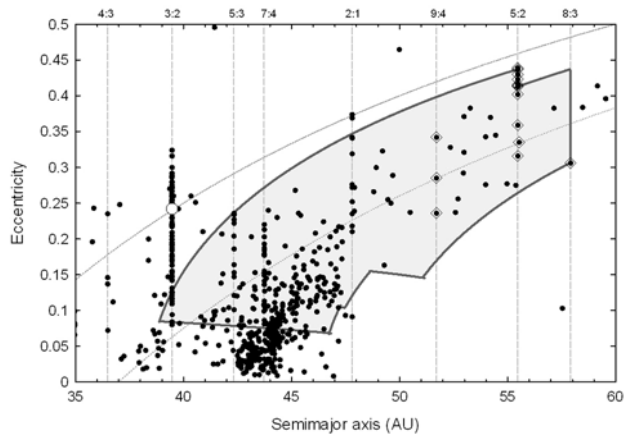

Figure 3. Orbital distribution of TNOs, the classical region (~37–48 AU), and resonant populations (aligned vertically). The orbits were averaged over 10 Myr after integration into the future. Original observational data were taken from the Lowell Observatory database on 12 September 2006. Only objects with long-arc orbits are plotted (≥2 oppositions). Vertical dashed lines indicate the positions of resonances with Neptune. Dotted curves represent the perihelia of 30 and 37 AU. Pluto is shown as a white large circle. The 9:4-, 5:2-, and 8:3-resonant TNOs are enclosed with diamonds. The enclosed region defines the eccentricities needed in an excited planetesimal disk to reproduce long-term members in the latter resonances, as derived in Lykawka & Mukai (2007a). The resemblance between the distribution of TNOs and the gray region suggests that current observations may reveal the relics of the ancient excited trans-Neptunian belt.

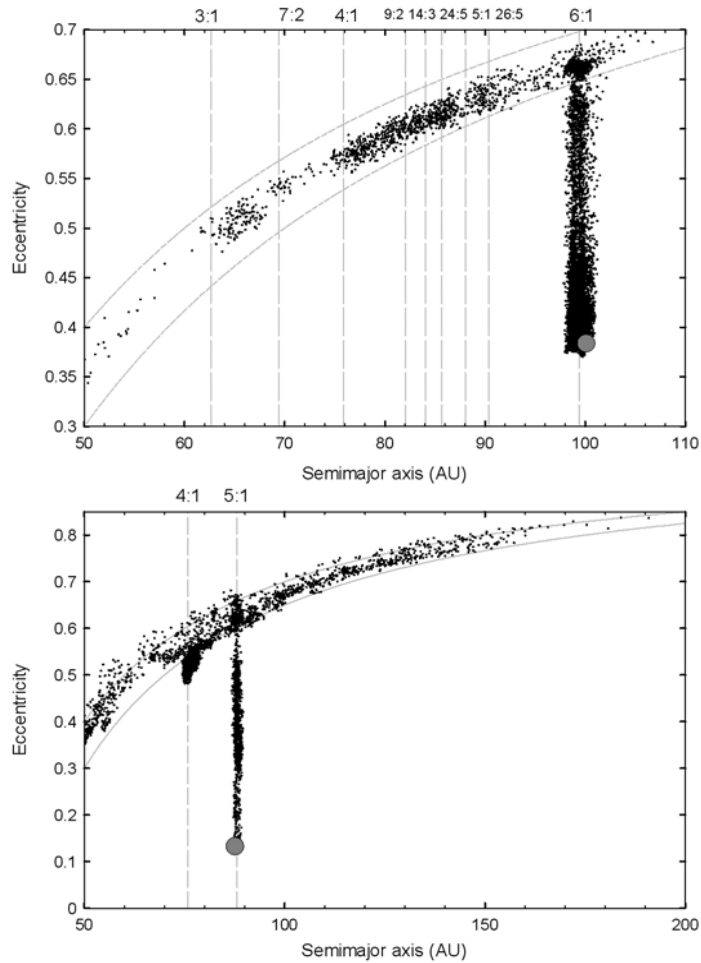

Figure 4. Orbital evolution of two scattered particles over 4 Gyr. Small dots represent the orbital evolution plotted every 1 Myr. Dashed curves represent the perihelia of 30 and 35 AU (upper and lower, respectively). The particles began on Neptune-crossing orbits and with $a < 50$ AU. The final orbital elements of the particles are indicated by gray circles. Long dashed vertical lines refer to the locations of relevant resonances where the particles spent most of their lifetimes during temporary trapping ($r$:1 and $r$:2 resonances played the greatest role). In particular, the particle spent more than 3 Gyr in the 6:1 resonance, reaching $q \approx 60$ AU and $i \approx 38°$ during Kozai interactions (top panel). Similarly, the other particle exhibited $q \approx 76$ AU and $i \approx 45°$ in the 5:1 resonance for more than 1 Gyr (bottom panel).

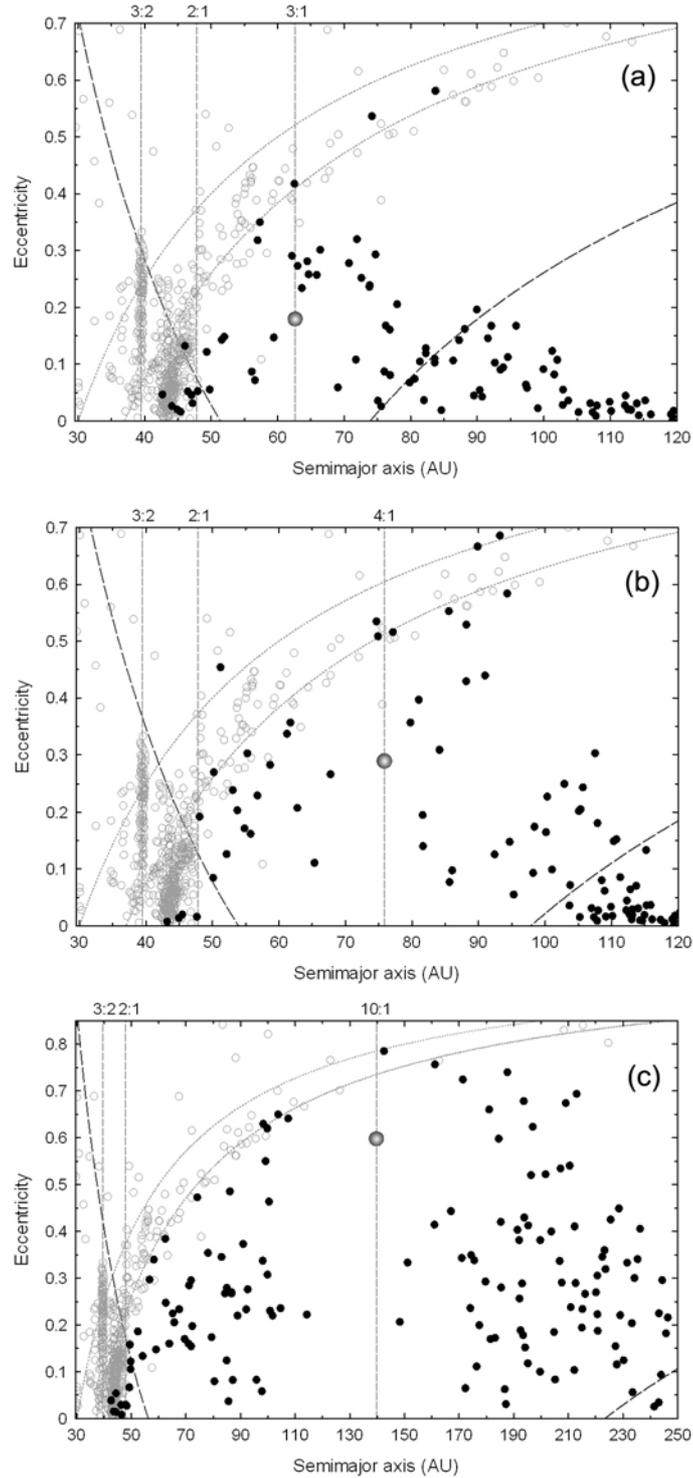

Figure 5. Orbital distribution of a cold planetesimal disk with static outer planets after 4 Gyr (black filled circles). The disk was initially set at 40–120 AU (panels a and b) and 40–240 AU (panel c). TNOs are illustrated for reference (gray open circles). Dotted curves represent the perihelia of 30 and 37 AU. Dashed vertical lines indicate the location of the 3:2, 2:1, 3:1, 4:1, and 10:1 resonances. The outer planet has (a) 0.15 $M_\oplus$ and $i_P \approx 10°$ (3:1 resonance); (b) 0.3 $M_\oplus$ and $i_P \approx 12°$ (4:1 resonance); (c) 1.0 $M_\oplus$ and $i_P \approx 12°$ (10:1 resonance) (gray spheres). Objects above the long-dashed curves could encounter the trans-Plutonian planet.

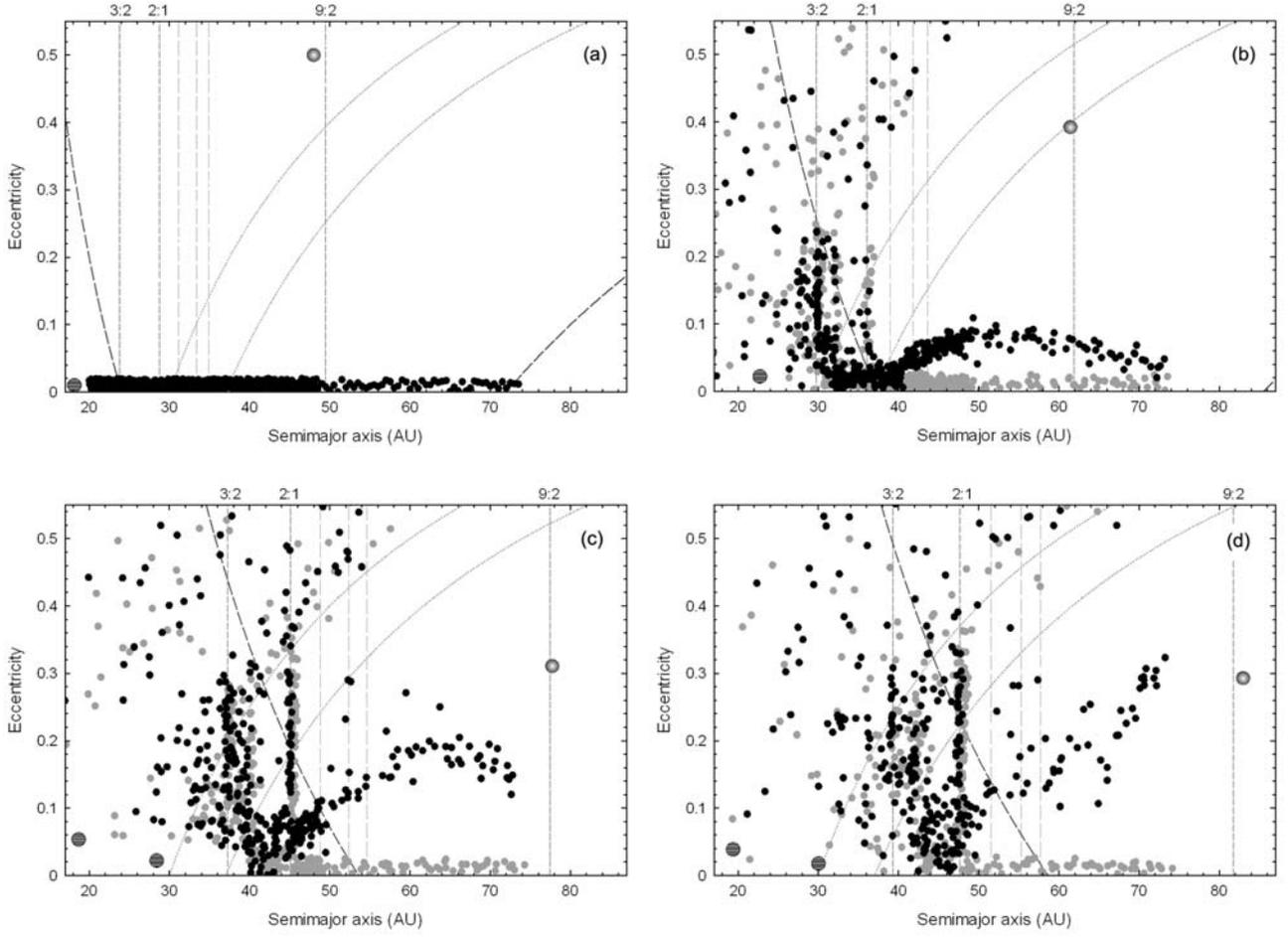

Figure 6. Orbital evolution of a planetesimal disk with a migrating outer planet at (a) 0, (b) 5, (c) 20, and (d) 50 Myr (black filled circles). For reference, the outcomes of these simulations without the planetoid are indicated by gray filled circles. Dotted curves represent the perihelia of 30 and 37 AU. Dashed vertical lines indicate the location of the 3:2, 2:1, and 9:2 resonances. Uranus and Neptune are indicated by hatched circles. The trans-Plutonian planet (0.4 $M_\oplus$; $i_P \approx 10°$) migrated from $a_P = 48.5$ to 83 AU. Objects above the long-dashed curves could encounter the trans-Plutonian planet. Planet migration followed equation (1) with $\tau = 10$ Myr. No resonant populations beyond 50 AU formed, in particular in the 9:4, 5:2, and 8:3 resonances (long-dashed vertical lines; see text for details).

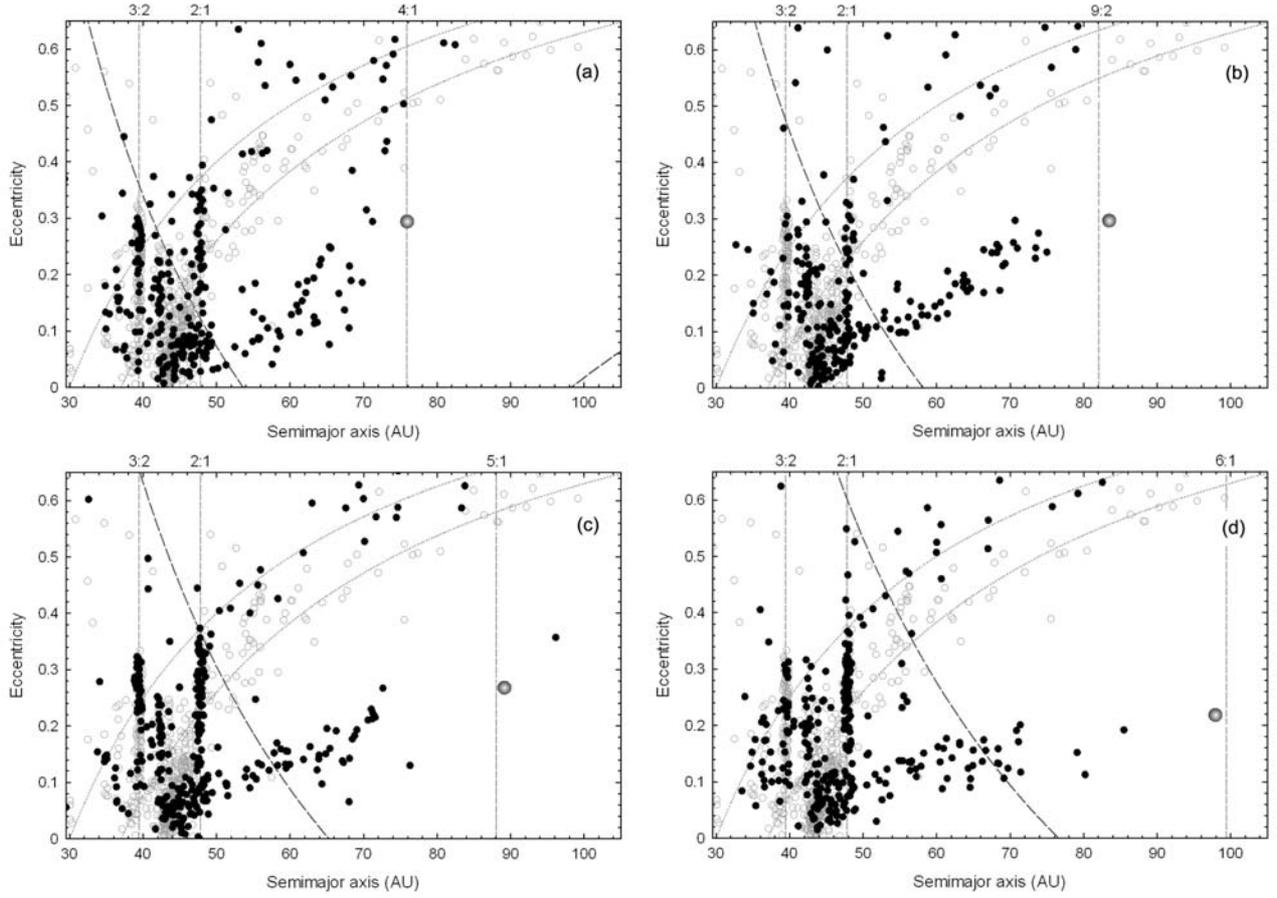

Figure 7. Orbital distribution of planetesimal disks with migrating outer planets after the end of planet migration at 100–125 Myr (black filled circles). The disks were initially in cold orbital conditions, as shown in Fig. 6a. TNOs are illustrated for reference (gray open circles). Dotted curves represent the perihelia of 30 and 37 AU. Dashed vertical lines indicate the location of the 3:2, 2:1, and resonances near the outer planet. The trans-Plutonian planet has (a) 0.3, (b) 0.3, (c) 0.4, and (d) 0.5 $M_\oplus$, respectively. All outer planets possess $i_P \approx 10°$. Objects above the long-dashed curves could encounter the trans-Plutonian planet. The orbital excitation in the classical region and the formation of an edge at 48 AU are evident. The formation of resonant objects was also accomplished, although resonant populations beyond 50 AU were absent (see text for details).

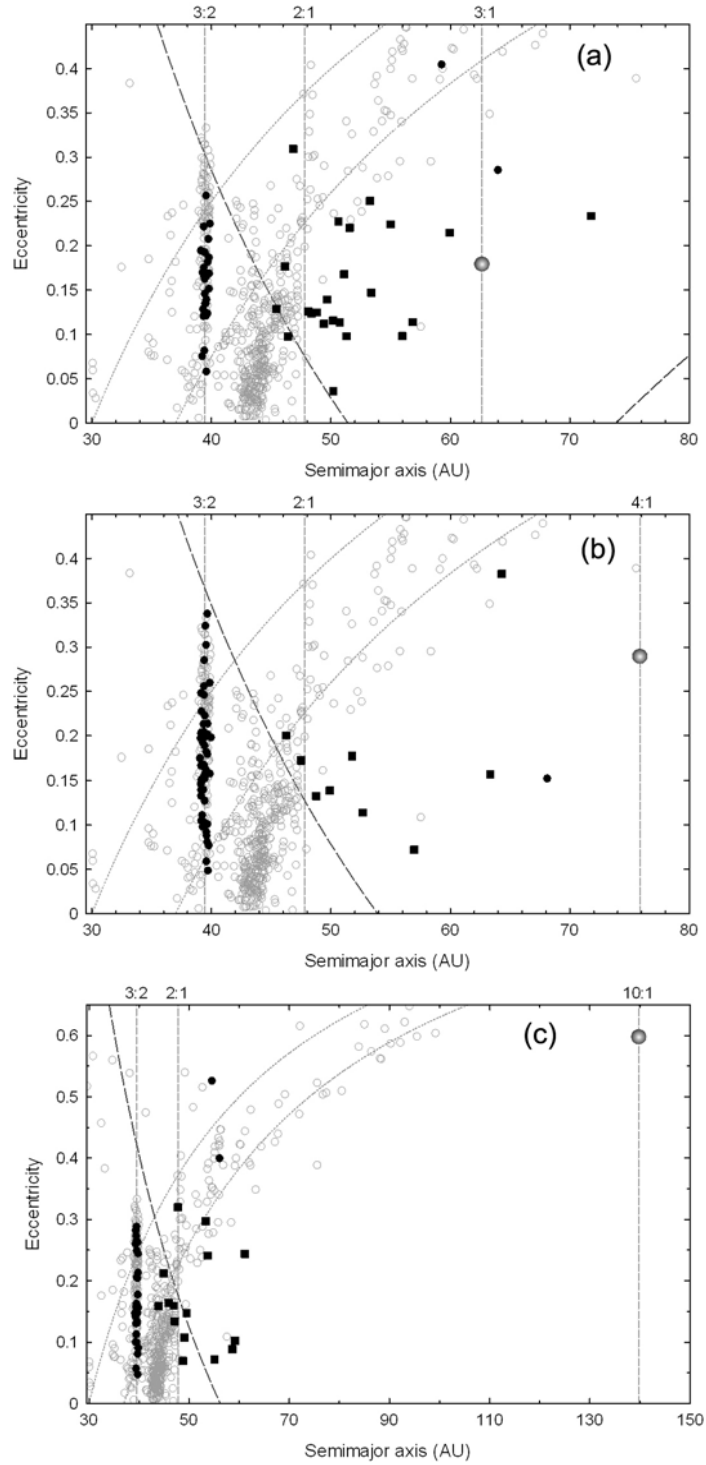

Figure 8. Orbital distribution of particles in the 3:2 and 2:1 resonances with resident outer planets after 4 Gyr (black filled circles and squares, respectively). The initial orbital elements of 3:2 and 2:1 resonant populations were $0.1 < e < 0.34$ and $0.07 < e < 0.4$ ($i < 25°$). TNOs are illustrated for reference (gray open circles). Dotted curves represent the perihelia of 30 and 37 AU. Dashed vertical lines indicate the locations of the 3:2, 2:1, 3:1, 4:1, and 10:1 resonances. The outer planet has (a) 0.15 $M_\oplus$ and $i_P \approx 10°$ (3:1 resonance); (b) 0.3 $M_\oplus$ and $i_P \approx 12°$ (4:1 resonance); (c) 1.0 $M_\oplus$ and $i_P \approx 12°$ (10:1 resonance) (gray spheres). Objects above the long-dashed curves could encounter the trans-Plutonian planet. Large-$e$ 3:2-resonant bodies were depleted and the entire 2:1 population was devastated after 4 Gyr.

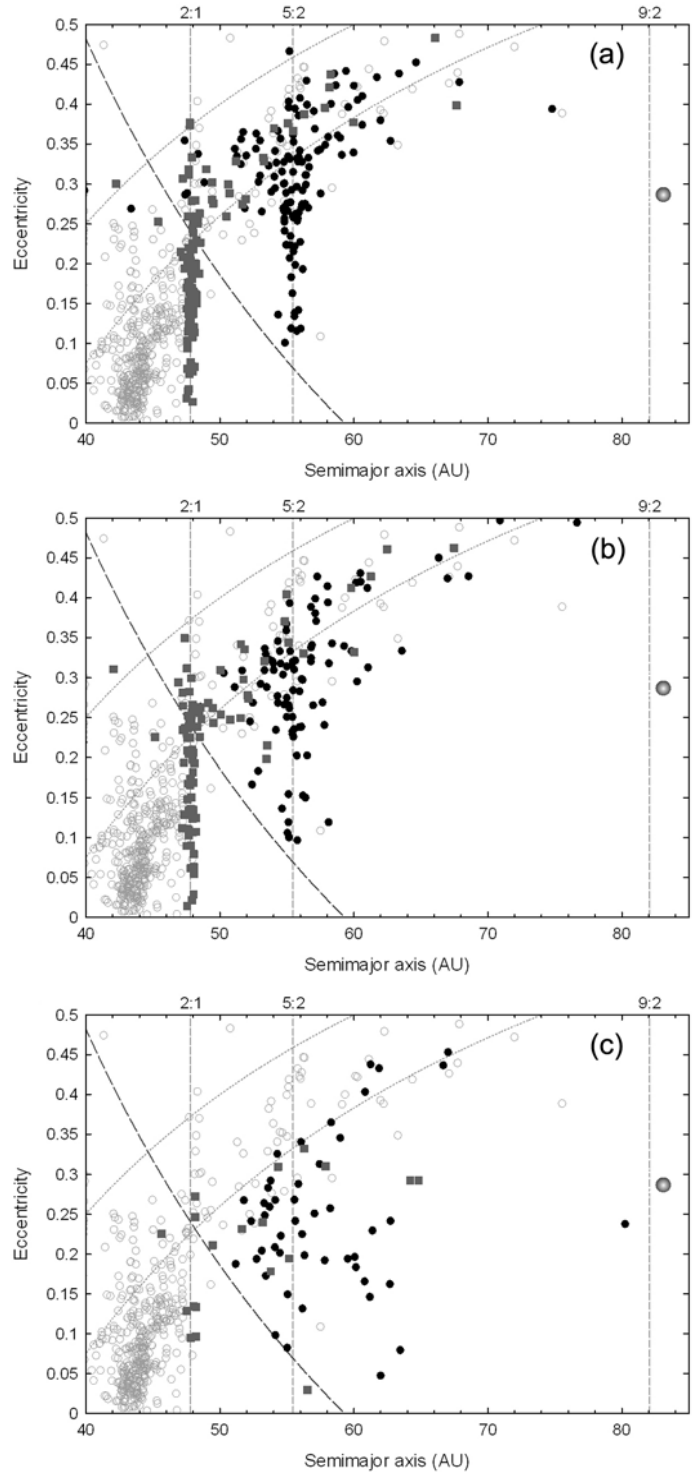

Figure 9. Orbital distribution of particles in the 2:1 and 5:2 resonances with a resident outer planet after (a) 500 Myr, (b) 1 Gyr, and (c) 4 Gyr (filled squares and circles, respectively). The initial orbital elements of 2:1- and 5:2-resonant populations were $0.07 < e < 0.4$ and $0.35 < e < 0.45$ ($i < 25°$). TNOs are illustrated for reference (gray open circles). Dotted curves represent the perihelia of 30 and 37 AU. Dashed vertical lines indicate the locations of the 2:1, 5:2, and 9:2 resonances. The outer planet has 0.3 $M_\oplus$ and $i_P \approx 11°$ and is near the 9:2 resonance (gray sphere). Objects above the long-dashed curves could encounter the trans-Plutonian planet. The entire 5:2-resonant population was destroyed and large-$e$ 2:1 bodies were severely depleted within 1 Gyr. After 4 Gyr, both resonant populations are practically devastated and in conflict with observations.

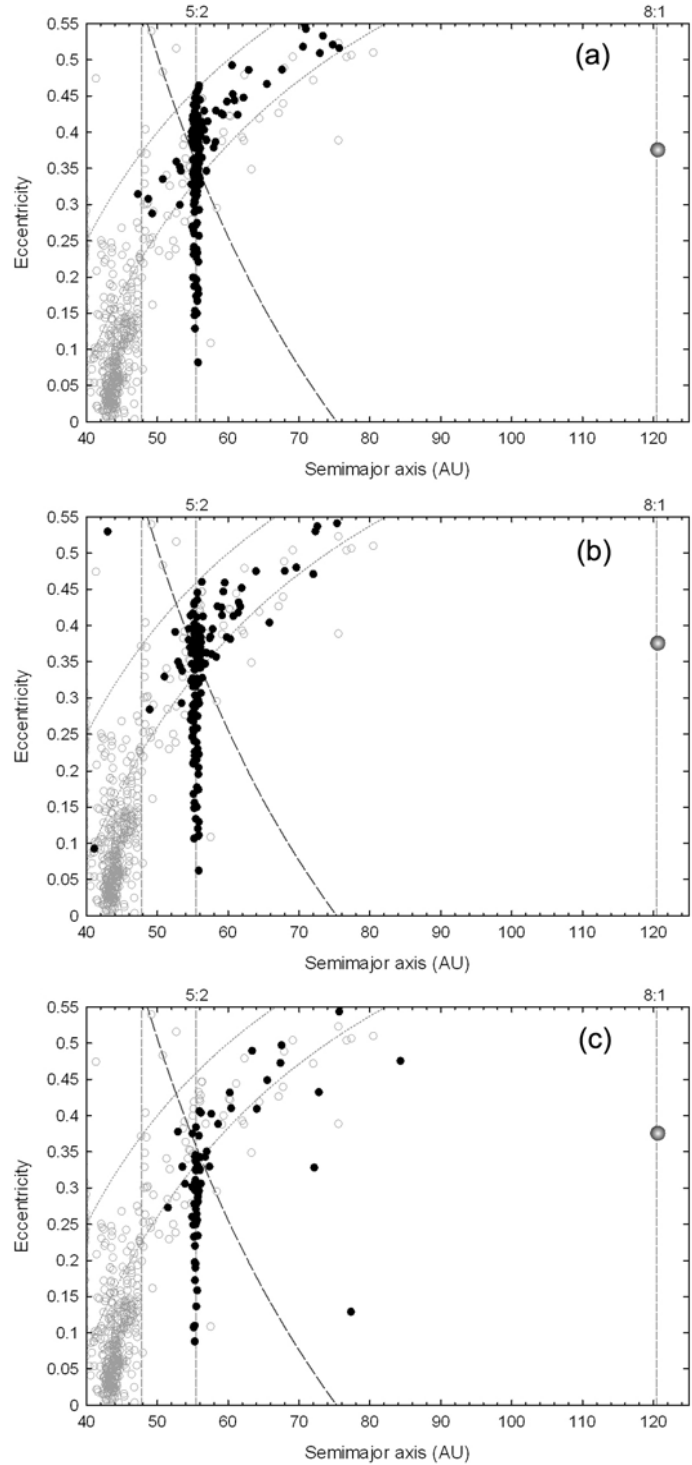

Figure 10. Orbital distribution of particles in the 5:2 resonance with a resident outer planet after (a) 500 Myr, (b) 1 Gyr, and (c) 4 Gyr (filled circles). The initial orbital elements of 5:2-resonant populations were $0.35 < e < 0.45$ ($i < 25°$). TNOs are illustrated for reference (gray open circles). Dotted curves represent the perihelia of 30 and 37 AU. Dashed vertical lines indicate the locations of the 5:2 and 8:1 resonances. The outer planet has 0.4 $M_\oplus$ and $i_P \approx 25°$ and is near the 8:1 resonance (gray sphere). Objects above the long-dashed curves could encounter the trans-Plutonian planet. The 5:2-resonant population was quite depleted over 4 Gyr (80%). Only a few members remained with $Q > q_P$.

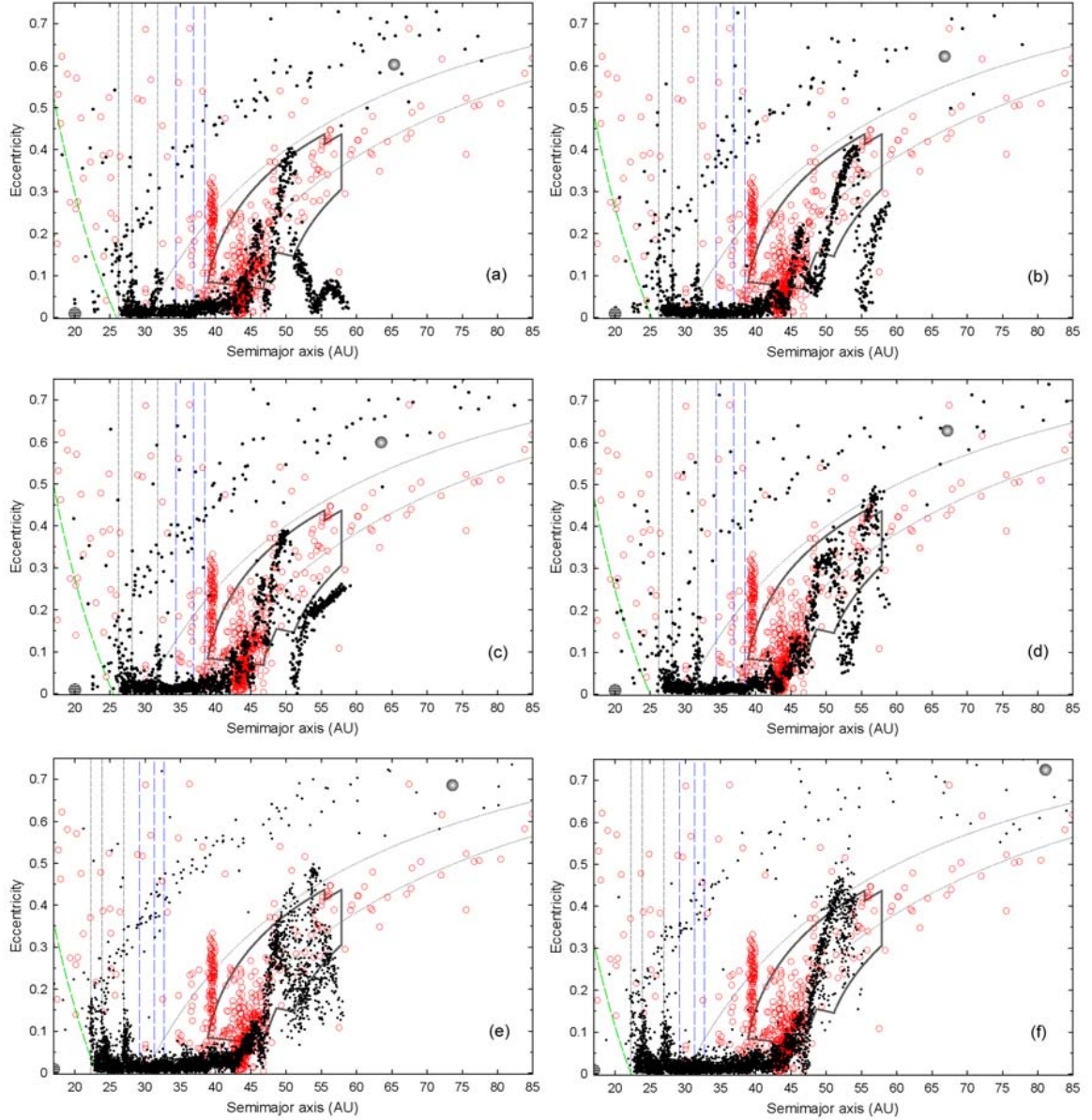

Figure 11. Orbital evolution of planetesimal disks with scattered outer planets after (a–c) 60 Myr, (d) 80 Myr, (e) 90 Myr, and (f) 110 Myr before planet migration (black circles). The disks were initially in cold orbital conditions from 17 to (a–d) 59 AU, (e) 57 AU, and (f) 54 AU. TNOs are illustrated for reference (open circles). Dotted curves represent the perihelia of 30 and 37 AU. Dashed and long-dashed vertical lines indicate the primordial locations of the 3:2, 5:3, 2:1, and 9:4, 5:2, and 8:3 resonances, respectively. In these reference simulations, Neptune is at (a–d) 20 AU and (e and f) 17 AU, and is indicated by a hatched circle. The outer planet has 0.4 $M_\oplus$, $i_P \approx 10°$, and evolves at $a_P \sim 60$–70 AU (gray sphere). In the bottom panels, the planetoid is more massive (0.5 $M_\oplus$) and evolves at $a_P \sim 75$–80 AU. Objects above the long-dashed curves could encounter the trans-Plutonian planet. The enclosed region defines the conditions needed in an excited planetesimal disk to reproduce long-term TNOs in the 9:4, 5:2, and 8:3 resonances according to Lykawka & Mukai (2007a) (see text for details). Objects did not exhibit appreciable radial changes. The planetoid is very effective in exciting the outer trans-Neptunian belt (45–50 AU) and truncating the disks. In addition, the orbital excitation obtained from the simulations in the 40–50 AU region is strikingly similar to that currently observed for TNOs in the same region.

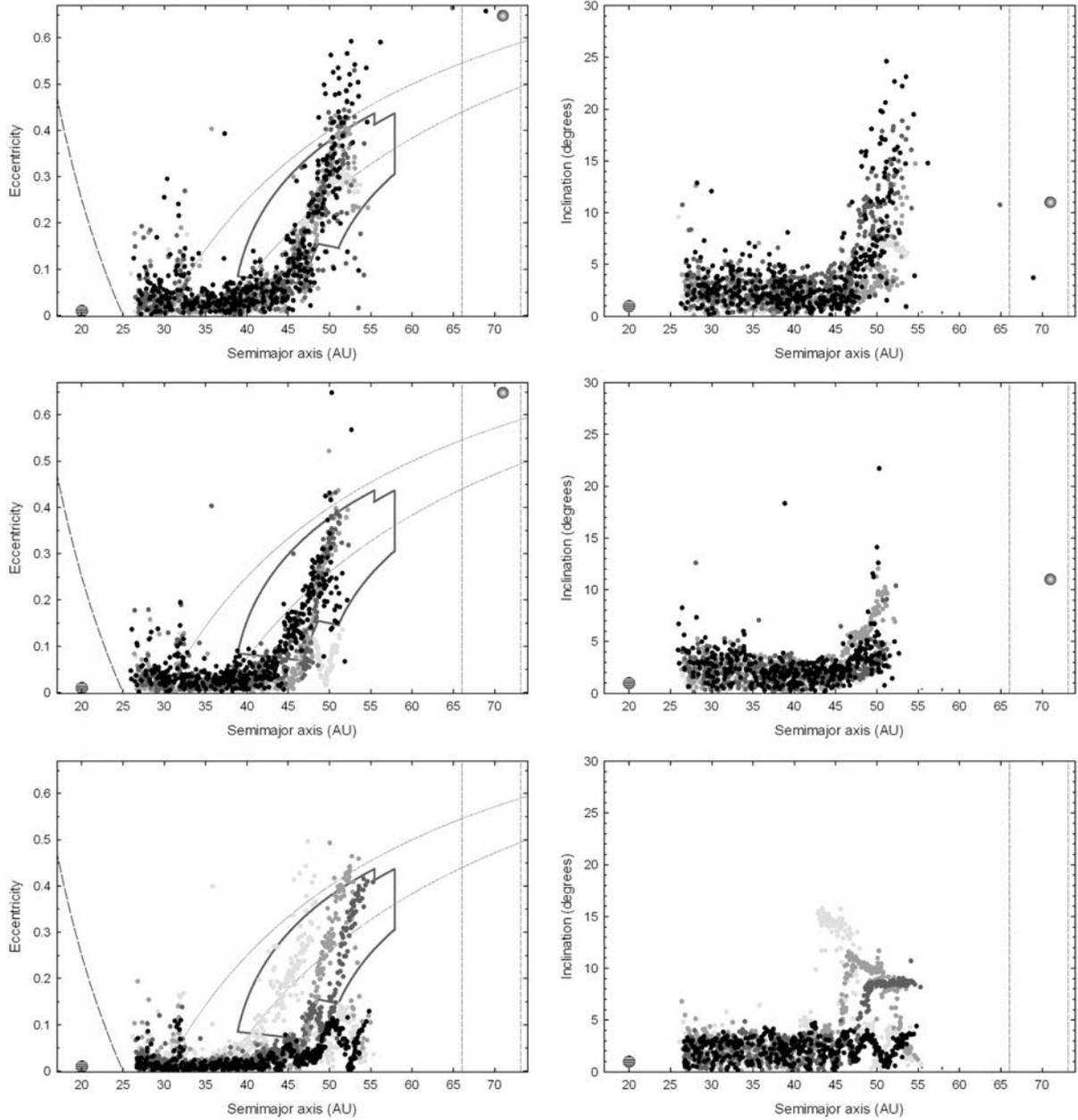

Figure 12. Orbital excitation of planetesimal disks with scattered outer planets (gray spheres), before planet migration (filled circles). The disks were initially in cold orbital conditions ($e \approx 0$ and $i \approx 0$). Dotted curves represent the perihelia of 30 and 37 AU (left panels). The dashed vertical line indicates the primordial location of the 6:1 resonance. In these reference simulations, Neptune is at 20 AU and is indicated by a hatched circle. The enclosed region defines the conditions needed in an excited planetesimal disk to reproduce long-term TNOs in the 9:4, 5:2, and 8:3 resonances according to Lykawka & Mukai (2007a) (see text for details). Objects did not exhibit appreciable radial changes. In all panels, the light to dark gray color variation represents increasing numbers for each varying quantity of interest. Top panels: Perturbation of a planetoid with $a_P = 71$ AU and 0.7 $M_\oplus$ at four distinct timescales $t = 20, 50, 80,$ and 150 Myr. The planetesimal disk extended to 53 AU. Middle panels: Perturbation of a planetoid with $a_P = 71$ AU at $t = 50$ Myr for four distinct masses 0.2 $M_\oplus$, 0.5 $M_\oplus$, 0.7 $M_\oplus$, and 1.0 $M_\oplus$. The planetesimal disk extended to 51 AU. Bottom panels: Perturbation of a planetoid with 0.5 $M_\oplus$ at $t = 50$ Myr for four distinct initial semimajor axes $a_P = 57, 66, 75,$ and 85 AU. The planetesimal disk extended to 54 AU.

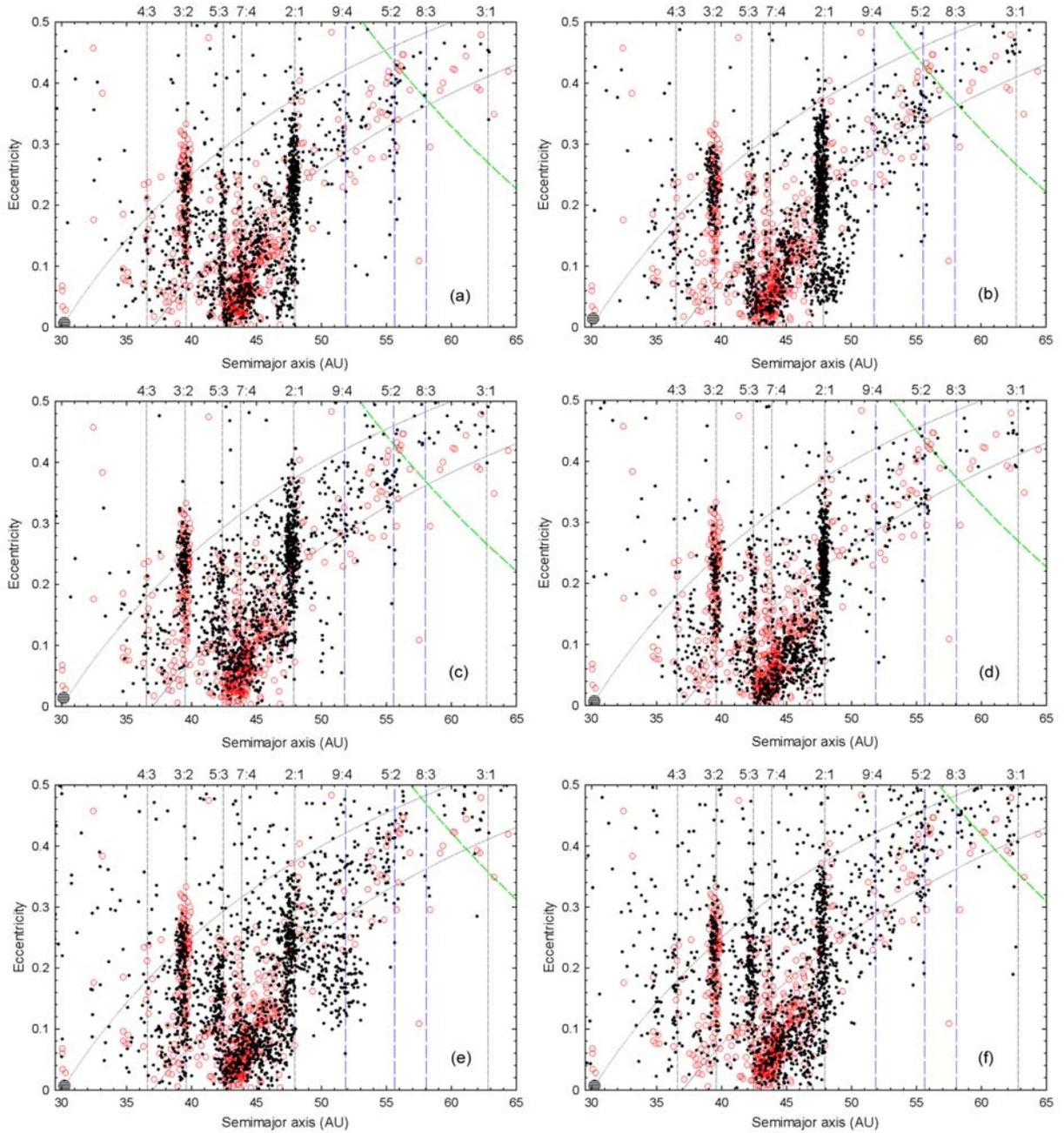

Figure 13. Orbital distribution of previously stirred planetesimal disks with migrating outer planets at the end of planet migration (black circles). The total migration time was 100 Myr. The initial orbital conditions of each disk were taken from the end states of pre-migration runs for the same disks, but taken up to ~51–54 AU (as shown in Fig. 11). TNOs are illustrated for reference (open circles). Dotted curves represent the perihelia of 30 and 37 AU. Vertical dashed lines indicate the position of resonances with Neptune, which is indicated by a hatched circle. In panels a–d, the trans-Plutonian planet (0.4 $M_⊕$) was transported to $a_P ≈ 100$ AU ($e_P ≈ 0.2$ and $i_P ∼ 29$–$36°$), following the location of the 6:1 resonance, while in the bottom panels, the planetoid (0.5 $M_⊕$) acquired $a_P ≈ 130$ AU ($e_P ≈ 0.345$ and $i_P ∼ 40$–$43°$) near the location of the 9:1 resonance. Objects above the long-dashed curves could encounter the trans-Plutonian planet. The orbital distribution from the simulations matches several of the observed structures in the trans-Neptunian region: resonant populations (including those in the 9:4, 5:2, and 8:3 resonances), excited classical region, outer edge at 48 AU, lack of low-$e$ TNOs beyond 45 AU, and the detached population.

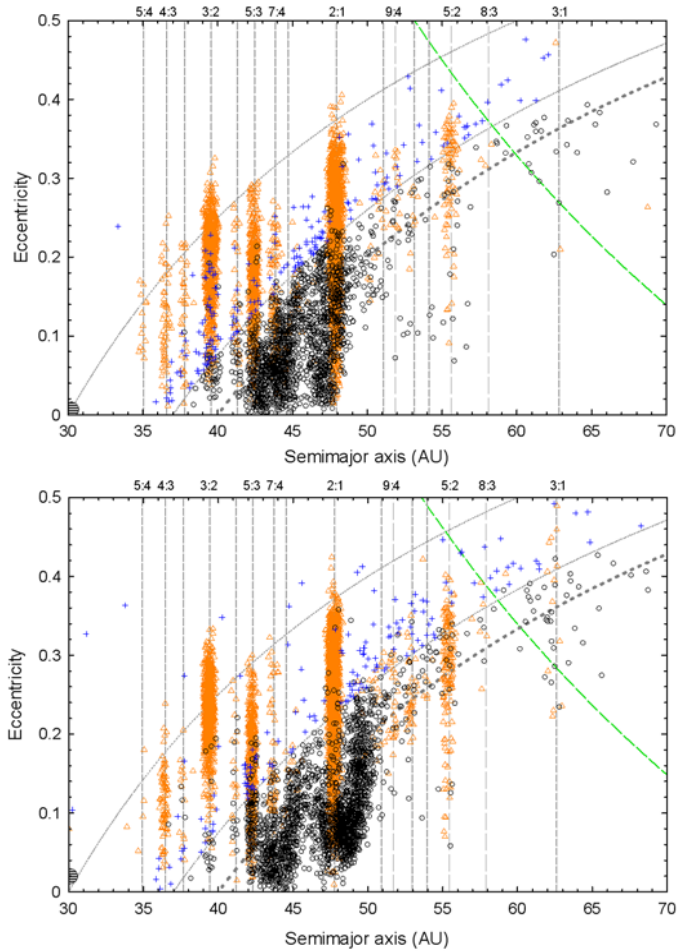

Figure 14. Orbital distribution of planetesimal disks of 51 and 54 AU radius from two reference large-scale simulations after 4 Gyr (top and bottom panels). Three groups are shown: nonresonant (circles), scattered (crosses), and resonant (triangles) objects. Dotted curves represent the perihelia of 30, 37, and 40 AU. Dashed vertical lines indicate the locations of resonances with Neptune. The trans-plutonian planet (0.4 $M_\oplus$) acquired $a_P \approx 100$ AU, $e_P \approx 0.2$ and $i_P \sim 29$–$36°$. Objects above the long-dashed curves could encounter the trans-Plutonian planet. The four main trans-Neptunian populations are reproduced: classical (at ~37–50 AU), resonant (aligned vertically), scattered (mostly at $q < 37$ AU), and detached TNOs ($q > 40$ AU). The excitation in the classical region and the trans-Neptunian belt outer edge at $a \approx 48$ AU are also obtained.

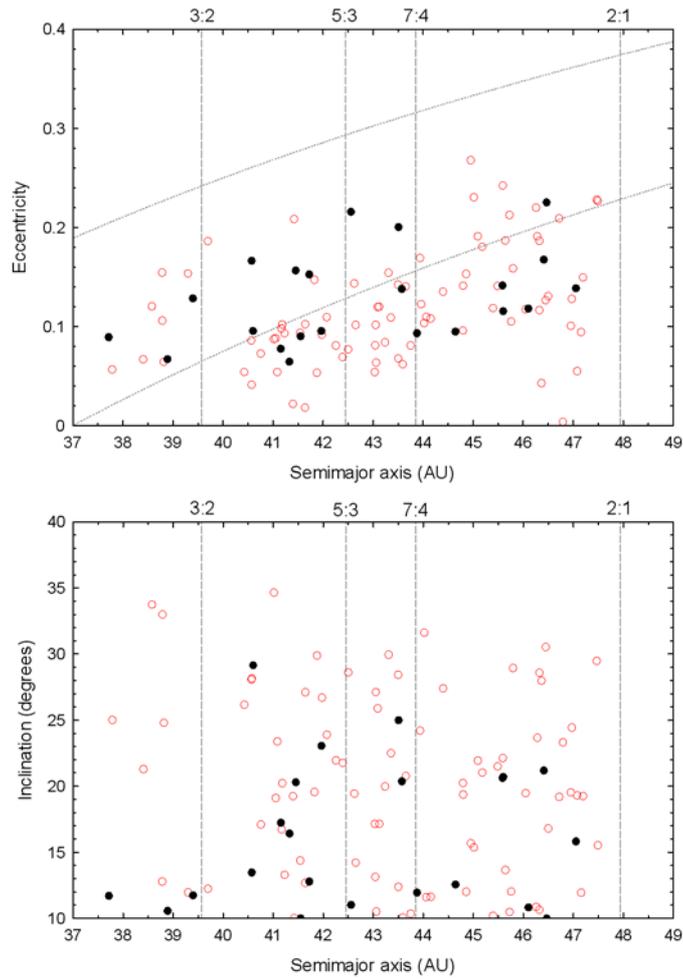

Figure 15. Comparison of orbital distributions of classical objects with $i > 10°$ and $q > 33$ AU between the model (black circles) and observations (open circles). Original observational data were taken from the Lowell Observatory database on 12 September 2006. Only TNOs with long-arc orbits are plotted ($\geq 2$ oppositions). Vertical dashed lines indicate the positions of resonances with Neptune. Dotted curves represent the perihelia of 30 and 37 AU (upper panel). The results represent outcomes after the end of planet migration at 100 Myr for a reference simulation including the trans-Plutonian planet.

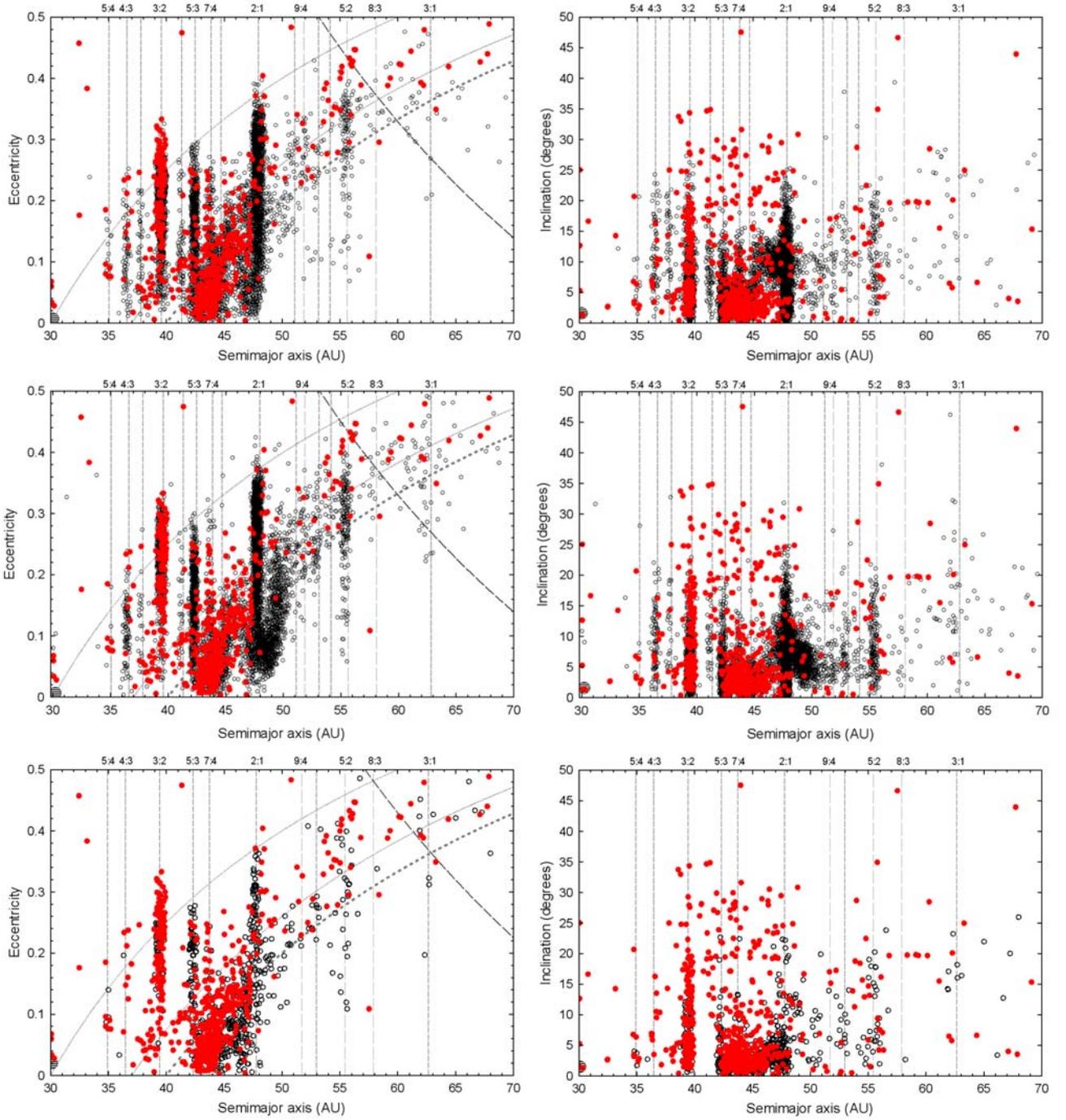

Figure 16. Comparison of orbital distributions between the model (open circles) and observations (filled circles). The results represent outcomes after 4 Gyr for three reference large-scale simulations (two using high resolution) of planetesimal disks of 51 AU (panels a and b) and 54 AU (panels c–f) radius. Dotted curves represent the perihelia of 30, 37, and 40 AU (left panels). Dashed vertical lines indicate the locations of resonances with Neptune. In panels a–d, the trans-Plutonian planet has 0.4 $M_\oplus$, and acquired $a_P \approx 100$ AU, $e_P \approx 0.2$ and $i_P \sim 29$–36°, while in the bottom panels, the planetoid (0.5 $M_\oplus$) acquired $a_P \approx 130$ AU ($e_P \approx 0.345$ and $i_P \sim 40$–43°) near the location of the 9:1 resonance. Objects above the long-dashed curves could encounter the trans-Plutonian planet. Observational biases favor discovery of TNOs at closer distances; hence, 3:2 resonants are overrepresented. Except for the lack of $i > 15$–20° classical objects, the model reproduces the observations very well.

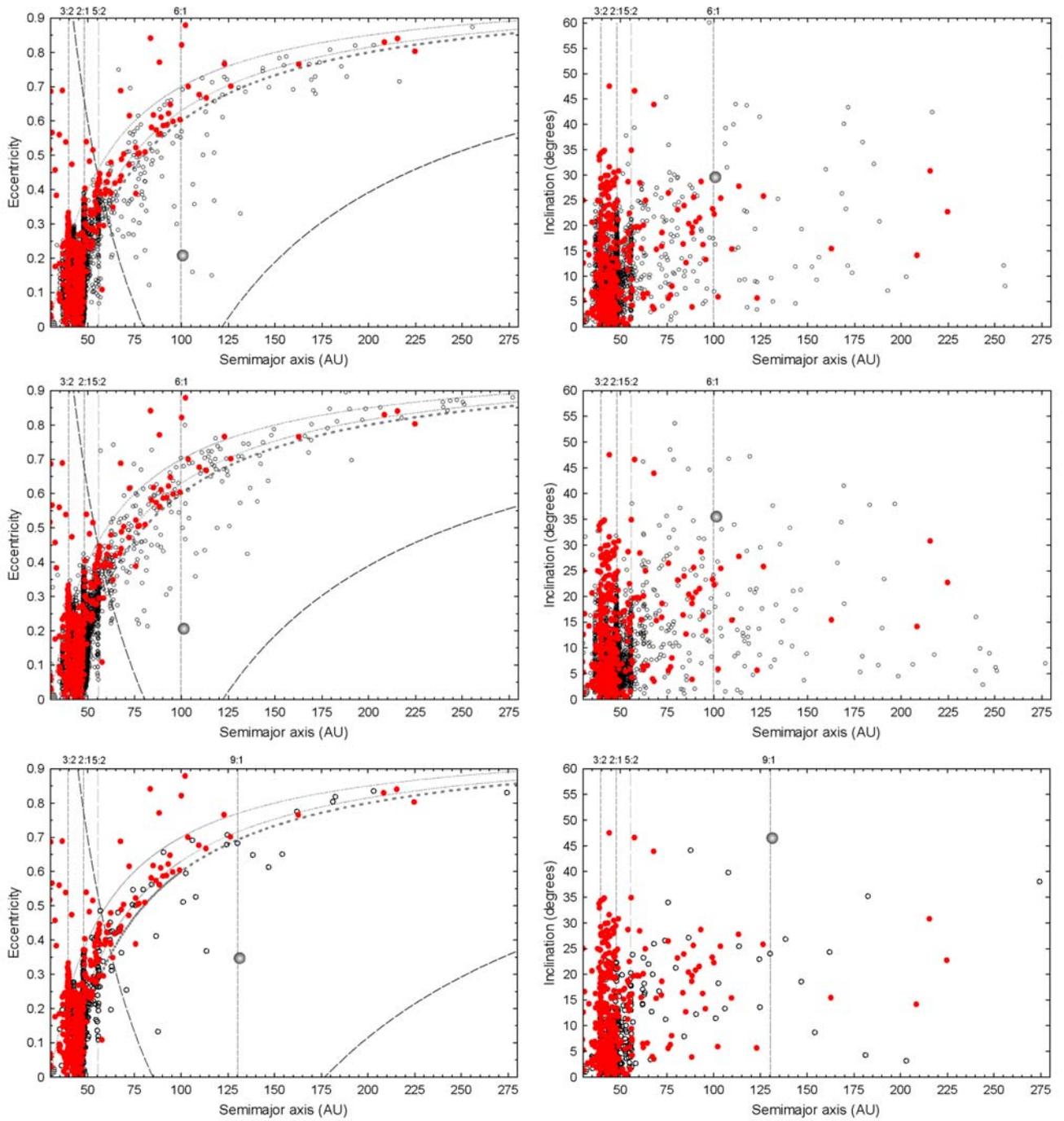

Figure 17. The same outcomes shown in Figure 16, but extended to 280 AU. A detached population with bulk 40 AU < $q$ < 60 AU was formed. Because of severe observational biases, the observed detached population is underrepresented.

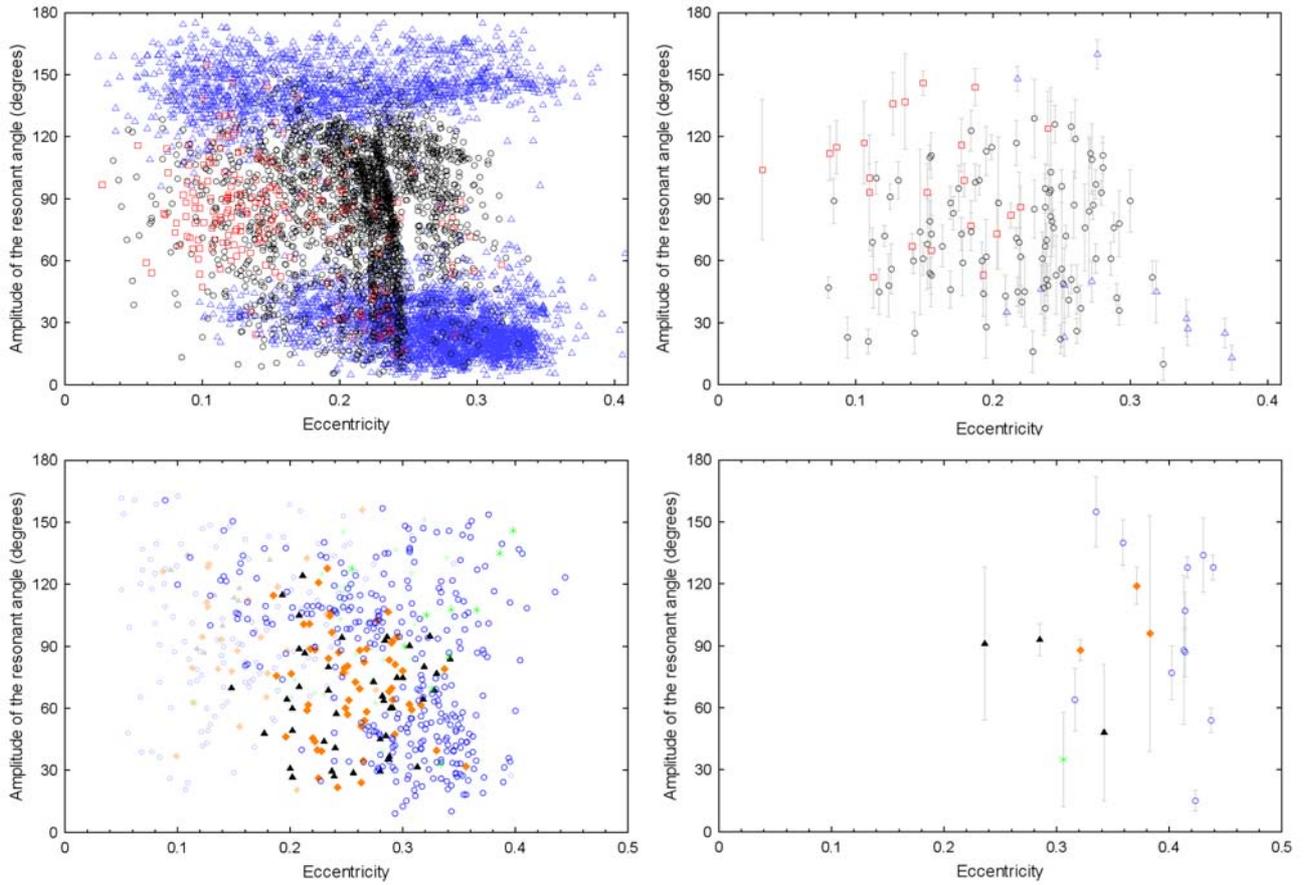

Figure 18. Relation between averaged eccentricities and amplitudes of the resonant angle for some resonances according to the main simulations after 4 Gyr (left panels). For comparison, the distributions for resonant TNOs are also shown (right panels). Top panels: 3:2 (circles), 7:4 (squares), and 2:1 (triangles) resonant bodies. Bottom panels: 9:4 (triangles), 7:3 (diamonds), 5:2 (circles), and 8:3 (stars) resonant bodies. The same symbols in light color refer to resonant bodies that were located beyond ~51–54 AU during capture in the simulations. In general, the agreement of the results with observations is fairly good.

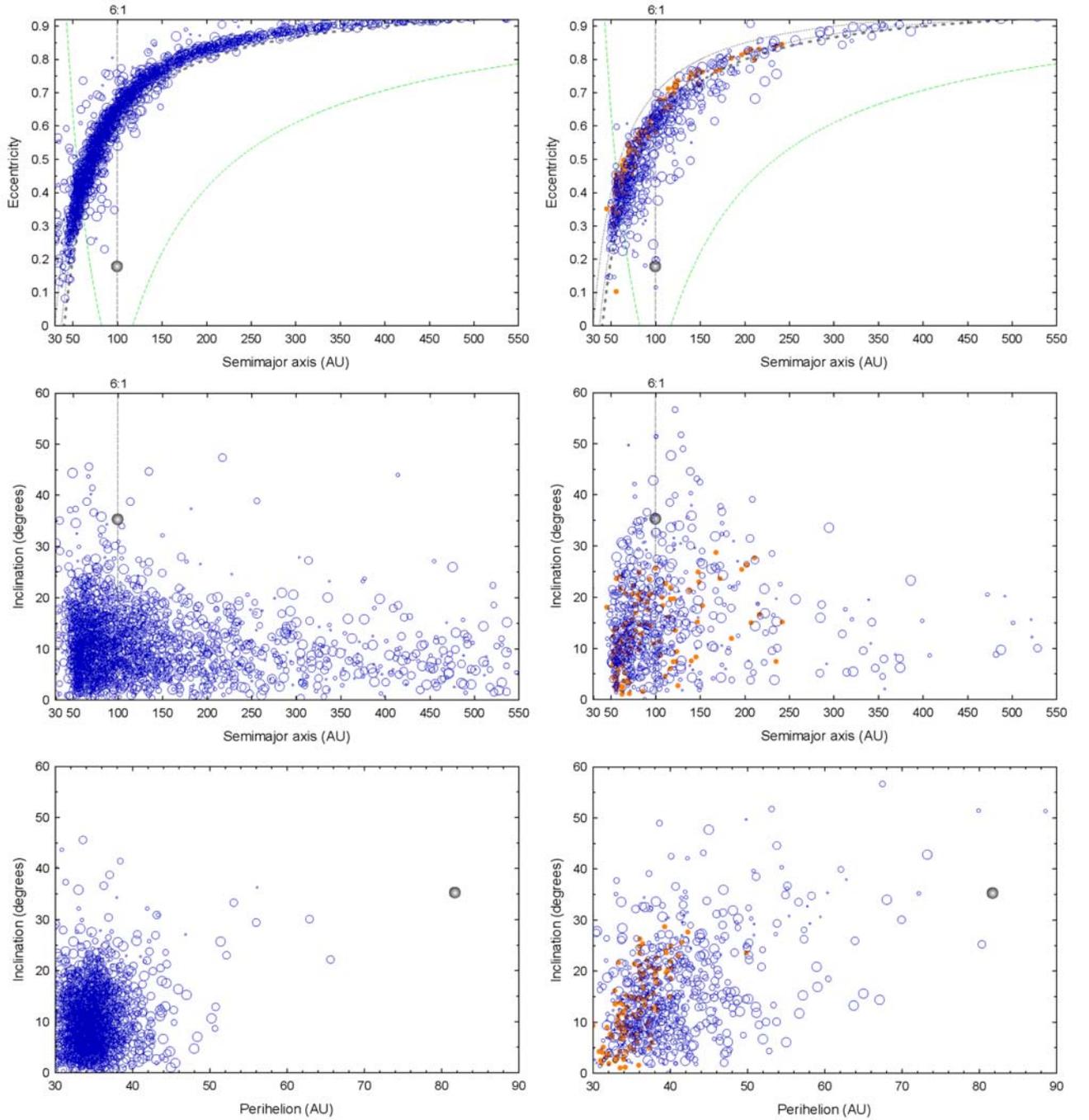

Figure 19. Orbital distributions of scattered objects after 700 Myr (left panels) and 4 Gyr (right panels) (open circles) and the dependence on the outer planet's mass. Dotted curves represent the perihelia of 30, 37, and 40 AU (top panels). A dashed vertical line indicates the location of the 6:1 resonance (top and middle panels). The trans-Plutonian planet is near the 6:1 resonance at $a_P \approx 100$ AU ($e_P \approx 0.18$ and $i_P \approx 35°$) (gray sphere). Objects above the long-dashed curves could encounter the trans-Plutonian planet. We used five distinct masses for the planet, i.e., 0.1, 0.2, 0.3, 0.4, and 0.5 $M_\oplus$, with the size of the open circles representing increasing masses, respectively. For reference, the outcomes of scattered objects after 4 Gyr without an outer planet are indicated by filled circles (right panels).

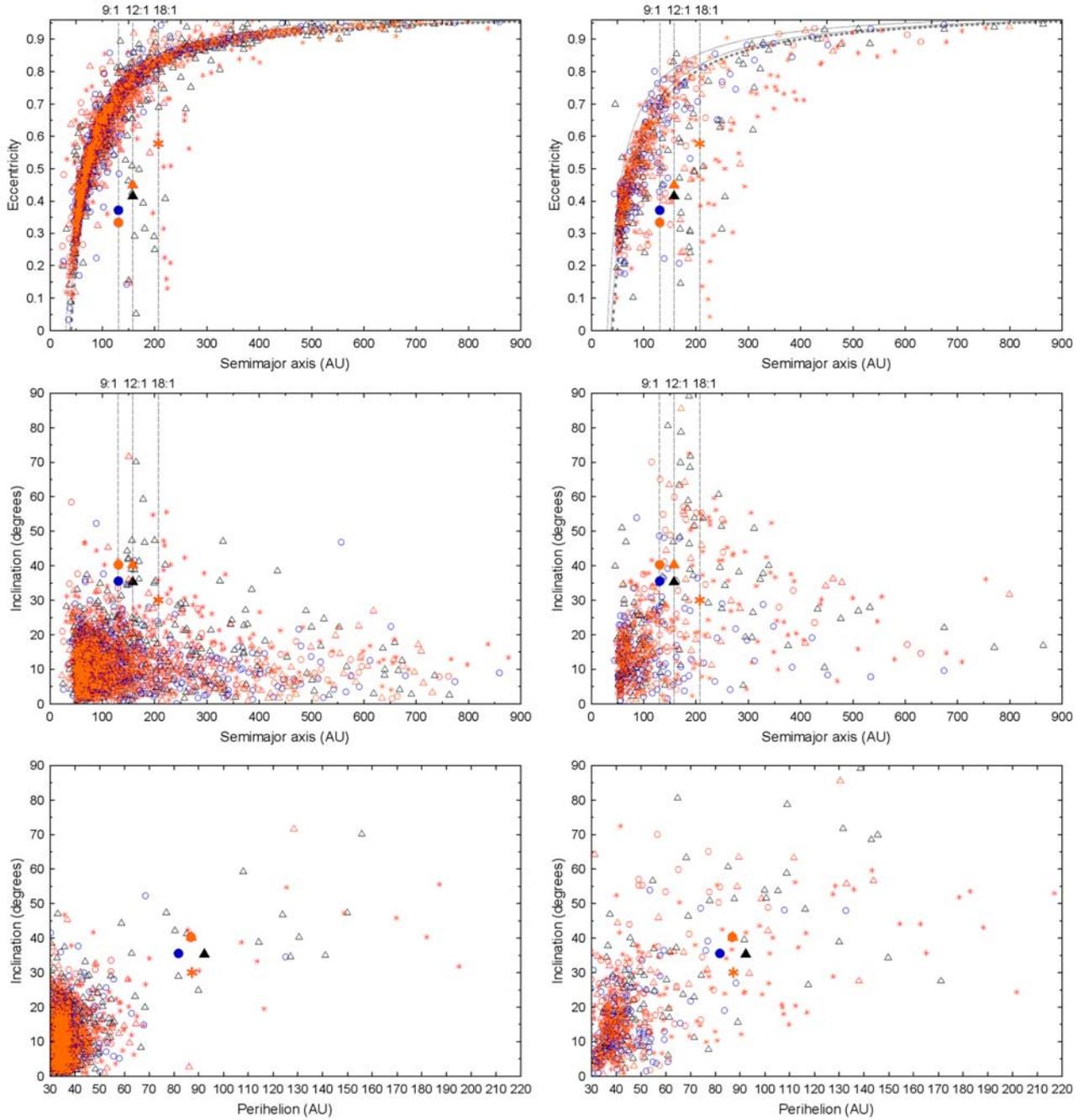

Figure 20. Orbital distributions of scattered objects after 700 Myr (left panels) and 4 Gyr (right panels), for outer planets located near the 9:1 (circles), 12:1 (triangles), and 18:1 (stars) resonances. We used three distinct masses for the planet, that is, 0.5, 0.7, and 1.0 $M_\oplus$, represented by blue, orange, and black symbols, respectively. Dotted curves represent the perihelia of 30, 37, and 40 AU (top panels). Dashed vertical lines indicate the locations of the 9:1, 12:1, and 18:1 resonances (top and middle panels). The trans-Plutonian planets have similar eccentricities and inclinations (big symbols). Objects above the long-dashed curves could encounter the trans-Plutonian planet.

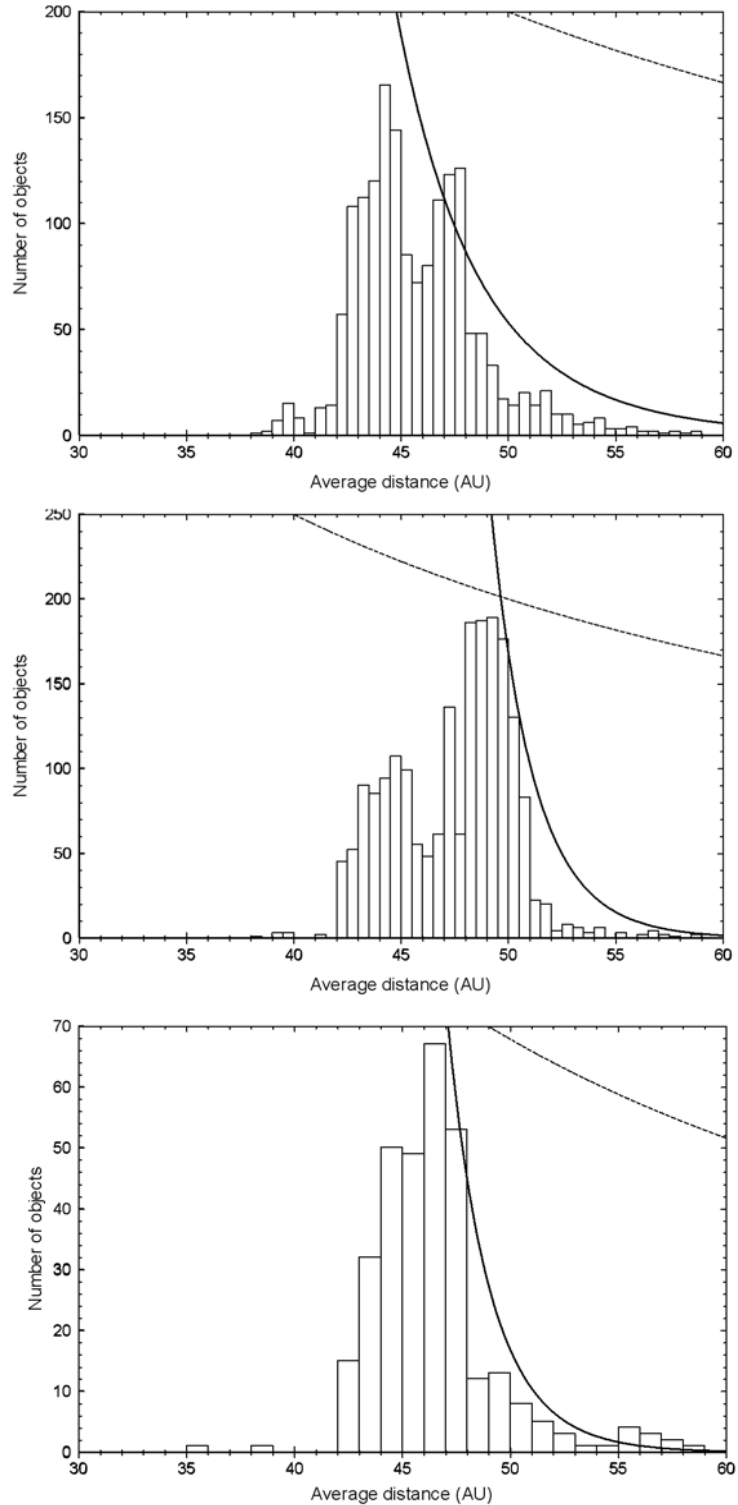

Figure 21. Radial distribution of objects that survived 4 Gyr from three reference large-scale simulations (those shown on the left panels of Figs. 16 and 17). Only nonresonant objects are shown. The disks were originally 51 (top panel) and 54 AU (middle and bottom panels), following an $R^{-1}$ decay in the top and middle panels and an $R^{-1.5}$ decay in the bottom panel (dashed curves). After 4 Gyr, the decay around 45–50 AU is much more abrupt (solid curve), in excellent agreement with unbiased radial studies of the trans-Neptunian belt (e.g., see Morbidelli & Brown, 2004, and references therein).

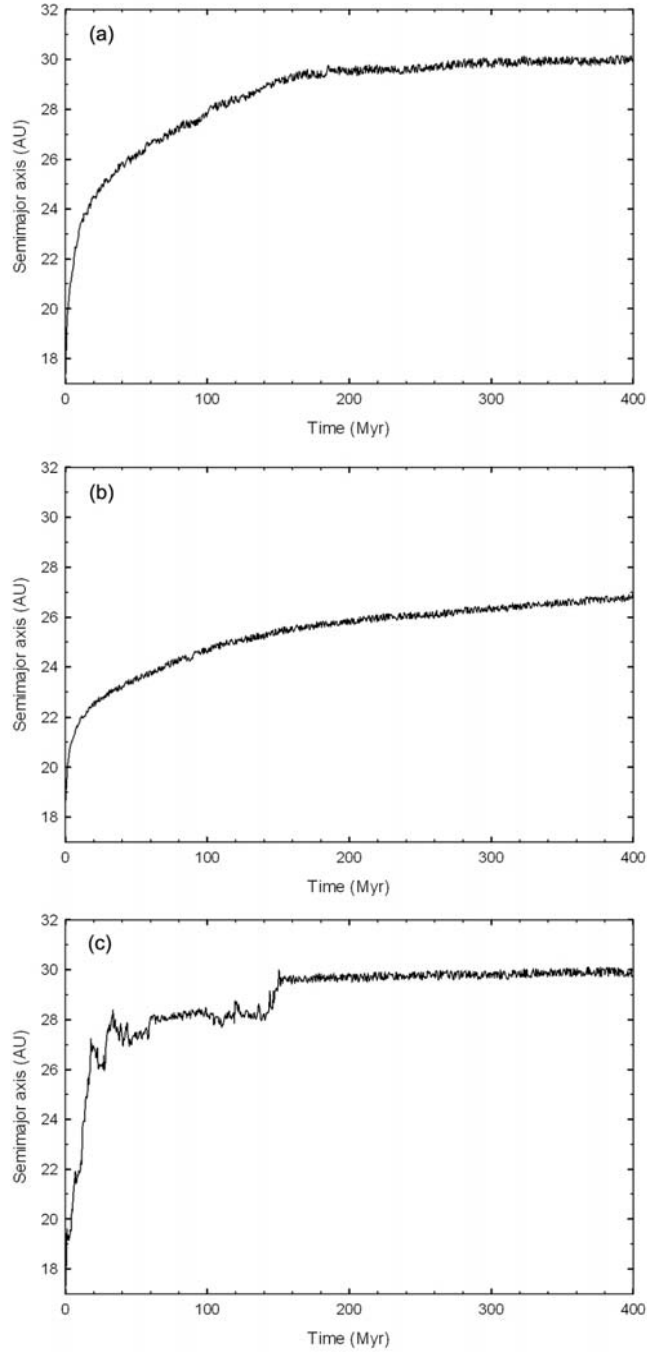

Figure 22. Evolution of Neptune in self-consistent simulations of planet migration using massive planetesimal disks. In all cases, we used 10000 equal-mass planetesimals. (a) The disk was set at 10–60 AU with 75 $M_\oplus$ following an $R^{-1.5}$ distribution (1 MMSN). Only the four giant planets were included in the simulation. Neptune started at 17.1 AU. (b) The disk was set at 20–60 AU with 27 $M_\oplus$ following an $R^{-1.5}$ distribution (0.5 MMSN). The four giant planets and a 0.5-$M_\oplus$ planetoid were included in the simulation. Neptune and the planetoid started at 18.5 AU and 20 AU, respectively. (c) The disk was set at 10–50 AU with 35 $M_\oplus$ following an $R^{-2}$ distribution (0.5 MMSN). The four giant planets and ten 1.0-$M_\oplus$ planetoids (representing massive planetesimals) were included in the simulation. Neptune started at 17.1 AU. The planetoids were randomly distributed within 22–31 AU. One of the massive planetesimals collided with Neptune at 59.4 Myr. The stochastic behavior of Neptune was due to scattering encounters with other massive bodies. At the end of 975 Myr, two of the planetoids remained in the scattered disk.

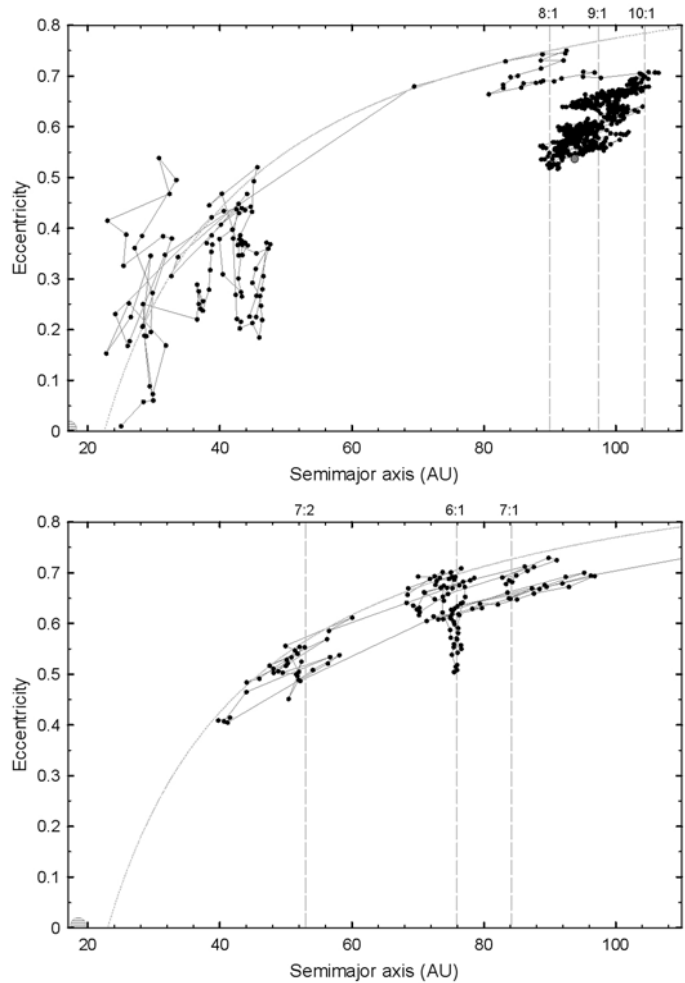

Figure 23. Evolution of scattered planetoids in self-consistent simulations using massive planetesimal disks with the presence of the four giant planets. Dots connected with lines represent the orbital evolution of planetoids for every 0.5 Myr. Vertical dashed lines indicate the locations of resonances with Neptune. The dotted curve represents the perihelion of Neptune at a given time during its migration. In all cases, we used 10 000 equal-mass planetesimals. Top panel: The disk was set at 10–50 AU with 18 $M_\oplus$ following an $R^{-2}$ distribution (0.3 MMSN). Neptune and the planetoid (1.0 $M_\oplus$) started at 17.1 AU and 25 AU, respectively. After being quickly scattered, the planetoid spent several tens of Myr around the 8:1–10:1 resonances when Neptune was at ~21.5–23.5 AU. Bottom panel: The disk was set at 20–60 AU with 54 $M_\oplus$ following an $R^{-1.5}$ distribution (1 MMSN). Neptune and the planetoid started at 18.5 AU and 40 AU, respectively. The planetoid spent 30 Myr in the 6:1 and neighboring resonances. During Kozai resonance interactions, the planetoid's inclination reached up to 36°.

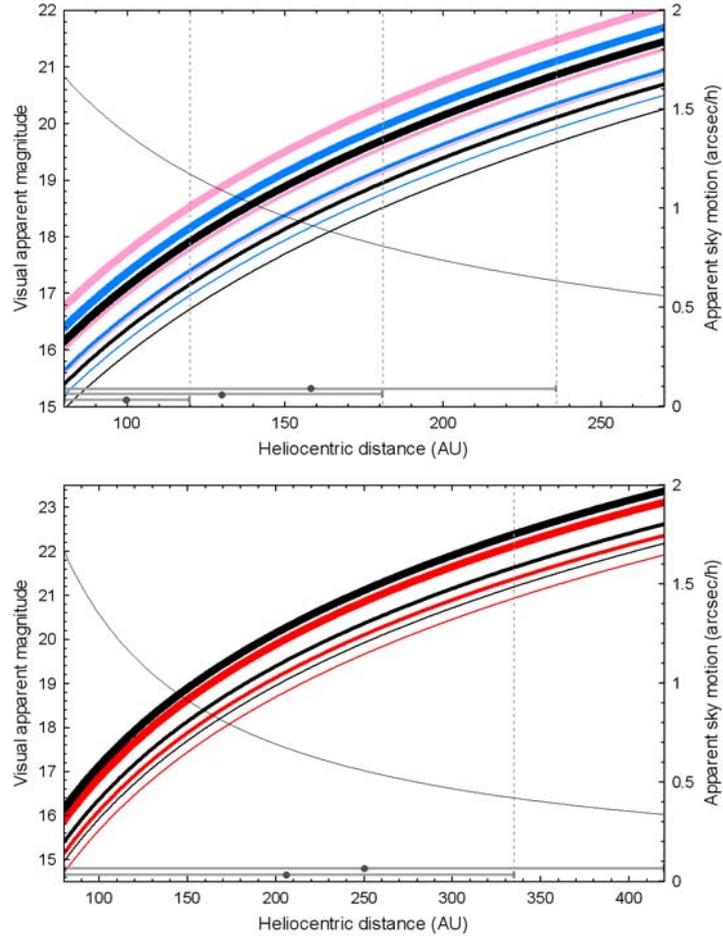

Figure 24. Apparent magnitudes of an outer planet in distant orbits. Pink, blue, and black curves represent outer planets with 0.3, 0.5, and 0.7 $M_\oplus$ (top panel), respectively, and black and red curves indicate outer planets with 0.7 and 1.0 $M_\oplus$ (bottom panel), respectively (mean density $\rho = 2$ g cm$^{-3}$). Thinner curves indicate higher albedos for the assumed values of 0.1, 0.2, and 0.3. Greater mean densities shift the curves upwards as a consequence of slightly smaller diameters. For example, when $\rho = 3$ g cm$^{-3}$, the apparent magnitudes become approximately 0.3 mag darker. A decreasing curve is also shown, representing the apparent sky motion of an outer planet as a function of heliocentric distance. The variation in a typical planet's heliocentric distance, as exemplified in Figs. 16 and 17, is plotted as a horizontal gray line with a dot representing the planetoid's semimajor axis (top panel). Two other hypothetical orbits are shown assuming $q_P = 80$ AU, $a_P \approx 130$ AU (near the 9:1 resonance), and $a_P \approx 158$ AU (near the 12:1 resonance). The planetoid's aphelion would be about 120, 180, and 270 AU for those orbits, respectively. For other possible orbital configurations, such as more distant hypothetical orbits at $a_P \approx 207$ AU (near the 18:1 resonance), or $a_P \approx 250$ AU (near the 24:1 resonance—the uppermost limit for $r$:1 resonances to work in this scenario), the planetoid's aphelion could reach about 335 or 420 AU (bottom panel). Finally, if the outer planet has a moderately large eccentricity, it will spend more time near aphelion during its orbit, resulting in smaller apparent rates and fainter apparent magnitudes.

**Table 1**
**Preliminary Simulations (SIM) – Outer Planets in Selected Resonances in the Scattered Disk**

| SIM | Planetoid location | $M_P$ [a] ($M_⊕$) | $i_P$ [a] (°) | $q_P$ [a] (AU) | Cold disk [b] | 3:2 survival [c] | 2:1 survival [c] | 5:2 survival [c] | Scattered disk [d] |
|---|---|---|---|---|---|---|---|---|---|
| 1 | $a_P$=47.8AU (2:1) | 0.01 | 5 | 45 | 1 | | | | |
| 2 | | 0.025 | 5 | 45 | 1 | | | | |
| 3 | | 0.05 | 5 | 45 | 1 | | | | |
| 4 | $a_P$=62.6AU (3:1) | 0.1 | 10 | 37-53 | 29 | 2 | 2 | | 2 |
| 5 | | | 20 | 50-51 | 2 | | | | |
| 6 | | | 30 | 50-51 | 2 | | | | |
| 7 | | 0.15 | 10 | 48-53 | 14 | 4 | 4 | | 4 |
| 8 | $a_P$=75.8AU (4:1) | 0.1 | 10 | 37-53 | 21 | | | | |
| 9 | | | 20 | 50-53 | 1 | | | | |
| 10 | | | 30 | 50-53 | 1 | | | | |
| 11 | | 0.2 | 10 | 37-53 | 19 | 1 | 1 | | 1 |
| 12 | | | 20 | 50-53 | 3 | | | | |
| 13 | | | 30 | 50-53 | 3 | | | | |
| 14 | | 0.3 | 10 | 40-53 | 23 | 7 | 8 | | 7 |
| 15 | | | 20 | 50-53 | 2 | 1 | 3 | | 2 |
| 16 | | | 30 | 50-53 | 2 | | 1 | | |
| 17 | | | 40 | 50-53 | 1 | | 1 | | |
| 18 | | 0.4 | 10 | 50-51 | 1 | | | | |
| 19 | | 0.5 | 10 | 50-51 | 1 | | | | |
| 20 | $a_P$=82.0AU (9:2) | 0.3 | 10 | 59 | | 1 | 1 | 1 | |
| 21 | | 0.35 | 10 | 59 | | 1 | 1 | 1 | |
| 22 | | 0.4 | 10 | 59 | | 1 | 1 | 1 | |
| 23 | $a_P$=88.0AU (5:1) | 0.25 | 10 | 47-53 | 8 | | | | |
| 24 | | 0.4 | 10 | 50-53 | 10 | | | | |
| 25 | $a_P$=93.8AU (11:2) | 0.4 | 15 | 70 | | 1 | 1 | 1 | |
| 26 | | | 40 | 70 | | 1 | 1 | 1 | |
| 27 | | 0.5 | 15 | 70 | | 1 | 1 | 1 | |
| 28 | | | 40 | 70 | | 1 | 1 | 1 | |
| 29 | $a_P$=99.4AU (6:1) | 0.1 | 36 | 82 | | | | | 1 |
| 30 | | 0.2 | 36 | 82 | | | | | 1 |
| 31 | | 0.3 | 36 | 82 | | | | | 1 |
| 32 | | 0.4 | 36 | 82 | | 1 | 1 | 1 | 1 |
| 33 | | | 25 | 52 | | | | | 1 |
| 34 | | 0.5 | 36 | 82 | | | | | 2 |
| 35 | | | 46 | 81 | | | | | 1 |
| 36 | | | 41 | 91 | | | | | 1 |
| 37 | $a_P$=120.4AU (8:1) | 0.4 | 25 | 75 | | | | 1 | |
| 38 | | | 40 | 75 | | | | 1 | |
| 39 | | 0.5 | 25 | 75 | | | | 1 | |
| 40 | | | 40 | 75 | | | | 1 | |
| 41 | $a_P$=130.2AU (9:1) | 0.4 | 36 | 82 | | | | | 1 |
| 42 | | 0.5 | 36 | 82 | | | | | 1 |
| 43 | | | 41 | 83 | | | 1 | 1 | 1 |
| 44 | | | 41 | 91 | | | | | 1 |
| 45 | | 0.7 | 20 | 87 | | | | | 1 |
| 46 | | | 40 | 87 | | | | | 1 |
| 47 | $a_P$=139.7AU (10:1) | 0.2 | 10 | 40-56 | 18 | | | | |
| 48 | | 0.5 | 10 | 53-56 | 7 | | | | |
| 49 | | 0.7 | 36 | 87 | | | | | 1 |
| 50 | | 1.0 | 10 | 55-56 | 8 | 2 | 4 | | 2 |
| 51 | | | 20 | 55-56 | 1 | | 1 | | |
| 52 | | | 30 | 55-56 | 1 | | 1 | | |
| 53 | | | 40 | 55-56 | 1 | | 1 | | |
| 54 | | 1.5 | 10 | 56 | 1 | | | | |
| 55 | $a_P$=148.9AU (11:1) | 0.5 | 10 | 70-71 | 3 | | | | |
| 56 | | 0.7 | 40 | 87 | | | | | 1 |
| 57 | | 1.0 | 10 | 70-71 | 6 | | | | |
| 58 | | 1.5 | 10 | 70-71 | 2 | | | | |

| | | | | | | | |
|---|---|---|---|---|---|---|---|
| 59 | $a_P$=157.8AU (12:1) | 0.5 | 21 | 83 | | | 1 |
| 60 | | 0.5 | 40 | 82 | | | 1 |
| 61 | | 0.7 | 40 | 87 | 1 | 1 | 1 |
| 62 | | 1.0 | 35 | 92 | | | 1 |
| 63 | | | 50 | 92 | | | 1 |
| 64 | $a_P$=206.7AU (18:1) | 0.7 | 30 | 87 | | | 1 |
| 65 | | 1.0 | 30 | 90 | | | 1 |
| 66 | $a_P$=250.4AU (24:1) | 0.5 | 30 | 86 | | | 1 |
| 67 | | 0.7 | 30 | 88 | | | 1 |
| 68 | | 1.0 | 30 | 92 | | | 1 |

[a] $M_P$ = mass in Earth masses, $i_P$ = inclination, and $q_P$ = perihelion distance of the outer planet. A range of perihelia was used in some cases.

[b] Number of runs using a disk of particles uniformly distributed in cold orbital conditions (near-circular and very low-$i$ orbits). The disk for the 2:1 resonance contained 301 particles inside 35–55 AU. For close resonances (3:1 to 5:1), the disk was set at 40–120 AU and was modeled with 100 particles. For farther resonances, a larger disk at 40–240 AU was used with 250 particles. All runs were followed for 4 Gyr.

[c] Number of runs to check the survival of specific stable resonant populations with an outer planet over 4 Gyr. We used pre-recorded long-term resonant members, namely 166, 147, and 260 particles inside the 3:2 ($0.10 < e < 0.34$), 2:1 ($0.07 < e < 0.40$), and 5:2 ($0.35 < e < 0.45$) resonances (all $i < 25°$).

[d] Number of runs using a disk of particles distributed initially on Neptune-encountering orbits ($q < 35$ AU). In most cases the disk was set between 50 and 500 AU. All runs were followed for 4 Gyr.

**Table 2**
**Preliminary Simulations (SIM) – Migrating Outer Planets**

| SIM | Final planetoid location | $M_P$ [a] ($M_\oplus$) | $\tau$ [b] (Myr) | $q_P$ [c] (AU) | Cold disk [d] | Extension over Gyr [e] | Hot classicals [f] | Excited disk [g] |
|---|---|---|---|---|---|---|---|---|
| 1 | $a_P \approx 62.5$AU (3:1) | 0.1 | 1 | 47-48 | 10 | | | |
| 2 | | | 5 | 47-52 | 12 | 1 | | |
| 3 | | | 10 | 47-50 | 15 | 3 | | |
| 4 | | 0.15 | 10 | 47-48 | 8 (8) | 1 (1) | | |
| 5 | $a_P \approx 76$AU (4:1) | 0.2 | 1 | 50 | 7 | | | |
| 6 | | | 5 | 50 | 9 | 1 | | |
| 7 | | | 10 | 50 | 9 | 3 | 1 | |
| 8 | | 0.3 | 5 | 56-65 | 9 | | 3 | |
| 9 | | | 10 | 52-65 | 20 (8) | 7 (4) | 2 | |
| 10 | $a_P \approx 82$AU (9:2) | 0.3 | 5 | 59 | 1 | | | |
| 11 | | | 10 | 57-59 | 2 | | | |
| 12 | | | 15 | 57-60 | 5 (2) | 1[*] | 2 | |
| 13 | | 0.35 | 10 | 59-60 | 4 (2) | 1[*] | 2 | |
| 14 | | | 15 | 60 | 1 | | | |
| 15 | | 0.4 | 5 | 59 | 1 | | 1 | |
| 16 | | | 8 | 59 | 1 | | | |
| 17 | | | 10 | 59-60 | 4 (2) | 1[*] | 3 | |
| 18 | $a_P \approx 88$AU (5:1) | 0.3 | 15 | 66 | 1 | | | |
| 19 | | | 20 | 59 | 1 | | | |
| 20 | | 0.35 | 10 | 66 | 1 | | | |
| 21 | | | 15 | 66 | 1 | | | |
| 22 | | 0.4 | 10 | 66 | 1 | | 1 | |
| 23 | | | 15 | 66 | 1 | | 1 | |
| 24 | $a_P \approx 94$AU (11:2) | 0.4 | 10 | 71-75 | 2 | | | 1 |
| 25 | | | 15 | 71 | 1 | | | |
| 26 | | 0.5 | 10 | 71 | 1 | | | |
| 27 | | | 15 | 71 | 1 | | | |
| 28 | $a_P \approx 100$AU (6:1) | 0.3 | 10 | 59 | 1 | | | |
| 29 | | | 15 | 59 | 1 | | | |
| 30 | | | 20 | 59 | 1 | | | |
| 31 | | 0.4 | 10 | 75 | 1 | | | 3 |
| 32 | | | 15 | 77 | 2 | | 1 | |
| 33 | | 0.5 | 10 | 59-77 | 3 | | 1 | |
| 34 | | | 15 | 59-77 | 3 | | | |
| 35 | | | 20 | 59 | 1 | | | |
| 36 | $a_P \approx 105$AU (13:2) | 0.4 | 10 | 76 | 1 | | | 2 |
| 37 | $a_P \approx 110$AU (7:1) | 0.3 | 20 | 62 | 1 | | | |
| 38 | | 0.4 | 10 | 75-77 | 2 | | | 5 |
| 39 | | 0.7 | 5 | 53-84 | 3 | | | |
| 40 | | | 10 | 53-84 | 4 | | | |
| 41 | | | 15 | 62 | 1 | | | |
| 42 | | | 20 | 62 | 1 | | | |
| 43 | $a_P \approx 120$AU (8:1) | 0.4 | 10 | 75 | 1 | | | 1 |
| 44 | $a_P \approx 130$AU (9:1) | 0.4 | 10 | 77 | 1 | | | 1 |
| 45 | $a_P \approx 140$AU (10:1) | 1.0 | 1 | 53 | 9 | | | |
| 46 | | | 5 | 53 | 8 | 1 | | |
| 47 | | | 10 | 53 | 20 (6) | 11 (2) | | |

[a] Mass of the outer planet in Earth masses.

[b] Migration timescale (eq. [1]).

[c] Outer planet perihelion acquired at the end of migration.

[d] Total number of runs using a disk of particles uniformly distributed in cold orbital conditions (near-circular and very low-$i$ orbits). The disk consisted of two regions set within 20–48.5 AU with 220–445 particles and within 49–78 AU with 50–105 particles. The planetoid's default inclination was

10°. When 20° or 30° were used, the number of extra runs is indicated in parentheses. All runs were followed for 100–125 Myr.

[e] Total number of runs with cold disks extended to 4 Gyr. The numbers in parentheses indicate the cases in which $i_P = 20°$ or 30°. A few simulations were followed for 1 Gyr only (marked with *).

[f] Number of runs to investigate the formation of hot classical objects using an approach similar to that of Gomes (2003b).

[g] Number of runs using an excited disk of particles representing observations in the 42–58 AU region. All runs were followed for 100–125 Myr.

**Table 3**

**Survival of Resonant TNOs and Their Maximum Eccentricities with the Presence of Outer Planets in the Scattered Disk after 4 Gyr**

| Planetoid location | $M_P$ [a] ($M_⊕$) | $i_P$ [a] (°) | $q_P$ [a] (AU) | 3:2 resonance Fraction [b] (%) | 3:2 resonance Maximum $e$ [b] | 2:1 resonance Fraction [b] (%) | 2:1 resonance Maximum $e$ [b] | 5:2 resonance Fraction [b] (%) | 5:2 resonance Maximum $e$ [b] |
|---|---|---|---|---|---|---|---|---|---|
| $a_P≈62.5$AU (3:1) | 0.1 | 10 | 50 | 21 | 0.259 (49.7AU) | 0 | - | | |
| | | 10 | 50 | 14 | 0.245 (49.1AU) | 0 | - | | |
| | 0.15 | 10 | 50 | 22 | 0.224 (48.3AU) | 0 | - | | |
| | | 10 | 50 | 9 | 0.236 (48.8AU) | 0 | - | | |
| | | 10 | 50 | 20 | 0.274 (50.3AU) | <<1 | - | | |
| | | 10 | 50 | 14 | 0.258 (49.6AU) | <<1 | - | | |
| $a_P≈76$AU (4:1) | 0.2 | 10 | 53 | 69 | 0.311 (51.7AU) | 0 | - | | |
| | 0.3 | 10 | 50 | 35 | 0.278 (50.4AU) | 0 | - | | |
| | | 10 | 50 | 15 | 0.265 (49.9AU) | 0 | - | | |
| | | 10 | 52 | | | 0 | - | | |
| | | 10 | 53 | 45 | 0.297 (51.2AU) | 0 | - | | |
| | | 10 | 53 | 65 | 0.341 (52.9AU) | <<1 | - | | |
| | | 10 | 53 | 35 | 0.268 (50.0AU) | 0 | - | | |
| | | 10 | 53 | 41 | 0.262 (49.8AU) | 0 | - | | |
| | | 10 | 53 | 43 | 0.321 (52.1AU) | 0 | - | | |
| | | 20 | 52 | | | 0 | - | | |
| | | 20 | 53 | 37 | 0.251 (49.4AU) | 0 | - | | |
| | | 20 | 53 | | | <<1 | 0.255 (*) (60.0AU) | | |
| | | 30 | 52 | | | 0 | - | | |
| | | 40 | 52 | | | 1 | 0.137 (54.3AU) | | |
| $a_P≈82$AU (9:2) | 0.3 | 10 | 59 | 79 | 0.336 (52.6AU) | 4 | 0.256 (60.0AU) | 1 | 0.373 (*) (76.2AU) |
| | 0.35 | 10 | 59 | 74 | 0.320 (52.0AU) | 3 | 0.213 (58.0AU) | 1 | 0.316 (*) (73.0AU) |
| | 0.4 | 10 | 59 | 75 | 0.332 (52.5AU) | 3 | 0.174 (56.1AU) | 0 | - |
| $a_P≈94$AU (11:2) | 0.4 | 15 | 70 | | | 48 | 0.436 (68.6AU) | 3 | 0.372 (76.1AU) |
| | | 40 | 70 | | | 57 | 0.414 (67.6AU) | 3 | 0.326 (73.5AU) |
| | 0.5 | 15 | 70 | | | 47 | 0.395 (66.7AU) | 2 | 0.356 (75.2AU) |
| | | 40 | 70 | | | 44 | 0.351 (64.6AU) | 2 | 0.272 (70.5AU) |
| $a_P≈100$AU (6:1) | 0.4 | 36 | 82 | 85 | 0.321 (52.0AU) | 52 | 0.395 (66.7AU) | 49 | 0.459 (80.8AU) |
| $a_P≈120$AU (8:1) | 0.4 | 25 | 75 | | | | | 16 | 0.388 (77.0AU) |
| | | 40 | 75 | | | | | 23 | 0.449 |
| | 0.5 | 25 | 75 | | | | | 10 | 0.382 (80.4AU) |
| | | 40 | 75 | | | | | 19 | 0.418 (76.7AU) |
| $a \approx 130$AU (9:1) | 0.5 | 41 | 83 | | | 55 | 0.377 (65.8AU) | 54 | 0.461 (78.6AU) (81.0AU) |
| $a_P \approx 140$AU (10:1) | 1.0 | 10 | 55 | 15 | 0.319 (52.0AU) | 0 | - | | |
| | | 10 | 55 | 25 | 0.284 (50.7AU) | 0 | - | | |
| | | 10 | 56 | | | 0 | - | | |
| | | 10 | 56 | | | <<1 | - | | |
| | | 20 | 56 | | | 0 | - | | |
| | | 30 | 56 | | | 0 | - | | |
| | | 40 | 56 | | | 0 | - | | |
| $a \approx 158$AU (12:1) | 0.7 | 40 | 87 | | | 52 | 0.395 (66.7AU) | 62 | 0.476 (81.8AU) |
| No outer planet [c] | - | - | - | 100 | 0.330 (52.5AU) | 100 | 0.399 (66.8AU) | 100 | 0.440 (79.9AU) |
| Observations [d] | - | - | - | - | 0.324 (52.2AU) | - | 0.374 (65.7AU) | - | 0.439 (79.8AU) |

[a] $M_P$ = mass in Earth masses, $i_P$ = inclination, and $q_P$ = perihelion distance of the outer planet.

[b] Fraction of stable resonant bodies that survived 4 Gyr. We estimated an error of ±5% to account for the stochastic long-term evolution of resonant bodies after the inclusion of the planetoid in the system. The aphelion distance associated with the maximum eccentricity is indicated in parentheses. In a few cases, survivors with high eccentricity were either captured or had their eccentricity excited by resonance dynamics near the end of the run (marked with *).

[c] Results of simulations for the stability of 3:2-, 2:1-, and 5:2-resonant TNOs over 4 Gyr in which only the four giant planets were included.

[d] Averaged values were taken from LM07b. Original orbital data are from the Lowell Observatory database.

**Table 4**

**Main Simulations (SIM) – Migrating Outer Planets Using the Hybrid Migration Model**

| SIM | $a_{N0}$ [a] (AU) | $M_P$ [b] ($M_\oplus$) | $\Delta a_P$ [c] (AU) | $\Delta i_P$ [d] (°) | $\Delta q_P$ [e] (AU) | Cold disk [f] | Pre-migration perturbed disks [g] | Excited disk [h] |
|---|---|---|---|---|---|---|---|---|
| 1 | 30.1* | 0.4 | ≈100* | 10-36 | 35-76 | 1 | | 2 |
| 2 | 20 | 0.4 | 66.5-99.5 | ≈10* | 30-35 | 1 | | 2 |
| 3 | 20 | 0.4 | 66.5-99.5 | 10-36 | 30-79.5 | 2 | 13 (2) | 4 |
| 4 | 20 | 0.5 | 66.5-99.5 | 10-36 | 35-79.5 | 1 | 3 | 2 |
| 5 | 18 | 0.4 | 54.5-99.5 | 10-26 | 35-79.5 | 2 | | |
| 6 | 18 | 0.4 | 60.5-99.5 | 10-29 | 30-79.5 | 2 | 6 (2) | |
| 7 | 17 | 0.5 | 73.5-130 | 11-43 | 22-85 | 5 | 16 (2) | |

[a] $a_{N0}$ = initial semimajor axis of Neptune, before planet migration. * In SIM 1, the giant planets did not migrate.

[b] Mass of the outer planet in Earth masses.

[c] Approximate variation in the outer planet's semimajor axis during planet migration. * In SIM 1, there was no such variation.

[d] Variation in the outer planet's inclination during planet migration. * In SIM 2, there was no such variation.

[e] Variation in the outer planet's perihelion during planet migration.

[f] Total number of runs using a disk of particles uniformly distributed in cold orbital conditions (near-circular and very low-$i$ orbits). The disk consisted of inner and outer regions set within ~33–49 AU with 400 particles and within 49–110 AU with 95 particles. All runs were followed for 100–200 Myr.

[g] Total number of runs using perturbed disks as found at the end of the best runs of pre-migration excitation (see Section 6.1). The numbers in parentheses indicate simulations extended to 4 Gyr using the same disks, performed with several thousand particles.

[h] Number of runs using an excited disk of particles representing observations in the 42–58 AU region. All runs were followed for 100–200 Myr.

**Table 5**
**Resonant Populations Using 60-AU-sized Disks after 4 Gyr—Summary**

| Resonant population | $a_{res}$[a] (AU) | $N$[b] | $e$[c] | $i$[c] (°) | $N_{KR}$[b] | $f_{KR}$[b] (%) | Min $A$[d] (°) | Max $A$[d] (°) |
|---|---|---|---|---|---|---|---|---|
| 5:4 | 34.9 | 20 (20) | 0.11 | 5.6 | 1 | 5 | 20 | 126 (126) |
| 4:3 | 36.5 | 152 (152) | 0.11 | 9.7 | 18 | 12 | 10 | 142 (142) |
| 7:5 | 37.7 | 50 (48) | 0.13 | 12.4 | 15 | 30 | 14 | 113 (113) |
| 3:2 | 39.4 | 2184 (2157) | 0.22 | 5.3 | 535 | 24 | 5 | 150 (146) |
| 8:5 | 41.2 | 54 (34) | 0.15 | 7.9 | 9 | 17 | 23 | 137 (111) |
| 5:3 | 42.3 | 1193 (1163) | 0.18 | 2.8 | 69 | 6 | 7 | 164 (164) |
| 7:4 | 43.7 | 171 (134) | 0.14 | 5.7 | 59 | 35 | 14 | 155 (131) |
| 9:5 | 44.5 | 15 (12) | 0.10 | 2.8 | 0 | 0 | 34 | 145 (116) |
| 11:6 | 45.1 | 17 (11) | 0.16 | 4.4 | 0 | 0 | 26 | 135 (132) |
| 2:1 | 47.8 | 4122 (4088) | 0.26 | 4.1 | 223 | 5 | 60 | 175 (175) |
|  |  |  |  |  |  |  | 5 | 60 (60) |
| 13:6 | 50.4 | 12 (5) | 0.26 | 5.2 | 0 | 0 | 53 | 131 (77) |
| 11:5 | 50.9 | 25 (22) | 0.26 | 6.8 | 0 | 0 | 34 | 136 (129) |
| 9:4 | 51.7 | 49 (44) | 0.23 | 4.5 | 0 | 0 | 27 | 134 (134) |
| 7:3 | 53.0 | 75 (64) | 0.24 | 5.8 | 11 | 15 | 20 | 128 (128) |
| 5:2 | 55.4 | 397 (353) | 0.27 | 6.7 | 158 | 40 | 9 | 169 (161) |
|  |  | [277 (264)] | [0.30] | [8.2] | [118] | [43] |  |  |
| 8:3 | 57.9 | 30 (12) | 0.28 | 3.6 | 1 | 3 | 33 | 146 (128) |
|  |  | [9 (5)] | [0.33] | [5.8] | [1] | [11] |  |  |
| 11:4[*] | 59.1 | 17 (6) | 0.26 | 2.7 | 0 | 0 | 54 | 134 (98) |
| 3:1 | 62.6 | 206 (175) | 0.11 | 3.1 | 3 | 1 | 103 | 174 (174) |
|  |  | [20 (8)] | [0.41] | [17.8] | [1] | [5] |  |  |

[a] Resonance semimajor axis.

[b] $N$ = number of resonant particles. The number of particles locked in resonance over timescales >3 Gyr is indicated in parentheses. $N_{KR}$ = number of objects that experienced Kozai resonance and their fraction of the total resonant population, $f_{KR}$. When considering truncation of the initial disks at ~51–54 AU, the populations and fractions of KR bodies in affected resonances are given within brackets.

[c] Orbital elements represent the median of each population, where $e$ = eccentricity and $i$ = inclination. When considering truncation of the initial disks at ~51–54 AU, the values of $e$ and $i$ in affected resonances are given within brackets.

[d] Min $A$ and Max $A$ give the minimum and maximum vales of libration amplitudes (amplitudes of the resonant angle), respectively. When only particles with timescales >3 Gyr are taken into account, the maximum libration amplitude is indicated in parentheses. The error is approximately ±5°. In the case of 2:1-resonant objects, the second line represents asymmetric librators.

* 11:4-resonant bodies were obtained only in planetesimal disks extending beyond 55 AU.

**Table 6**
**Time Evolution of Scattered and Detached Populations with Resident Outer Planets in the Scattered Disk**

| Planetoid location | $M_P$ [a] ($M_\oplus$) | $i_P$ [a] (°) | $q_P$ [a] (AU) | $P_{scat}$ [b] (%) | | | $P_{det}$ [b] (%) | | | Ratio$_{SD}$ at 4Gyr [c] | Median $q_{det}$ [d] (AU) |
|---|---|---|---|---|---|---|---|---|---|---|---|
| | | | | t=200 Myr | t=700 Myr | t=4Gyr | t=200 Myr | t=700 Myr | t=4Gyr | | |
| $a_P \approx 100$AU (6:1) | 0.1 | 36 | 82 | 99.9 | 99.2 | 83.9 | 0.1 | 0.8 | 16.1 | 5.2 | 42.5 |
| | 0.2 | 36 | 82 | 99.8 | 98.3 | 68.3 | 0.2 | 1.7 | 31.7 | 2.2 | 42.5 |
| | 0.3 | 36 | 82 | 98.9 | 95.7 | 58.0 | 1.1 | 4.3 | 42.0 | 1.4 | 43.3 |
| | 0.4 | 36 | 82 | 99.0 | 91.0 | 47.2 | 1.0 | 9.0 | 52.8 | 0.9 | 45.7 |
| | 0.5 | 36 | 82 | 98.6 | 90.8 | 44.2 | 1.4 | 9.2 | 55.8 | 0.8 | 44.8 |
| | 0.5 | 41 | 91 | 98.7 | 92.0 | 46.2 | 1.3 | 8.0 | 53.8 | 0.9 | 44.2 |
| | 0.5 | 46 | 81 | 98.6 | 91.2 | 45.4 | 1.4 | 8.8 | 54.6 | 0.8 | 44.8 |
| $a \approx 130$AU (9:1) | 0.4 | 36 | 82 | 98.0 | 89.2 | 47.6 | 2.0 | 10.8 | 52.4 | 0.9 | 45.4 |
| | 0.5 | 36 | 82 | 96.8 | 87.9 | 50.0 | 3.2 | 12.1 | 50.0 | 1.0 | 49.2 |
| | 0.5 | 41 | 83 | 97.3 | 88.5 | 59.8 | 2.7 | 11.5 | 40.2 | 1.5 | 48.1 |
| | 0.5 | 41 | 91 | 97.9 | 89.8 | 54.2 | 2.1 | 10.2 | 45.8 | 1.2 | 43.5 |
| | 0.7 | 20 | 87 | 91.7 | 65.4 | 34.8 | 8.3 | 34.6 | 65.2 | 0.5 | 51.4 |
| | 0.7 | 40 | 87 | 96.6 | 84.2 | 39.7 | 3.4 | 15.8 | 60.3 | 0.7 | 47.4 |
| $a_P \approx 140$AU (10:1) | 0.7 | 36 | 87 | 95.1 | 77.9 | 35.7 | 4.9 | 22.1 | 64.3 | 0.6 | 57.0 |
| $a \approx 158$AU (12:1) | 0.5 | 21 | 83 | 92.7 | 72.1 | 37.8 | 7.3 | 27.9 | 62.2 | 0.6 | 54.6 |
| | 0.5 | 40 | 82 | 97.0 | 89.4 | 52.0 | 3.0 | 10.6 | 48.0 | 1.1 | 47.4 |
| | 0.7 | 40 | 87 | 95.4 | 85.0 | 51.2 | 4.6 | 15.0 | 48.8 | 1.0 | 48.3 |
| | 1.0 | 35 | 92 | 91.2 | 72.4 | 30.5 | 8.8 | 27.6 | 69.5 | 0.4 | 51.5 |
| | 1.0 | 50 | 92 | 94.9 | 80.5 | 46.0 | 5.1 | 19.5 | 54.0 | 0.9 | 47.4 |
| $a \approx 207$AU (18:1) | 0.7 | 30 | 87 | 94.1 | 80.0 | 38.4 | 5.9 | 20.0 | 61.6 | 0.6 | 56.2 |
| | 1.0 | 30 | 90 | 86.3 | 65.6 | 34.8 | 13.7 | 34.4 | 65.2 | 0.5 | 56.1 |
| $a \approx 250$AU (24:1) | 0.5 | 30 | 86 | 98.9 | 88.4 | 58.1 | 1.1 | 11.6 | 41.9 | 1.4 | 48.6 |
| | 0.7 | 30 | 88 | 95.3 | 79.1 | 43.0 | 4.7 | 20.9 | 57.0 | 0.8 | 53.9 |
| | 1.0 | 30 | 92 | 89.2 | 67.6 | 32.0 | 10.8 | 32.4 | 68.0 | 0.5 | 72.9 |
| No outer planet [e] | - | - | - | 99.9 | 99.0 | 93.2 | 0.1 | 1.0 | 6.8 | 13.7 | 40.6 |
| No outer planet - extra [f] | - | - | - | | | 88-95 | | | 5-12 | 7.33-19.0 | 41.6 |
| | | | | | | 80 | | | 20 | 4.0 | 41.4 |
| | | | | | | ~84-88 | | | ~12-16 | 5.25-7.33 | 41.5 |
| Observations apparent [g] | - | - | - | - | - | ≈90 | - | - | ≈10 | ≈9.0 | >41.3 |

[a] $M_P$ = mass in Earth masses, $i_P$ = inclination, and $q_P$ = perihelion distance of the outer planet.

[b] We evolved a disk of 2475 particles initially on Neptune-encountering orbits ($q < 35$ AU) set between 50 and 500 AU. $P_{scat}$ and $P_{det}$ represent the proportion of scattered and detached particles ($q > 39$ AU, $q > 39.5$ AU, and $q > 40$ AU) to the total population at 200 Myr, 700 Myr, and 4 Gyr, respectively (beyond 48 AU).

[c] Ratio of scattered to detached populations at the end of 4 Gyr (beyond 48 AU).

[d] Median perihelion distance of the detached population at 4 Gyr.

[e] Results of simulations in which only the four giant planets were included (without planet migration).

[f] Results of extra simulations in which only the four giant planets were included. The first line refers to cumulative results of runs with 29000 particles that started on Neptune-encountering orbits ($q < 35$ AU; $a < 50$ AU) and were evolved over 4 Gyr. The second line refers to a single run executed for 100 Myr

(planet migration) + 3.9 Gyr (long-term evolution) using 37200 particles. Neptune migrated from 20 to 30 AU and the disk was initially set at 19–30 AU in cold orbital conditions. The last line shows the combined results of both investigations.

[g] Computed with the identification of 72 scattered and 9 detached TNOs (LM07b). Long-term scattered disk resonant TNOs were excluded from this list (Lykawka & Mukai 2007a). Because of severe observational biases, the observed detached population is underrepresented so that the intrinsic ratio of scattered to detached TNOs is expected to be approximately ≤1.0 (e.g., Gladman et al. 2002). For the same reason, the median of this population should be >41.3 AU because detached TNOs with greater perihelia are discriminated against discovery.

Table 7

**Best *r*:1 Resonance Candidates for the Outer Planet**

| Resonance | Current semimajor axis (AU) | $Q_P$ [a] (AU) | $d_P$ [b] (AU) | $a_{N0}$ [c] (AU) |
|---|---|---|---|---|
| 6:1 | 99.4 | 118.8 | ≈101.3 ($e_P$≈0.195) | 19-24 |
| 7:1 | 110.1 | 140.4 | ≈114.3 ($e_P$≈0.274) | 17-21 |
| 8:1 | 120.4 | 160.9 | ≈127.2 ($e_P$≈0.336) | 15-20 |
| 9:1 | 130.2 | 180.5 | ≈139.9 ($e_P$≈0.386) | 15-18 |
| 10:1 | 139.7 | 199.5 | ≈152.5 ($e_P$≈0.428) | 15-17 |
| 11:1 | 148.9 | 217.9 | ≈164.9 ($e_P$≈0.463) | 15-16 |
| 12:1 | 157.8 | 235.6 | ≈177.0 ($e_P$≈0.493) | 15 |
| 13:1 | 166.4 | 252.8 | ≈188.9 ($e_P$≈0.519) | <15 |
| 14:1 | 174.8 | 269.6 | ≈200.5 ($e_P$≈0.542) | <15 |

[a] Aphelion of a hypothetical planet in the resonance assuming perihelion $q_P$ = 80 AU.

[b] Average distance of a hypothetical planet in the resonance assuming perihelion $q_P$ = 80 AU, calculated using $d_P = a_P\left(1 + \dfrac{e_P^2}{2}\right)$, where $a_P$ = semimajor axis and $e_P$ = eccentricity of the outer planet.

[c] Initial semimajor axis of Neptune for the resonance to be within ~60-80AU before migration.